\documentclass[fleqn,10pt]{wlscirep}
\usepackage[utf8]{inputenc}
\usepackage[T1]{fontenc}
\usepackage{xcolor}
\usepackage{lineno}
\usepackage{framed}
\usepackage{xr}
\usepackage{xr-hyper}

\usepackage[final]{changes}
\title{Disaster-induced behavioral change restructures social networks toward bonding ties}

\author[1,2]{Vaidehi Raipat}
\author[3]{Daniel Aldrich}
\author[1,2,*]{Takahiro Yabe}
\affil[1]{Department of Technology Management and Innovation, Tandon School of Engineering, New York University, Brooklyn, NY 11201, USA}
\affil[2]{Center for Urban Science and Progress, Tandon School of Engineering, New York University, Brooklyn, NY 11201, USA}
\affil[3]{Department of Political Science and Policy School, Northeastern University, Boston, MA 02115, USA}
\affil[*]{corresponding author: takahiroyabe@nyu.edu}

\begin{abstract}
Population displacement following environmental shocks reshapes the spatial organization of social interactions, often fragmenting existing ties and weakening community cohesion. 
Although social capital is widely recognized as a key determinant of resilience, its dynamic restructuring after disruption remains poorly quantified. 
Here, we develop a spatially embedded, dynamic network framework that operationalizes social capital as a network of repeated encounter opportunities inferred from large-scale mobility data. We construct temporal co-presence networks at third places to track how socio-spatial networks reorganize under disruption. 
We apply this framework to communities affected by the 2021 Marshall Fire in Colorado. 
We find that disaster-induced displacement leads to substantial contraction of socio-spatial networks, with mean weighted degree decreasing by 48\%. 
To isolate underlying mechanisms, we develop two counterfactual models: a random node removal model and a behaviour-informed model in which individuals are removed based on their estimated propensity to evacuate. 
Both counterfactuals predict substantially lower connectivity (37\% and 28\% lower mean weighted degree, respectively)
than observed, indicating that post-disaster connectivity remains systematically higher than expected based on displacement behavior alone. 
Structural analysis of the network reveals that 
this residual connectivity is disproportionately concentrated among bonding ties between sociodemographically similar individuals, while bridging ties are comparatively fragile. 
Furthermore, interaction becomes increasingly located around third places, suggesting that these places act as spatial anchors for the persistence of social ties under disruption. 
Together, these findings provide a first empirical view of how behavioral responses to disruption shape community resilience through the reorganization of social networks. 
\end{abstract}

\begin{document}

\flushbottom
\maketitle
%  Click the title above to edit the author's information and abstract
%
\thispagestyle{empty}

\section*{Introduction}

% ===== intro organization
% para 1: problem: social networks are important for xyz, but much less is know about how they restructure after disasters
% para 2: social capital is a key determinant of resilience and has been studied extensively. however, these are treated statically and how they dynamically evolve when beahviroal patterns are disrupted is less well known
% para 3: with novel mobility / social network (eg Facebook) data they have been studied using co-presence in space. general temporal differences have been studied but not particularly before/after environmental disasters. how networks restructure, what spatial anchors drive change, etc. are not understood.  
% para 4: in this study, we ... (short summary of what we did + found)

Social networks are central to recovery and resilience of disaster affected communities because they shape how information, trust, resources, and support circulate within communities under stress \cite{aldrich2012building, chetty2022social, sadri2017role, aldrich2015social}. Disasters not only damage physical and economic infrastructure; they also disrupt the social relations through which communities coordinate response, recovery, and adaptation \cite{yabe2020understanding, cutter2008place}. 
Although many studies identify social capital as a key determinant of community resilience, recovery, and collective action ~\cite{aldrich2012building, chetty2022social, sadri2017role, aldrich2015social}, much less is known about how social networks evolve and restructure because of disaster induced population dynamics such as displacement, return, relocation and migration \cite{fussell2014recovery, yabe2020understanding}. 
% Environmental shocks may therefore alter not only who remains in place, but also who connects with whom, and the geographical or place based conditions through which interaction opportunities are restructured.
In particular, it remains unclear whether disruption fragments social structure uniformly, or whether it induces systematic changes in the composition and spatial embedding of social ties. 

\added{A central tension in social capital research concerns whether social capital should be understood as a stock, that is, a durable attribute of communities that can be measured at a point in time, or as a process that is continuously produced and reproduced through ongoing social practices \cite{aldrich2015social}. Most empirical work in disaster research has adopted the stock perspective, treating social capital as a pre-existing community resource that either protects populations or fails to do so \cite{chetty2022social,fussell2014recovery, kyne2020capturing}. Survey-based measures, aggregate census proxies, and composite indices such as the Social Capital Index (SoCI) \cite{kyne2020capturing} have advanced this approach significantly, enabling cross-sectional comparisons across counties and identifying structural correlates of resilience. However, the stock perspective cannot capture how social capital is actively sustained, eroded, or reorganized when the routines and spatial patterns that produce social ties are disrupted. Chamlee-Wright \cite{chamlee2010cultural} argued that social capital is not passively stored but is maintained through the cultural and economic practices of everyday life: practices that disasters fundamentally disrupt.}
The increasing availability of large-scale mobility data~\cite{blondel2015survey} now makes it possible to directly observe patterns of co-presence and infer interaction opportunities at fine spatial and temporal resolution \cite{eagle2009inferring,genois2018can,sekara2014strength}. 
Recent studies have used such data to uncover experienced inequalities and place-based dependencies in urban encounters \cite{nilforoshan2023human,moro2021mobility,yabe2023behavioral,tizzoni2014use}. Yet, in the context of environmental shocks, it remains unclear how these encounter networks evolve through disruption: whether connectivity fragments or selectively persists, whether surviving ties become concentrated within specific types of relationships, and how behavioral shifts reshape the places through which social capital is produced and maintained.

\added{Building on a network-based perspective, we conceptualize social capital as the structure of social ties that emerges 
from repeated opportunities for encounter in shared spaces \cite{aldrich2012building, chamlee2010cultural, burt2000network, lin2017building}. Three complementary theoretical traditions support this operationalization. First, Granovetter’s \cite{granovetter1973strength} work on the strength of ties established that tie strength is a function of the frequency and duration of contact, emotional intensity, and reciprocal exchange, that is, properties that are directly shaped by physical co-presence. Repeated spatial and temporal overlap at shared locations increases the probability that latent ties become activated and that weak acquaintanceships develop into meaningful social connections \cite{eagle2009inferring, crandall2010inferring}. Second, Feld’s \cite{feld1981focused} focus theory of social ties argues that social relationships are organized around shared foci of activity: voluntary associations, recreational spaces, and other settings that structure who encounters whom. Third places, in this framework, are precisely such foci: they are neither home nor work, but discretionary social settings that organize the opportunity structure for tie formation \cite{oldenburg1999great}. Third, Coleman’s \cite{coleman1988social} theory of closure predicts that dense, repeated interactions in bounded settings generate trust and norms of reciprocity. Co-presence at the same third place over multiple occasions creates exactly the conditions Coleman theorized: individuals encounter each other’s contacts, triadic structures form, and the monitoring and norm enforcement that sustain cooperative behavior become possible. Together, these three mechanisms, propinquity-driven tie strengthening, focus-organized encounter opportunities, and closure-generated trust, provide a rigorous theoretical foundation for inferring social tie structures from co-presence patterns at third places.} \textcolor{black}{While the rapid growth of digital communication has enabled information exchange and remote emotional support across distance, the primary aim of this study is to understand how social interaction opportunities reorganize spatially when disasters redistribute populations, disrupt daily routines, and fundamentally alter mobility patterns. We therefore focus specifically on physical social ties, as they represent spatially embedded opportunities for interaction that emerge through repeated co-presence within shared environments. Such in-person encounters are particularly important during periods of disruption and recovery because physical co-presence uniquely facilitates spontaneous interaction, weak-tie formation, community visibility, collective efficacy, trust formation, local mutual aid, and place attachment \cite{granovetter1973strength, coleman1988social , putnam2000bowling, oldenburg1999great, klinenberg2018palaces, sampson1997neighborhoods}.}

In this study, we build on this theoretical framework by using proximity-based social networks inferred from large-scale mobility data, to examine how environmental shocks restructure social networks, focusing on the following key gaps. 
First, there is limited understanding of how displacement and spatial disruption translate into changes in social network structure, including the decay, persistence, or selective reinforcement of ties over time. 
This is particularly important because bonding ties, which are rooted in homophily and local closure, and bridging ties, which connect individuals across social or structural boundaries, may respond differently to disruption \cite{putnam2000bowling, aldrich2012building, mcpherson2001birds, coleman1988social, granovetter1973strength}. 
Second, the role of place in stabilizing or reshaping social networks after disasters remains poorly understood. Social infrastructure such as libraries, cafés, parks, and community centers provide opportunities for repeated co-presence, and have increasingly been used to infer latent social ties and characterize the spatial organization of social networks \cite{crandall2010inferring, eagle2009inferring, cho2011friendship}. However, their role in restructuring social networks after disruptions has not been systematically and quantitatively analyzed. Addressing these gaps is essential for understanding recovery not only as the return of populations, but as the reorganization of social encounter networks through which social capital is produced and maintained.

We construct a dynamic, spatially embedded network of social capital by inferring individual-level socio-spatial ties from repeated co-presence at third places before and following the 2021 Marshall Fire in Colorado \cite{tizzoni2014use, cattuto2010dynamics, stopczynski2014measuring, wang2011human, crandall2010inferring, eagle2009inferring}. 
\textcolor{black}{Importantly, this framework does not assume that every observed co-presence constitutes a direct friendship tie. Rather, repeated co-presence at shared third places captures structured opportunities for social interaction through which familiarity, weak ties, trust, and community embeddedness may emerge over time \cite{eagle2009inferring, milgram1972familiar, cattuto2010dynamics, stopczynski2014measuring, crandall2010inferring}. 
Additionally, robustness analyses using substantially stricter recurrence criteria (10\,m 20\,m, 30\,m, 40\,m and 50\,m spatial thresholds, 5\,min and 10\,min temporal thresholds), including repeated co-presence across multiple spatially distinct third places (Supplementary Section 1.5), produce highly similar structural and spatial patterns, suggesting that the inferred networks capture persistent and socially embedded interaction opportunities rather than primarily incidental overlap.}
We restrict the analysis to individuals observed in the area before the disaster, such that the post-disaster network reflects changes within the original exposed population.
We compare observed post-disaster networks against two counterfactuals: a displacement-controlled random-removal model and a displacement behavior-informed model that incorporates heterogeneous displacement propensity. 
We find that socio-spatial networks contract sharply following the disaster, yet remain systematically more robust than expected under either counterfactual. This residual robustness is disproportionately concentrated within bonding ties, whereas bridging ties are significantly more fragile. 
We further show that post-disaster interactions become selectively concentrated in a limited subset of third places, including community-oriented, recreational, and food-related venues, consistent with a place-based mechanism through which repeated co-presence reinforces existing, socially embedded ties after disruption. 
These results remain robust across all spatial and temporal thresholds.
Together, these results provide a dynamic account of social capital in which resilience depends not only on the volume of connectivity, but on how social ties are selectively preserved and spatially reorganized following disruption. 
These findings provide a foundation for designing interventions that support community recovery and resilience by sustaining the encounter structures through which social ties are maintained and rebuilt.

\section*{Results}

We construct individual-level, spatially embedded, evolving temporal social networks, inferred from large-scale mobility data derived from anonymized GPS trajectories provided by Cuebiq, and based on co-location at third places (See Supplementary Material sections 1.1 and 1.2 for details about mobility data). We first identify user stops at \textit{third places} which include cafes, parks, libraries, and community centers, following established definitions of socially neutral spaces that facilitate repeated interaction \cite{oldenburg1999great, soja2008thirdspace}. Third places are selected from the Global Places and Geometry dataset provided by SafeGraph, which contains points of interest (POIs) and building footprints for non-residential locations (See Supplementary Material Section 1.3 for details on third place identification). 
We focus specifically on the reorganization of social-spatial ties among individuals who were present in the study area before the disaster. The post-disaster network is therefore defined using only pre-disaster residents who remain observable post-disaster rather than incorporating in-migrants or newly observed entrants during recovery. This design enables us to isolate how disruption reshapes the existing interaction structure, including tie persistence, decay, and selective rewiring within the pre-disaster population.

We spatially join user stops to third-place POI polygons to identify the spatial context of potential interaction opportunities. We define a repeated co-presence tie between two users based on their co-location within shared third places, and weight each tie by the frequency of repeated co-presence. Spatially, co-presence is identified when two users are located at the same point of interest (POI) within 30 meters, and with a temporal overlap of more than 5 minutes. Only spatial and temporal overlaps during weekends or after work hours during weekdays (18:00–23:00) are analyzed to focus on discretionary activities more likely to reflect social exposure. We also exclude POIs located within each user’s home and work Census Block Group (CBG) to remove home- and work-related co-presence (See Supplementary Material Section 1.4 for details on home and work identification). For each tie, edge weights are defined as the frequency of co-presence events between user pairs at shared third places over a one-month period (Fig.~\ref{fig:Figure 1}a). 

To test the robustness of this network construction, we test alternative spatial thresholds of \{10, 20, 30, 40, and 50\} meters and minimum temporal overlap thresholds of \{5, 10, and 15\} minutes. We find that the resulting network structure is stable across these thresholds (See Supplementary Section 1.5, Figure S2, S3). We therefore use 30 meters and 5 minutes as the baseline spatial and temporal thresholds in the main analysis, while results using alternative spatial and temporal thresholds are reported in Supplementary Section 1.5.  

% To further test whether We also construct another, more restricted network to evacuate if out selected baseline (30\,m and 5\,min) network primarily emphasizes repeated interaction opportunities occurring across multiple spatially distinct social contexts. 
To further assess whether our co-location-based tie definition (within 30 meters within same POI, and at least 5 minutes overlap) captures repeated opportunities for social interaction, rather than isolated or incidental physical encounters, we construct an additional, more restricted network.
In this specification, we retain only those dyads that are repeatedly co-located across at least two distinct POIs whose centroids are at least 100 meters apart (see Supplementary Material Section 1.5.2, Figures S4, S5, S6). This stricter definition is intended to identify dyads whose repeated co-presence occurs across multiple spatially distinct social contexts, making purely incidental encounters less likely. We find that, across both specifications, the dominant POIs and POI subcategories supporting repeated co-presence ties remain highly stable. This consistency provides additional confidence that our baseline network captures robust patterns of repeated social exposure, and we therefore use this definition for the main analysis.

We apply this framework to communities affected by the 2021 Marshall Fire in Colorado to construct individual-level spatio-temporal social networks before and after the disaster. Nodes represent individuals, and weighted edges encode the intensity of repeated co-presence at third places, yielding a sequence of weighted, undirected interaction networks over time. Monthly networks are constructed for two pre-disaster periods (October and November 2021) and two post-disaster periods (January and February 2022), with the disaster occurring between 30 December 2021 and 2 January 2022.

\subsection*{Disaster induced displacement contracts socio-spatial networks}
 
\begin{figure}[t]
    \centering
    \includegraphics[width=0.9\linewidth]{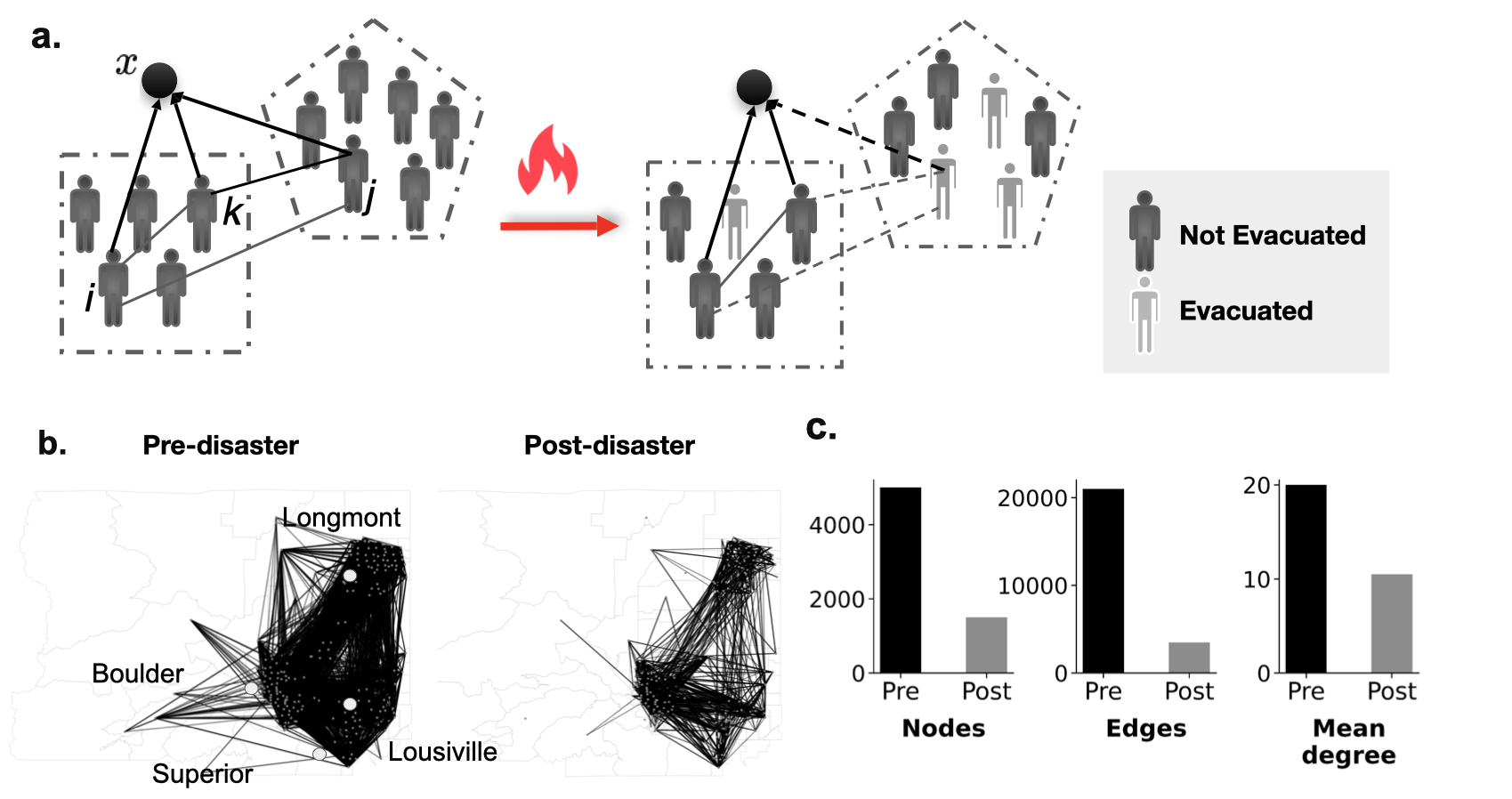}
    \caption{{\textbf{Disaster-induced displacement contracts socio-spatial co-presence networks.} a) Schematic of the inferred co-presence network: individuals (nodes) are connected by weighted ties representing repeated spatio-temporal co-location at shared third places. b) Empirical networks for Marshall Fire–affected communities in Colorado, visualized using the centroids of users’ home Census Block Groups (CBGs), illustrating pre-disaster structure (October–November 2021) and post-disaster structure (January–February 2022); the disaster occurred from 30 Dec 2021 to 2 Jan 2022. The post-disaster network shows pronounced contraction relative to the pre-disaster baseline. c) Summary network statistics for the pre- and post-disaster networks. Compared with the pre-disaster period, the post-disaster network contains fewer active individuals (4,816 versus 1,538 nodes), fewer co-location ties (20,971 versus 3,241 edges), and lower mean weighted degree (19.94 (95\% confidence interval: [19.50, 20.40]) versus 10.33 (95\% confidence interval: [9.88, 10.82])), indicating a sharp decline in repeated interaction opportunities after displacement. 
}}
    \label{fig:Figure 1}
\end{figure}

We quantify network dynamics using core structural measures, including the number of nodes, edges, and weighted degree. Degree is defined as the number of distinct social contacts connected to an individual, while weighted degree captures the cumulative intensity of interactions, representing opportunities to strengthen social capital through repeated contact, $k_i = \sum_j w_{ij}$. Importantly, repeated and temporally overlapping co-presence at shared third places is not treated as direct evidence of friendship. Rather, we conceptualize it as an exposure process that increases the probability of tie formation or maintenance by creating structured opportunities for interaction \cite{crandall2010inferring, granovetter1973strength}. Comparing these metrics across pre- and post-disaster networks, we find that post-disaster networks contract sharply and nonlinearly, with substantial node loss driven by displacement and corresponding declines in connectivity and centrality. Mean weighted degree decreases from $k = 19.94 $ (95\% confidence interval: [19.50, 20.40]) to $k = 10.33$ (95\% confidence interval: [9.88, 10.82]) , as shown in Figures \ref{fig:Figure 1}b and \ref{fig:Figure 1}c. 

We further compute the clustering coefficient, defined as the fraction of closed triads around a node, $CC_i = \frac{2T_i}{k_i(k_i - 1)}$, where $T_i$ denotes the number of triangles involving node $i$. This metric captures triadic closure, reflecting the extent to which an individual’s contacts are also connected, supporting trust, norm enforcement, and mutual support \cite{coleman1988social}. We also compute closeness centrality, defined as the inverse of the sum of shortest path distances from node $i$ to all other nodes, $CLC_i = \frac{1}{\sum_{j} d(i,j)}$, which reflects an individual’s embeddedness within the network and their potential to access information and support through short network paths \cite{freeman1978centrality}. These values are reported later with counterfactual networks in Figure \ref{fig:Figure 2}d, and similarly indicate a sharp post-disaster contraction. From $G_{\mathrm{pre}}$ to $G_{\mathrm{post}}$, clustering coefficient declines from 0.36 (95\% confidence interval: [0.35, 0.36]) to 0.26 (95\% confidence interval: [0.25, 0.27]) and closeness centrality from 0.45 (95\% confidence interval: [0.44, 0.45])  to 0.29 (95\% confidence interval: [0.28, 0.29]) .

\subsection*{Socio-spatial networks remain more connected than predicted by displacement behavior}

To evaluate whether post-disaster network contraction can be explained by displacement alone, we construct two complementary counterfactual baselines: (i) a displacement-controlled random node removal model and (ii) a behavior-informed node removal model. The random node removal model removes nodes from the pre-disaster network while conditioning on the observed displacement proportion within each census block group (CBG), thereby preserving the spatial heterogeneity of population loss but randomizing which individuals are removed within each CBG. Comparing the observed post-disaster network with this displacement-controlled null shows that realized connectivity remains substantially higher than expected under random node loss ($G_{\mathrm{random}}$: $k = 6.43$ (95\% confidence interval: [5.87, 7.05]), $CC_i = 0.19$ (95\% confidence interval: [0.18, 0.22]), $CLC_i = 0.16$ (95\% confidence interval: [0.14, 0.18])).

To build the behavior-based node removal mode, we estimate each individual’s displacement propensity using a logistic regression in which the dependent variable is the observed displacement indicator ($E_i = 1$ if displaced, $0$ otherwise). Displacement is modeled as a function of pre-disaster centrality (${C}_i^{\mathrm{pre}}$) and mobility behavior, including exploration rate and repeat visitation ($X_i^{\mathrm{pre}}$). We control for distance-related variables $D_i$, including the distance of the third place to disaster and average distance traveled to visit a third place, and socio-demographic attributes ${SDM}_i$: $\Pr(E_i = 1) \sim \mathrm{logit}^{-1}\!\left(\beta_{0} + \beta_{1} C_{i} + \beta_{2} D_{i} + \beta_{3} SDM_{i} + \beta_{4} X_{i}\right)$. These coefficients are shown in \ref{fig:Figure 2}a and, full results are reported in Supplementary Material Table S1. Conditional on socio-demographic attributes, individuals with weaker pre-disaster connectedness, measured by lower weighted degree, and lower embeddedness, measured by lower closeness centrality, are more likely to be displaced. Individuals whose third-place activity is less routine-reinforcing, reflected by lower concentration within similar place categories, and who travel farther to access third places also exhibit higher displacement propensity.

We then use the fitted values, $\hat{p}_i = \Pr(E_i = 1)$, as propensity scores to construct a behavior-informed counterfactual model of node loss. As in the displacement-controlled random model, this counterfactual preserves the observed number of displaced individuals within each census block group (CBG). However, instead of removing nodes at random, it samples displaced individuals with probabilities proportional to their predicted displacement propensity, $\hat{p}_i$. Specifically, nodes are removed without replacement within each CBG so that the number of removed nodes matches the actual displacement counts, while individuals with higher predicted displacement probabilities are more likely to be selected. The resulting counterfactual network is constructed by applying this behavior-informed node removal process to the pre-disaster graph. 

The behavior-informed counterfactual preserves more connectivity than the random baseline ($G_{\mathrm{behavior}}$: $k = 7.32$ (95\% confidence interval: [6.68, 8.01), $CC_i = 0.210$ (95\% confidence interval: [0.190, 0.231]), $CLC_i = 0.190$ (95\% confidence interval: [0.166, 0.214])), but remains substantially less connected than the observed post-disaster network. Figure \ref{fig:Figure 2}c compares the spatial structure of the observed post-disaster network with the behavior-informed counterfactual, while Figure \ref{fig:Figure 2}d compares centrality measures across the pre-disaster, observed post-disaster, and counterfactual networks. The gap between the observed post-disaster and the behavior-informed counterfactual indicates that observable behavioral, spatial, and socio-demographic factors are insufficient to explain the persistence of social connectivity. Instead, this residual connectivity points to higher-order structural or place-based mechanisms underlying selective network resilience. 
We test the robustness of this result across combinations of five spatial and two temporal co-location thresholds and find that the observed post-disaster graph remains systematically more connected than both counterfactual baselines under all specifications. This confirms that the main result is robust to both narrower and broader spatial and temporal definitions of co-location (See Supplementary Material Section 2, Figures S7 - S10). 

\begin{figure}[!t]
    \centering
    \includegraphics[width=.95\linewidth]{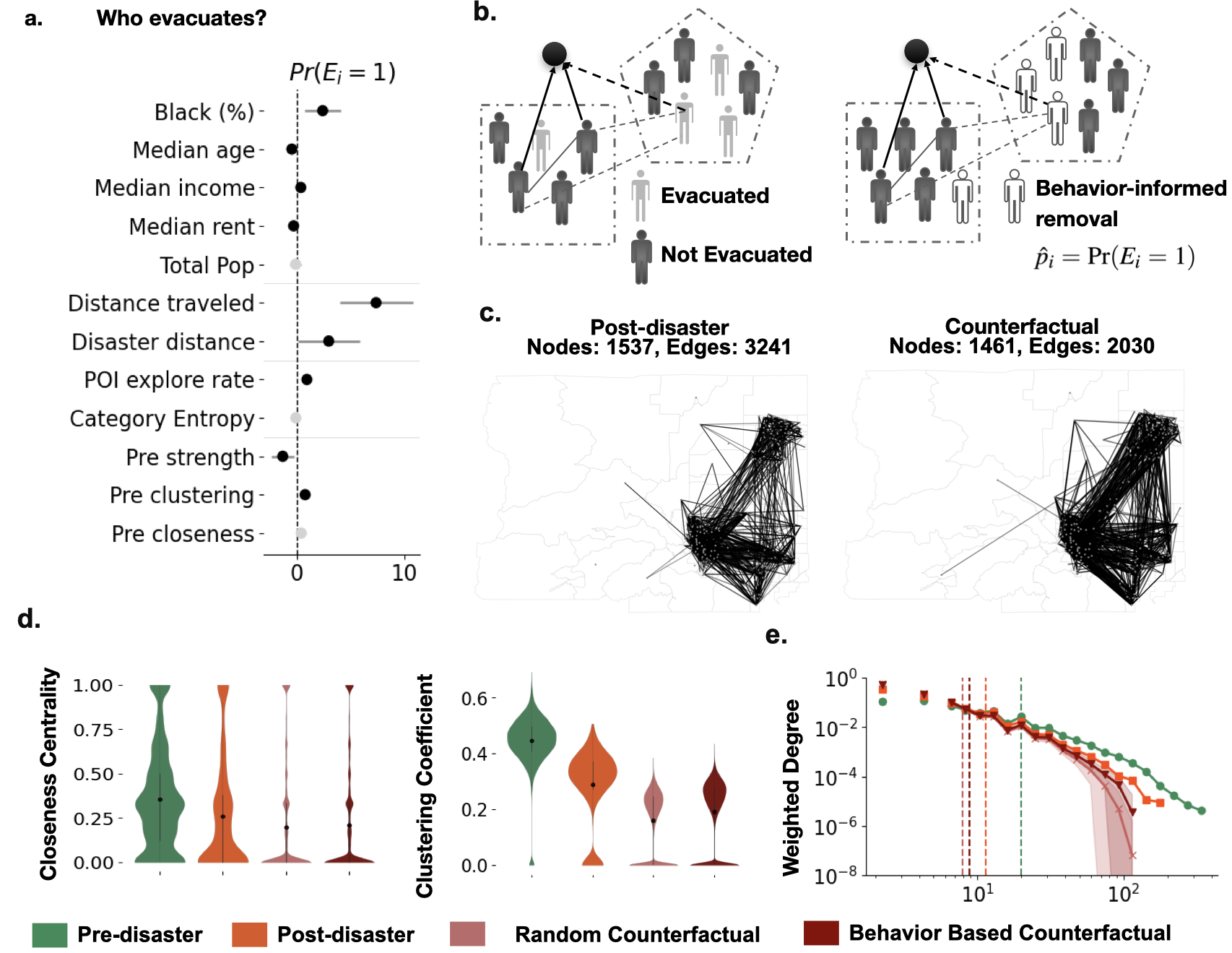}
    \caption{{\textbf{Disaster-induced network contraction is not proportional to displacement and cannot be explained by observable behavior alone.} a) Logistic regression coefficients for individual displacement propensity used as probabilities to construct the behavior-informed counterfactual. Lower pre-disaster connectedness, greater travel distance to third places, and less routine-reinforcing third-place activity are associated with higher displacement probability.   
    b) Schematic of the behavior-informed counterfactual, in which node removal is sampled from the pre-disaster network using fitted displacement propensities while matching observed displacement counts within census block groups (CBGs). 
    c) Observed post-disaster network and behavior-informed counterfactual network, visualized using the centroids of users’ home CBGs. 
    d) Node-level distributions of closeness centrality and clustering coefficient across the pre-disaster, post-disaster, random counterfactual, and behavior-informed counterfactual networks. 
    e) Node level distributions of weighted degree across the pre-disaster, post-disaster, random counterfactual, and behavior-informed counterfactual networks.
    Relative to both counterfactual baselines. The observed post-disaster network exhibits higher connectivity (weighted degree), greater local closure (clustering coefficient), and stronger embeddedness (closeness centrality), indicating that displacement and observable behavioral sorting alone are insufficient to explain the persistence of social connectivity after disaster.
}}
    \label{fig:Figure 2}
\end{figure}

\subsection*{Bonding ties persist while bridging ties disproportionately decay}

We next examine whether the residual connectivity, not explained by either random or behavior-informed counterfactual, is heterogeneously distributed across different types of social ties. Social capital theory distinguishes between two fundamental types of ties: bonding ties, which are cohesive and typically form among socially similar individuals, and bridging ties, which connect individuals across heterogeneous social groups and enable access to novel information and resources \cite{coleman1988social, burt2005brokerage, granovetter1973strength}. \added{Bonding ties are constituted by homophily, the systematic tendency of individuals to form and maintain ties with others who share their demographic characteristics, social position, and identity \cite{putnam2000bowling, mcpherson2001birds}. In the Putnam–Coleman tradition, this similarity is not incidental to bonding capital but definitional: it is precisely the shared identity and mutual recognition among similar others that generates the trust, reciprocity, and norm enforcement through which bonding ties produce their characteristic benefits \cite{coleman1988social}.}

We quantify homophily between user pairs using cosine similarity over socio-demographic attributes, specifically race and income. Individual-level attributes are not directly observed in the GPS location data, therefore, we probabilistically assign these attributes to users based on the demographic distributions of their home census block groups (CBGs). To account for uncertainty in this imputation, we bootstrap this assignment procedure 100 times and compute a distribution of homophily values for each dyad. A tie is classified as bonding if its mean homophily exceeds the median homophily across all ties in the pre-disaster network (see Methods).

\begin{figure}[!th]
    \centering
    \includegraphics[width=\linewidth]{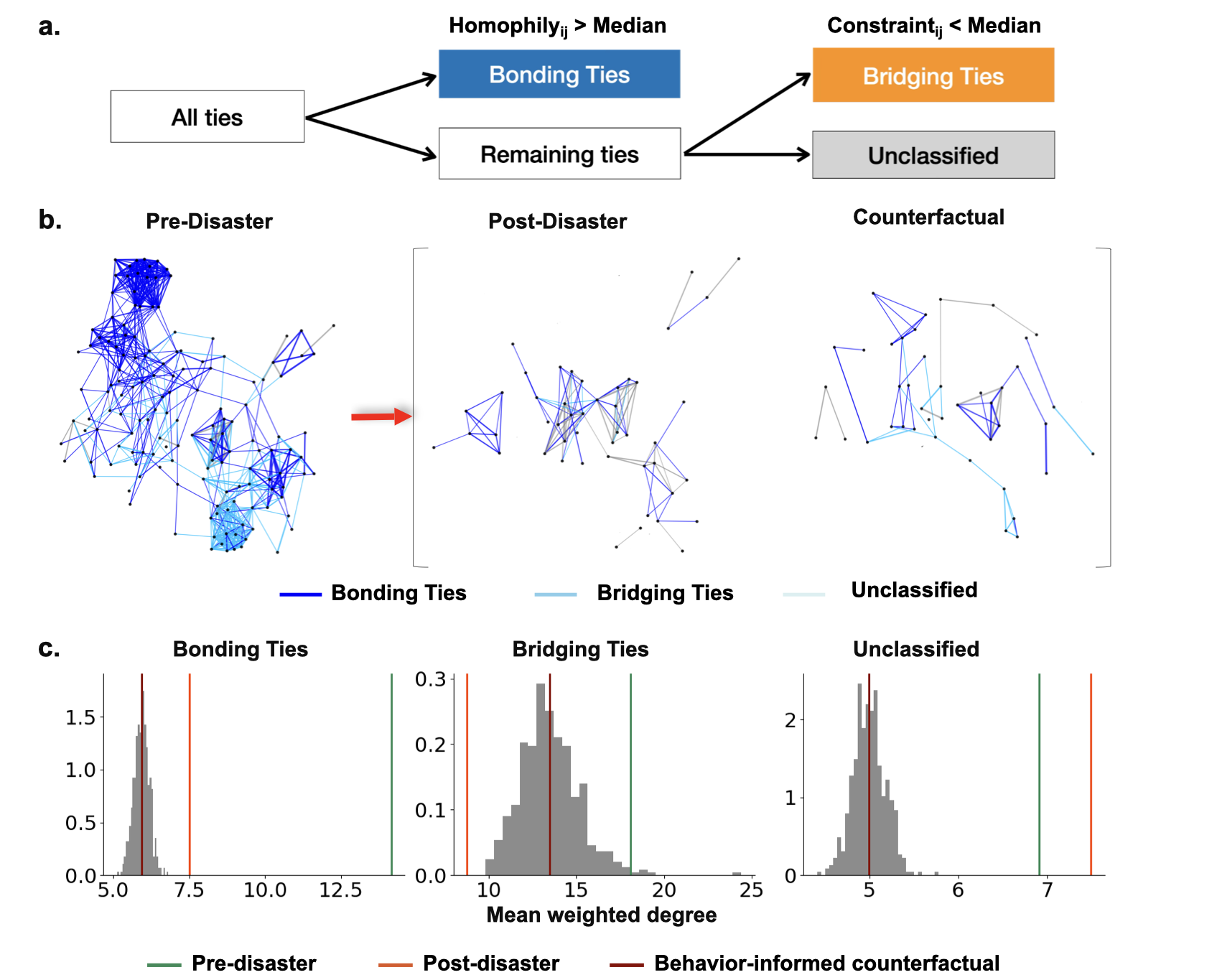}
    \caption{{\textbf{Bonding ties exhibit greater post-disaster robustness than bridging ties.} a) Schematic illustration of tie-type classification. Ties are first labeled bonding if mean dyadic homophily exceeds the median pre-disaster homophily. Among the remaining ties, edges with below-median edge-level local constraint are labeled bridging; the remaining are unclassified. b) Pre-disaster, post-disaster, and behavior-informed counterfactual networks visualized by tie type. Bonding ties dominate the observed network across phases, and the post-disaster network retains a larger bonding component than predicted by the behavior-informed counterfactual. c) Mean weighted degree distributions across 500 behavior-based counterfactual runs for the bonding, bridging, and unclassified subnetworks. For bonding ties, the counterfactual distributions are centered below the observed post-disaster values, indicating that these subnetworks remain more connected after the disaster than expected under the behavior-informed counterfactual. By contrast, for bridging ties, the counterfactual distribution lies above the observed post-disaster value, suggesting weaker-than-expected post-disaster bridging connectivity. Together, these results indicate that residual post-disaster robustness is concentrated disproportionately in bonding ties. 
}}
    \label{fig:Figure 3}
\end{figure}

\added{Bridging ties derive their distinctive value from spanning structural holes in the network between otherwise disconnected clusters through which novel information, diverse perspectives, and non-redundant resources flow \cite{burt2005brokerage}. A tie’s bridging potential is therefore not a property of the individuals it connects but of its structural position within the broader network: ties that lie on many shortest paths between otherwise distant nodes, and that connect individuals whose remaining contacts are not redundant with each other, occupy the brokerage positions that Burt \cite{burt2005brokerage} theorized as the primary source of social capital’s competitive and informational advantages. We operationalize this structural conception using} the local constraint metric, which measures the extent to which a tie is embedded within a redundant local neighborhood \cite{burt2005brokerage}. For an ordered pair $(u,v)$, local constraint is defined as $\ell(u,v)=\left(p_{uv}+\sum_{w\in N(v)} p_{uw}p_{wv}\right)^2$, where $p_{uv}=\frac{w_{uv}}{\sum_k w_{uk}}$, $w_{uv}$ denotes the weight of the edge between $u$ and $v$, and $\sum_k w_{uk}$ denotes the total weighted degree of node $u$, i.e., the sum of its edge weights across all neighbors. Because the graph is undirected, we compute a symmetric edge-level local constraint by averaging the two directional values for each edge. Low constraint indicates higher brokerage potential, as interactions are less confined to closed triadic structures. Among non-bonding ties, an edge is classified as bridging if its local constraint falls below the mean edge-level local constraint in the pre-disaster network and both of its endpoints (nodes $u$ and $v$) have a degree of at least 3. All remaining non-bonding ties are categorized as unclassified ties (Fig.~\ref{fig:Figure 3}a). 

We split the pre-, post-, and behavior-informed counterfactual networks into bonding, bridging, and unclassified edges, and compute network statistics separately for each tie type.
Across all networks, bonding ties are more numerous than bridging ties, but their structural trajectories differ markedly after the disaster. In the pre-disaster network, bridging ties have the highest mean weighted degree ($k = 17.95$, 95\% CI: 17.28–19.03), followed by bonding ties ($k = 13.91$, 95\% CI: 13.79–14.51) and unclassified ties ($k = 6.92$, 95\% CI: 6.76–7.08). After the disaster, connectivity declines across tie types, but bonding ties remain relatively strong ($k = 7.49$, 95\% CI: 7.21–7.86), while bridging ties fall to a comparable level ($k = 8.91$, 95\% CI: 7.89–9.72). The behavior-informed counterfactual fails to reproduce this pattern: it underestimates the connectivity of bonding ties ($k = 5.93$, 95\% CI: 5.43–6.41) while overestimating the connectivity of bridging ties ($k = 13.48$, 95\% CI: 10.67–17.13), relative to the observed post-disaster network (Figure \ref{fig:Figure 3}c). This divergence indicates that bonding ties exhibit greater resilience and adaptability following disruption than can be explained by heterogeneous displacement alone. By contrast, bridging ties appear more fragile and more susceptible to overestimation in counterfactual reconstructions, suggesting that disaster-induced behavioral change selectively preserves more locally embedded and socially proximate connections.

% across all three networks. We find that bonding ties substantially outnumber bridging ties across all networks. However, their structural evolution differs across phases. Comparing mean weighted degree, the behavior-informed counterfactual fails to reproduce the observed structure of bonding ties, underestimating their connectivity, while overestimating the connectivity of bridging ties ($G_{\mathrm{pre}}$: $k_{\mathrm{bonding}} = 13.91$ (95\% confidence interval: [13.79, 14.51]), 
% $k_{\mathrm{unclassified}} = 6.92$ (95\% confidence interval: [6.76, 7.08]), 
% $k_{\mathrm{bridging}} = 17.95$ (95\% confidence interval: [17.28, 19.03]); 
% $G_{\mathrm{post}}$: $k_{\mathrm{bonding}} = 7.49$ (95\% confidence interval: [7.21, 7.86]), 
% $k_{\mathrm{unclassified}} = 7.51$ (95\% confidence interval: [7.08, 7.86]), 
% $k_{\mathrm{bridging}} = 8.91$ (95\% confidence interval: [7.89, 9.72]); 
% $G_{\mathrm{Behavior-informed-cf}}$: $k_{\mathrm{bonding}} = 5.93$ (95\% confidence interval: [5.43, 6.41]), 
% $k_{\mathrm{unclassified}} = 5.00$ (95\% confidence interval: [4.64, 5.34]), 
% $k_{\mathrm{bridging}} = 13.48$ (95\% confidence interval: [10.67, 17.13])) (Figure\ref{fig:Figure 3}c). This indicates that bonding ties exhibit greater resilience and adaptability following disruption than predicted by behavior-driven mechanisms alone, whereas bridging ties are more fragile and more susceptible to overestimation in counterfactual reconstructions. 

To test the robustness of these insights to ho we define bonding and bridging ties, we replicated these results for 5 spatial and 2 temporal co-location thresholds and additionally use various combinations of homophily (50, 70, and 85 percentiles) and constraint (25, 35, and 50th percentiles) values for categorizing ties into bonding and bridging, respectively. We find that bonding ties show higher post disaster robustness under both stricter as well as broader thresholds, however bridging ties do not exhibit similar dissimilarity in robustness under stricter spatial thresholds due to data sparsity at smaller spatial scales (see Supplementary Material, Section 3, Figures S12 - S41).  

\added{The differential resilience of bonding and bridging ties under disruption can be understood through three complementary theoretical mechanisms. First, Putnam \cite{putnam2000bowling} distinguished bonding capital as the resource people draw on to “get by” providing emotional support, mutual aid, and immediate assistance while bridging capital enables people to “get ahead” through access to novel information and diverse resources. Under acute stress, the functional demands of crisis sharply elevate the value of “getting by”: displaced individuals need shelter, food, emotional comfort, and logistical help, all of which flow most readily through trusted, close, and familiar contacts. This creates a functional selection pressure that prioritizes bonding encounters over bridging ones.
Second, the homophily principle \cite{mcpherson2001birds} predicts that sociodemographically similar individuals tend to occupy similar residential locations, frequent similar establishments, and face similar constraints on mobility. When a disaster contracts everyone’s activity space, homophilous pairs are mechanically more likely to converge at the same reduced set of nearby venues, simply because they share spatial routines, residential proximity, and socioeconomic circumstances that shape venue choice. Bridging ties, by contrast, often depend on encounters in more distant or specialized locations, such as workplaces, cross-neighborhood organizations, or culturally diverse commercial districts, that are more likely to be disrupted or rendered inaccessible by displacement.
Third, bonding ties carry stronger norms of obligation and reciprocity than bridging ties \cite{aldrich2012building, coleman1988social}. Dense, closed networks enforce social expectations of mutual support: individuals feel duty-bound to maintain contact with family, close friends, and neighbors who share their identity and circumstances. 
% Weak ties, by definition, lack this normative enforcement mechanism.
When the costs of maintaining contact increase, as they do when displacement disrupts routines and increases travel distances, individuals rationally shed their least obligatory connections first. 
% The bridging-tie decay we observe is thus the network-structural signature of this social obligation gradient under conditions of constrained mobility.
}

\added{However, the disproportionate persistence of bonding ties is not unambiguously positive for community resilience. Studies \cite{portes1998social,portes2014downsides} have documented the ``downside of social capital”: tightly bonded networks can become exclusionary, hoarding resources within in-groups while restricting information flows to out-group members. In disaster contexts, this Janus-faced quality of social capital \cite{aldrich2012building, aldrich2015social} means that the bonding consolidation we observe may simultaneously sustain mutual aid within homophilous clusters and impede the cross-group coordination, diverse information access, and equitable resource distribution that bridging ties facilitate. Communities that lose bridging ties disproportionately risk becoming more insular, less adaptive, and more vulnerable to the reinforcement of pre-existing social and economic inequalities during recovery \cite{lin2017building, putnam2000bowling}. The selective erosion of bridging ties documented here thus carries implications not only for network structure but for the equity of recovery outcomes.
}

\subsection*{Third places anchor the persistence of bonding ties after disaster}

Building on the finding that residual connectivity is not explained by either counterfactual models and is disproportionately concentrated within bonding ties, we next test if this robustness is associated with a place-based restructuring mechanism whereby more frequented third places or anchor institutions sustain encounter opportunities in the aftermath of a disaster. Consistent with theories of closure and selective resilience under disruption \cite{almaatouq2020adaptive, jackson2008social}, we hypothesize that post disaster encounter opportunities are specifically concentrating in particular third places that function as local anchor institutions. 

% To evaluate this, we compute standardized residual changes in interaction potential intensity at each third place POI relative to a behavior-informed counterfactual, defined as $z_{i}=\frac{\log(\mathrm{post}_i)-\mu(\log(\mathrm{rand}_i))}{\sigma(\log(\mathrm{rand}_i))}$, 
% where $\mu(\cdot)$ and $\sigma(\cdot)$ denote the mean and standard deviation in the behavior-informed counterfactual. We compute this for all ties, bonding ties, bridging ties and unclassified ties. We then fit three linear models of the form: $z_{i}=\beta_0+\beta_1 D_i+\beta_2 X_i+\beta_3 SDM_i+ \gamma_c +\epsilon_i$, where where $D_i$ captures distance-related variables (distance to disaster and average travel distance), $X_i$ denotes pre-disaster visitation behavior, $SDM_i$ represents socio-demographic attributes, and $\gamma_c$ are POI category fixed effects. We show the key regression coefficient results for bonding ties in Figure \ref{fig:Figure 4}, and report the corresponding regression models in the Supplementary Material Table S4. We also visualize $z_{i}$ on an interactive Map in Supplementary Material Section 4.1.

\begin{figure}[t]
    \centering
    \includegraphics[width=\linewidth]{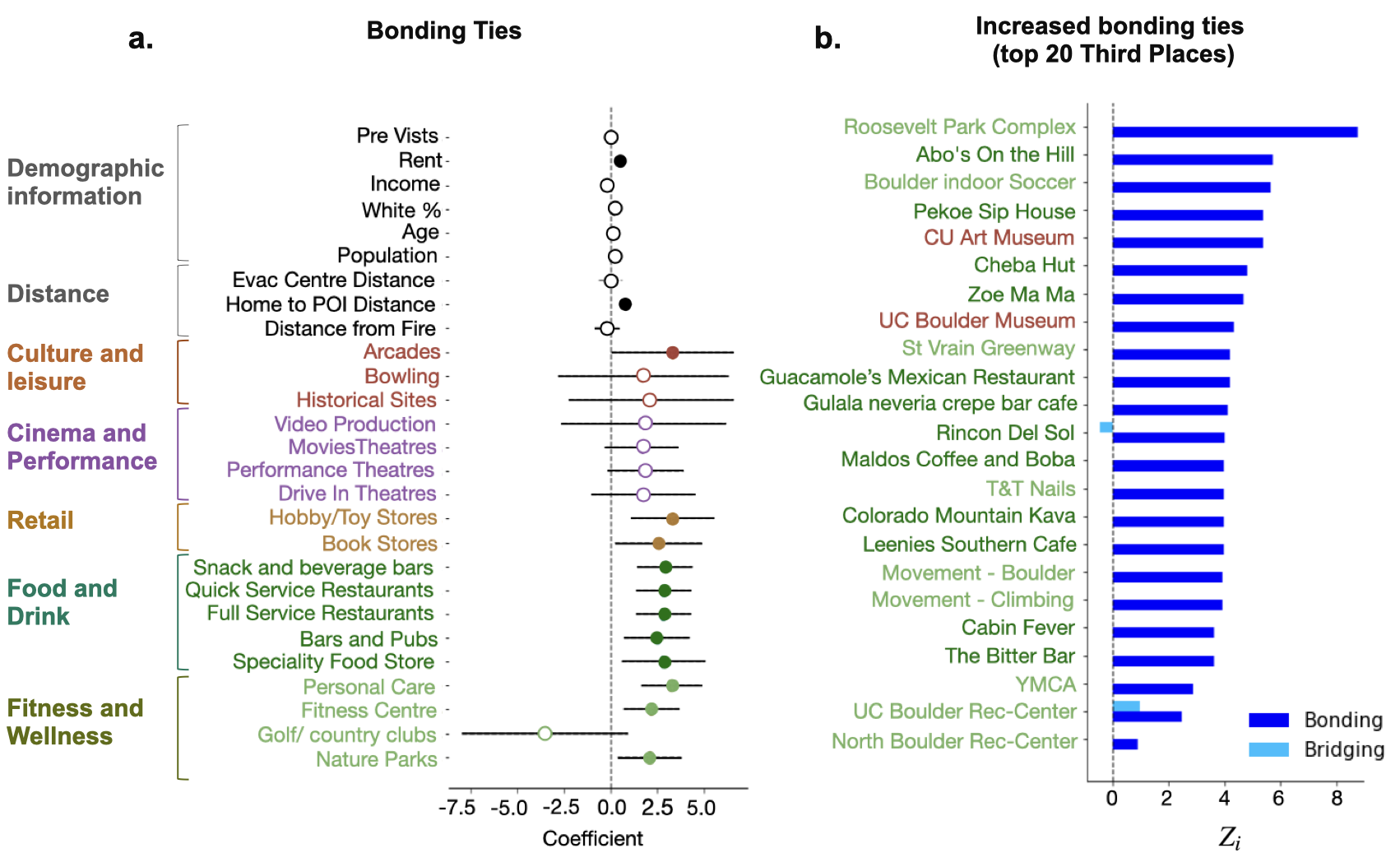}
    \caption{{\textbf{Third places anchor the persistence of bonding ties after disaster.} a) POI-level regression coefficients for bonding ties, where the dependent variable is the standardized residual change in interaction intensity relative to the behavior-informed counterfactual, such that positive values indicate observed post-disaster interaction above counterfactual expectations. The largest positive deviations are concentrated in recreation and leisure spaces, restaurants and bars and cultural venues. b) Top 20 individual third places ranked by residual increase in bonding-tie interaction intensity. Across these locations, bonding-tie residuals are consistently larger than bridging-tie residuals, indicating that post-disaster interaction becomes selectively concentrated in places that reinforce existing, socially similar ties rather than generalized across all forms of social contact.
}}
    \label{fig:Figure 4}
\end{figure}

To evaluate which third places exhibit greater post-disaster interaction potential than expected in the counterfactual model, we compute standardized residual changes in interaction potential intensity at each third-place POI relative to the behavior-informed counterfactual. Let $Y_i^{\mathrm{post}}$ denote the observed post-disaster interaction potential intensity at POI $i$, and let $Y_{i,r}^{\mathrm{cf}}$ denote the corresponding value in counterfactual realization $r$. We define the standardized residual as
$z_i =
\frac{
\log\left(Y_i^{\mathrm{post}}\right) -
\mu_r\left[\log\left(Y_{i,r}^{\mathrm{cf}}\right)\right]
}{
\sigma_r\left[\log\left(Y_{i,r}^{\mathrm{cf}}\right)\right]
},
$
where $\mu_r(\cdot)$ and $\sigma_r(\cdot)$ denote the mean and standard deviation across counterfactual realizations. We compute $z_i$ separately for all ties, bonding ties, bridging ties, and unclassified ties. We then fit linear models of the form
$
z_i \sim \beta_0 + \beta_1 D_i + \beta_2 X_i + \beta_3 S_i + \gamma_c
$
, where $D_i$ captures distance-related variables, including distance to the disaster perimeter and average travel distance to the POI; $X_i$ denotes pre-disaster visitation behavior; $S_i$ represents socio-demographic attributes of visitors; and $\gamma_c$ denotes POI category fixed effects. Figure~\ref{fig:Figure 4} reports the key regression coefficients for bonding ties, while the full model results are provided in Supplementary Table~S4. We also visualize the spatial distribution of $z_i$ in the interactive map in Supplementary Section~4.1. 
\added{Because the POI-level z-score is defined relative to each location’s own pre-disaster baseline, it already adjusts for pre-existing bonding intensity and therefore helps distinguish post-shock reorganization from the continuation of stable pre-shock co-presence patterns.}

We find that, across most third-place categories, observed post-disaster interaction levels exceed those predicted by the counterfactual. This pattern is strongest and most consistent for the bonding ties. The the largest positive residuals are concentrated in meaningful subset of third places, including community oriented third places such as cultural venues (museums), fitness and recreational places and especially restaurants and other eating places. Additionally, some officially assigned evacuation centers  or functionally similar resource-sharing locations, such as YMCA-related sites, also exhibit selectively elevated interaction. These patterns indicate that post-disaster persistence of social connectivity is not randomly redistributed across third places, but is instead selectively reinforced through a limited set of places that sustain repeated co-presence, especially among already connected ties.

%$$E_{ij} = w_{ij} \;\big|\; N_{POI_{ij}} \geq 2$$

\added{Beyond the bonding–bridging dyad, a third dimension, linking social capital, captures vertical relationships between individuals or communities and institutions that hold formal authority, resources, and decision-making power \cite{kyne2026development, szreter2004health}. Linking ties connect residents to government agencies, service providers, and organizations that can mobilize resources at scale. In disaster contexts, linking capital is essential for accessing emergency assistance, navigating bureaucratic recovery processes, and securing institutional support \cite{aldrich2012building, aldrich2015social}. While the mobility data employed in this study are best suited to capturing the horizontal dimensions of bonding and bridging, several of the third places in our analysis, including YMCA facilities, university sites, and functionally designated evacuation centers, are institutional spaces that may generate linking ties by connecting displaced residents to organizational resources. Although we cannot fully operationalize linking social capital with co-presence data, recognizing this dimension situates our findings within the complete tripartite framework of social capital (i.e., bonding, bridging, and linking ties) and highlights an important direction for future measurement.}

Taken together, these findings provide evidence for a geographically grounded mechanism of selective resilience.  The observed place based reinforcement of existing social ties is consistent with literature, whereby homophilous and structurally embedded (bonding) ties are preferentially sustained through repeated co-presence in familiar or utilitarian locations, while bridging ties show weaker recovery patterns.

\added{These findings reveal a theoretically significant pattern: the social capital function of third places is disaster state-dependent. Under normal conditions, third places operate as what Klinenberg \cite{klinenberg2018palaces} termed “palaces for the people”: sites where cross-cutting encounters between residents of different backgrounds generate bridging social capital. Empirical work across 25 U.S. cities has confirmed that social infrastructure, including community spaces, parks, and places of worship, positively correlates with bridging social capital at the census-tract level \cite{fraser2024tale}. Our results suggest that these bridging ties formed at third places are disproportionately eroded under disruption. 
% our post-disaster results partially invert this relationship: the same categories of third places that ordinarily facilitate bridging encounters become, under disruption, the spatial anchors through which bonding ties are disproportionately sustained. 
This functional shift likely occurs because displacement contracts the population that frequents any given venue toward a more homogeneous, local, and familiar subset of the pre-disaster user base. 
% The physical infrastructure remains the same, but the social encounter it produces changes character as the diversity of its users narrows. 
This has direct implications for recovery planning: reopening third places is necessary because it reinforces bonding capital, but may not be sufficient for restoring bridging capital. The composition of who returns to those spaces, and how quickly diverse populations regain access to them, determines whether third places resume their bridging function or continue to reinforce bonding consolidation.}

\section*{Discussion}

% what are the theoretical contributions to science? how does this change what we know about social networks, social capital, and resiliency? what new questions does this "dynamic network formulation" of social capital enable us to ask next?  
% how does this connect to practical decision making/interventions/policies in disaster resilience? who's what decisions can this 6-page piece of knowledge change, in what way? or at least, how do you want it to change? what implications does that have? 
% limitations to consider when interpreting the results, related to the data, assumptions, modeling, etc. 
% any next steps, and open questions?

In this study, we develop a dynamic and geographically grounded theory of social capital using co-presence patterns in shared third places, rather than a static attribute of neighborhoods \cite{aldrich2012building, chamlee2010cultural, crandall2010inferring}.
We find that the Marshall Fire sharply contracted the social network, reducing the number of active individuals, ties, and repeated interaction opportunities. In comparison with two counterfactual networks: (i) an displacement-controlled random node removal model and (ii) a behavior-informed node removal model, we find that post-disaster network remained systematically more connected than expected under both counterfactuals. Although all tie types decrease after the disaster, bonding ties remain substantially more robust than predicted, whereas bridging ties are fewer, weaker, and more fragile. This implies that displacement appears to contract social contacts in ways that make familiar, trusted, and routinized ties more likely to persist more than weaker or spatially extended ties. Finally, we find more bonding ties form in third places than the counterfactual models, suggesting that certain third places function as `anchor infrastructure' that stabilize social contacts after disasters. 

Through these findings, our study makes three major contributions: (i) We advance social capital research by shifting the scope of analysis from static place-based stocks to dynamic social-spatial network reorganization. In this formulation, social capital is not only a property of neighborhoods or communities measured before or after disruption, but an evolving structure of ties whose persistence, decay, and rewiring depend on how people continue to encounter one another in space over time. (ii) The post disaster connectivity displays a complex pattern of reorganization that cannot be understood merely by compositional measures. Additionally, displacement appears to contract the social field in ways that make familiar, trusted, and routinized ties more likely to persist than weaker or more spatially extended ties. (iii) Our findings identify third places as an active mediator of post-disaster network transformation. Rather than serving merely as a backdrop to social recovery, specific third places appear to function as stabilizing infrastructures through which existing ties are sustained and concentrated. 

These findings reshape how we understand the relationship between social networks, social capital, and resilience. Much of the disaster literature has shown that communities with stronger social capital recover better, but has typically treated that social capital as a pre-existing community attribute~\cite{cutter2008place, yabe2022toward, doi:10.1177/0002764214550299} . Our findings suggest that, instead, \added{community resilience depends not on the pre-disaster stock of social capital alone, but on how social ties are selectively preserved, reorganized, and spatially redistributed after shock. This finding provides the micro-network mechanism underlying the macro-level evidence that social capital is the primary driver of differential disaster recovery, more consequential than economic resources, government aid, or physical infrastructure \cite{aldrich2012building}. This interpretation is consistent with recent evidence that the expanded 26-indicator Social Capital Index significantly predicts reduced disaster damages, injuries, and fatalities at the county level \cite{kyne2026development}, suggesting that the structural dimensions we observe reorganizing at the micro-network level have documented consequences for aggregate community outcomes.}
In particular, the post-disaster network is not simply a contracted version of the pre-disaster network. It is a reorganized structure with a different balance of bonding and bridging ties, with a different geographical distribution of social encounters. A community may retain dense, trusted, and supportive bonding ties while losing cross-group or brokerage-oriented ties that are also important for access to novel information, coordination across groups, and equitable recovery \cite{lin2017building,granovetter1973strength, coleman1988social, putnam2000bowling}. This restructuring is primarily concentrated in a limited subset of third places, including restaurants, museums, recreation sites, parks, and some resource-sharing or community-serving locations such as YMCA-related facilities.  This supports the broader idea that anchor institutions and social infrastructure can stabilize fragmented networks after disruption, showing that surviving spaces of encounter shape which forms of social capital are most likely to persist \cite{aldrich2023social, klinenberg2018palaces, laurian2025libraries}.

\added{Taken together, these findings reveal a temporal sequence in how social capital reorganizes after environmental shock, a sequence that cannot be captured by static, cross-sectional measurement. First, displacement contracts the overall network: fewer people, fewer ties, fewer encounter opportunities. Second, behavioral sorting explains part of this contraction: individuals with weaker pre-disaster embeddedness and less routine-reinforcing mobility are more likely to evacuate. Third, and most critically, the residual connectivity that persists beyond what behavioral sorting predicts is not randomly distributed across the network; it is concentrated within bonding ties at a limited subset of third places. This three-stage process of contraction, sorting, and selective bonding consolidation, implies that post-disaster resilience depends not merely on the volume of pre-existing social capital, but on the compositional shift from a balanced portfolio of bonding and bridging ties toward bonding dominance.
This compositional shift can be situated within Woolcock’s \cite{woolcock1998social} broader framework for understanding the relationship between social capital and development outcomes. Woolcock identified several configurations of social capital, distinguishing between communities with balanced bonding and bridging ties—which he associated with synergistic development outcomes—and communities dominated by dense in-group bonds without cross-cutting connections, which he characterized as “communitarian” configurations that can trap communities in low-information, low-resource equilibria. The post-disaster network we observe resembles precisely this shift from a synergistic toward a communitarian configuration. The short-term adaptive value of bonding consolidation is clear: it provides immediate mutual aid, emotional support, and trusted coordination. But Woolcock’s framework warns that if this configuration persists, communities risk reduced access to the diverse information, institutional resources, and cross-group coordination that bridging and linking ties provide. Recovery, in this framing, requires not only the return of displaced populations but the active reconstruction of the bridging encounter structures through which diverse social capital is produced.}

\added{Our findings also provide the first dynamic network evidence for a process that has previously been documented only through cross-sectional or qualitative methods. Aldrich \cite{aldrich2012building} demonstrated that communities with stronger pre-disaster social capital recover faster and more completely, and the updated Social Capital Index \cite{kyne2026development} has shown that county-level social capital significantly predicts reduced disaster damages, injuries, and fatalities. The present study reveals the micro-mechanism underlying those macro-level findings: social capital’s protective value operates through the selective persistence of encounter structures at specific places, producing a reorganized network whose bonding–bridging composition shapes the community’s capacity for coordinated, equitable recovery. Norris and colleagues’ \cite{norris2008community} community resilience framework identified social capital as one of four core adaptive capacities, alongside economic development, information and communication, and community competence. Our results specify how that capacity evolves under stress: not as a uniform scaling-down of a pre-existing resource, but as a selective reorganization that strengthens some dimensions of community capacity (in-group support, trust, closure) while weakening others (cross-group coordination, information diversity, institutional access).}

\added{Methodologically, our approach addresses a recognized measurement gap in the social capital literature. Existing composite indices, including the SoCI8, \cite{chetty2022social} and the Rupasingha–Goetz index, measure structural conditions conducive to social capital - demographic similarity, organizational density, institutional employment - rather than social capital itself. These are valuable proxies for the capacity for tie formation, but they cannot observe whether ties are activated, maintained, or allowed to decay. Similarly, research on social infrastructure has demonstrated that the presence of community spaces, parks, and places of worship correlates with higher social capital \cite{klinenberg2018palaces, fraser2024tale}, but the causal mechanism through which physical places generate social ties has remained difficult to demonstrate at scale. Our co-presence network approach complements these traditions by capturing the behavioral manifestation of social capital: the actual patterns of repeated encounter through which ties are produced. Where the SoCI measures capacity and social infrastructure research measures opportunity, mobility-derived networks measure the revealed-preference patterns of co-presence that translate capacity and opportunity into observable interaction structure. This multi-level complementarity - structural conditions at the county level, physical infrastructure at the neighborhood level, and behavioral co-presence at the individual level - suggests a path toward integrated social capital measurement systems that capture the construct across its full causal chain.}

By identifying these social and behavioral mechanisms, and infrastructure that sustain or erode social capital during disaster induced population displacement, this research generates evidence-based insights for planning and designing interventions that accelerate and foster community resilience. These findings have practical implications for planning and disaster resilience policy. Given the fact that the robustness of social ties depends largely on a limited set of places that continue to host repeated co-presence, then recovery planning should pay greater attention to the maintenance, accessibility, and rapid reopening of those spaces. This applies not only to formal shelters or emergency facilities, but also to community-oriented everyday places that serve as encounter infrastructures during recovery  \cite{klinenberg2018palaces, aldrich2015social ,joshi2025corralling, aldrich2023social}. Local governments, planners, and community organizations could use these insights to prioritize social infrastructure alongside housing and physical infrastructure restoration, identify which neighborhood institutions are most likely to stabilize local social connectivity, and design interventions that reduce the loss of socially meaningful meeting spaces after disaster. 

These results should be interpreted within certain limitations. First, the inferred edges represent latent opportunities for social interaction derived from repeated co-presence rather than direct observations of friendship, support exchange, or trust. The network should therefore be interpreted as a social-spatial interaction structure, not a complete social relationship network. 
Second, race and income are not directly observed at the individual level, and bonding classification relies on probabilistic imputation from home-CBG demographic distributions. Although bootstrapping partially propagates this uncertainty, some measurement error in tie classification necessarily remains. In addition, because Boulder County is relatively affluent and demographically homogeneous, the resulting bonding versus non-bonding distinction is necessarily derived from variation within a comparatively narrow local distribution and may therefore capture fine-grained income and demographic gradations rather than the broader social differences typically emphasized in classic theories of bonding social capital. 
Third, the study focuses on one disaster case and one recovery context. The balance between bonding persistence, bridging fragility, and place-based stabilization may vary across disasters with different displacement intensity, urban form, institutional density, and recovery regimes.

These limitations point to several major avenues of research. One important direction is comparative work across multiple disasters and urban contexts to identify when post-disaster reorganization is dominated by bonding consolidation, bridging emergence, or broader fragmentation. The second is to examine how long these changes persist: whether selective reinforcement of bonding ties is primarily a short-run adaptive response or whether it produces longer-run restructuring of community networks, especially as people return or as new residents migrate into affected communities. The third is to integrate qualitative and survey-based approaches to validate and deepen these findings. Mobility-derived networks can capture repeated co-presence and changing encounter structures at population scale, but they cannot directly observe the quality, meaning, or activation of social ties. Interviews and survey-based measures would therefore help assess the validity of inferred network changes, clarify how residents experience the role of particular places in sustaining social support, and reveal dimensions of post-disaster social capital that remain unobserved through mobility traces.

\added{Additionally, our analysis focuses on the immediate post-disaster period, however, an important next step is to examine whether the observed bonding–bridging imbalance persists over a longer recovery period. Extending the analysis to later months, such as March–May 2022, would make it possible to assess whether the post-disaster social structure remains dominated by bonding ties or gradually becomes more heterogeneous over time. In particular, future work should examine how return and in-migration after the disaster reshapes local social networks. Key questions include how return and in-migration would affect the bonding capital within affected communities, whether their arrival contributes to a long-term increase in bridging ties, and whether any such increase is mediated through the same anchor third places that initially sustained bonding ties in the immediate aftermath. Addressing these questions would help clarify whether post-disaster recovery reproduces existing social structure or creates new opportunities for broader forms of social connection.}

\section*{Methods}
\subsection*{Co-location estimation}
We infer social interaction networks from repeated co-presence at third places using anonymized GPS trajectories from Cuebiq. 
Cuebiq is a location data intelligence company that collects anonymous, privacy-compliant location data of mobile devices using their software development kit (SDK) technology in mobile applications and privacy framework. Cuebiq processes data collected from mobile devices whose owners have actively opted in to share their location, and require all application partners to disclose their relationship with Cuebiq, directly or by category, in the privacy policy. With this commitment to privacy, the dataset contains location data for roughly 15 million daily active users in the United States. Individual level data analysis was done only within Cuebiq's Data Platform. All data analyzed in this study are aggregated to preserve privacy. We select non-residential points of interest (POIs) and building footprints from the SafeGraph Global Places and Geometry dataset. Third places are defined as socially neutral, discretionary spaces that plausibly facilitate repeated encounter, including cafés, parks, libraries, community centres, and other everyday gathering places \cite{oldenburg1999great}. To reduce mechanically induced co-location, we exclude POIs located within each individual’s home and work Census Block Group (CBG).

User stop points are spatially joined to third-place POI polygons to identify visits to shared social places. A co-location event between two users is recorded when they are observed within 30 meters of the same POI and their stop durations overlap for at least 5 minutes. To decide 30m and 5 m as our thresholds, we tried various thresholds before (see supplementary material for details). We further restrict the analysis to periods most likely to reflect discretionary social activity: weekends and weekday evening hours (18:00--23:00). This procedure yields place-based opportunities for social interaction rather than direct observations of friendship or communication. Consistent with prior work, repeated spatial and temporal co-occurrence is treated as evidence of elevated tie probability through structured encounter opportunities \cite{eagle2009inferring, cho2011friendship}.

For each monthly observation window $t$, we construct a weighted, undirected user--user network $G_t$, here nodes represent individuals and an edge $(i,j) \in E_t$ exists if users $i$ and $j$ are repeatedly co-located at shared third places during month $t$. $w_{ij}^{(t)} = \sum_{e \in \mathcal{E}_{ij}^{(t)}} 1,$, where $\mathcal{E}_{ij}^{(t)}$ denotes the set of all co-location events between users $i$ and $j$ in month $t$. Monthly networks are constructed for two pre-disaster periods (October and November 2021) and two post-disaster periods (January and February 2022), with the Marshall Fire occurring between 30 December 2021 and 2 January 2022.

\subsection*{Centrality Measures}
We characterize the structure of the inferred social-spatial networks using node- and edge-level measures that capture interaction intensity, though weighted degree strength, closure, embeddedness, and brokerage. These measures are computed for the full network both before and after disaster and eventually for the bonding, bridging, and unclassified subnetworks.

\subsubsection*{Weighted Degree Centrality}
For each node $i$, weighted degree centrality, or node strength, is defined as
$s_i = \sum_j w_{ij}$
where $w_{ij}$ is the monthly co-location frequency between users $i$ and $j$. This measure captures the cumulative intensity of repeated interaction opportunities available to an individual and is interpreted as the stock of direct social resources available through repeated co-presence \cite{granovetter1973strength, jackson2008social}.

\subsubsection*{Clustering Coefficient}
We compute the unweighted local clustering coefficient for node $i$ as
$CC_i = \frac{2T_i}{d_i(d_i-1)}$,
where $T_i$ is the number of triangles involving node $i$ and $d_i$ is the unweighted degree of node $i$, that is, the number of distinct neighbors of $i$. The clustering coefficient captures the extent of triadic closure, indicating whether a node’s contacts are also connected to one another. In the context of social capital, high clustering is interpreted as a signature of local closure, mutual support, and norm enforcement \cite{coleman1988social}.

\subsubsection*{Closeness Centrality}
Closeness centrality is used to measure how efficiently a node is embedded within the wider network. For node $i$, closeness is defined as
$CLC_i = \frac{1}{\sum_{j \neq i} d(i,j)}$
where $d(i,j)$ denotes the shortest-path distance between nodes $i$ and $j$. When weighted paths are used, edge lengths are defined as the inverse of interaction frequency, such that stronger ties correspond to shorter effective distances. Higher closeness indicates shorter average path lengths to other nodes and therefore greater potential to access information, support, and indirect contacts through the surrounding network structure \cite{freeman1978centrality, jackson2008social}.

\subsubsection*{Local Constraint}
To quantify brokerage at the dyadic level, we compute local constraint for an ordered pair $(u,v)$, this quantity is defined as $ell(u,v)=\left(p_{uv}+\sum_{w \in N(v)} p_{uw}p_{wv}\right)^2$, 
where $N(v)$ denotes the set of neighbors of node $v$, and $p_{uv}$ is the normalized mutual weight of the tie joining $u$ and $v$ \cite{burt2005brokerage}. Because our social network is undirected, we define a symmetric edge-level local constraint measure for edge $(i,j)$ by averaging the two directional values,
$ LC_{ij} = \frac{\ell(i,j) + \ell(j,i)}{2}.$
Lower values of $LC_{ij}$ indicate that the tie is less redundant within a tightly connected local neighbourhood and therefore more likely to span structural holes, whereas higher values indicate stronger local redundancy.

\subsection*{Random removal-based counterfactual model}
To test whether observed post-disaster network disintegration is proportional to displacement based induced population loss, we construct a displacement controlled random-loss counterfactual. Starting from the pre-disaster network, we remove nodes at random while matching the observed number of evacuated individuals within each Census Block Group (CBG). This preserves the spatial heterogeneity of population loss while randomizing which individuals are removed within each CBG. For each CBG $k$ and post-disaster period $\tau$, let $\Delta I_{k,\tau}$ denote the observed number of evacuated individuals. We generate a null ensemble of post-disaster networks $
P(G^{\mathrm{null}}_{k,\tau} \mid G_{k,\mathrm{pre}}, \Delta I_{k,\tau})$,
by sampling $\Delta I_{k,\tau}$ pre-disaster nodes uniformly without replacement from CBG $k$, independently across CBGs, and then taking the induced subgraph on the surviving nodes. 
% Repeating this procedure yields an empirical null distribution for each network statistic.
This procedure is repeated 500 times to generate an ensemble of randomized post-disaster networks and corresponding confidence intervals for the resulting network statistics.

This null model isolates the effect of compositional loss due purely to displacement counts while holding fixed the pre-disaster network structure and the spatial distribution of population loss. Observed deviations from this ensemble therefore indicate that post-disaster network change cannot be explained by random node removal alone.

\subsection*{Displacement behavior-informed counterfactual model}
We next construct a behavior-informed counterfactual that accounts for heterogeneity in displacement propensity. For each individual $i$, we estimate the probability of displacement using a logistic regression model in which the dependent variable is the observed displacement indicator,
$\Pr(E_i = 1) = \mathrm{logit}^{-1}\!\left(\alpha + \beta_1 C_i^{\mathrm{pre}} + \beta_2 D_i + \beta_3 SDM_i + \beta_4 X_i^{\mathrm{pre}} \right)$,
where $C_i^{\mathrm{pre}}$ denotes pre-disaster network position, $D_i$ captures distance-related variables including distance to the disaster and average travel distance to third places, $SDM_i$ denotes socio-demographic attributes, and $X_i^{\mathrm{pre}}$ captures pre-disaster mobility behaviour such as exploration rate, repeat visitation, and concentration within similar place categories.

Let $\hat{p}_i$ denote the fitted displacement propensity for individual $i$. Within each CBG $k$, we define sampling probabilities
$
\pi_i = \frac{\hat{p}_i}{\sum_{m \in V_k}\hat{p}_m},
$
for all pre-disaster nodes $i \in V_k$, where $V_k$ denotes the set of pre-disaster nodes in CBG $k$. We then remove exactly $\Delta I_{k,\tau}$ nodes from $V_k$ without replacement according to probabilities $\pi_i$, and define the behavior-informed counterfactual network as the induced subgraph on the surviving nodes. This yields an ensemble
$
P(G^{\mathrm{behavior}}_{k,\tau} \mid G_{k,\mathrm{pre}}, \Delta I_{k,\tau}, \hat{p}_i),
$
that preserves both the observed number of evacuees within each CBG and systematic heterogeneity in who is more likely to evacuate.

Compared with the random counterfactual, this model incorporates observed behavioural, geographic, and socio-demographic predictors of displacement, thereby providing a more realistic baseline for expected network change under selective node loss. Residual differences between the observed post-disaster network and this ensemble are interpreted as evidence of higher-order structural or place-based mechanisms not captured by individual displacement propensity alone.

\subsection*{Classifying bonding and bridging ties}
To distinguish bonding from non-bonding ties, we quantify dyadic homophily using socio-demographic similarity. Because individual-level race and income are not directly observed in the mobility data, we estimate homophily through repeated probabilistic assignment of latent socio-demographic attributes. For each bootstrap iteration, race and income are probabilistically assigned to each user on the basis of the demographic distributions of the user’s home Census Block Group (CBG).

Within each bootstrap iteration, homophily between users $i$ and $j$ is computed using cosine similarity, $H_{ij}^{(b)} = \frac{\mathbf{x}_i^{(b)} \cdot \mathbf{x}_j^{(b)}}{\|\mathbf{x}_i^{(b)}\| \, \|\mathbf{x}_j^{(b)}\|},
$, where  where $\mathbf{x}_i^{(b)}$ and $\mathbf{x}_j^{(b)}$ denote the bootstrap-specific race and income attribute vectors for users $i$ and $j$, respectively. Then we compute the median for this distribution and edge is classified as \textit{bonding} if its Homophily 
value exceeds the median homophily value across all edges in the pre-disaster network. This operationalization follows the interpretation of bonding ties as ties formed among socially similar individuals and rooted in homophily and local closure \cite{mcpherson2001birds, putnam2000bowling, aldrich2012building}.

\section*{Data Availability}
The data that support the findings of this study are available from Cuebiq through their Social Impact program, but restrictions apply to the availability of these data, which were used under the license for the current study and are therefore not publicly available. Information about how to request access to the data and its conditions and limitations can be found in \url{https://cuebiq.com/social-impact/}.  
Data access requests should be submitted through Cuebiq's Social Impact customer page \url{ https://cuebiq.com/demo/}, where the Sales team at Cuebiq may be contacted. Other data including the American Community Survey is available for download at \url{https://data.census.gov/}, and Tiger shapefiles can be downloaded from the US Census Bureau \url{https://www.census.gov/programs-surveys/geography/guidance/tiger-data-products-guide.html}.

% \section*{Code Availability}
% The analysis was conducted using Python. Code to reproduce the main results in the figures from the aggregated data is publicly available on GitHub \url{https://github.com/takayabe0505/socialhomophily}.

\section*{Code Availability}
The analysis was conducted using Python. Code to reproduce the main results in the figures from the aggregated data is publicly available on GitHub \url{https://github.com/VaidehiRaipat/Social-networks}. The mobility data used by this project are proprietary Cuebiq mobility data, therefore are not public, are not included here.

\bibliography{sample}

\section*{Acknowledgements}
We would like to thank Spectus who kindly provided us with the mobility dataset for this research through their Data for Good program. 
% Anonymized

\section*{Funding Declaration}
T.Y. acknowledges support by the National Science Foundation under Grant number 2343646.
% Anonymized

\section*{Author contributions statement}
All authors designed the algorithms, performed the analysis, developed models and simulations, and wrote the paper. Company data were processed by T.Y. and partially by V.R. T.Y. had access to aggregated (nonindividual) processed data. All authors reviewed the manuscript. 
% Anonymized

\section*{Ethical Approval}
This study is not related to human participants performed by any of the authors.

\section*{Informed Consent}
This article does not contain any studies with human participants performed by any of the authors.

\section*{Competing Interests}
The authors declare no competing interests.

\end{document}

\author{} 
\date{} 

\maketitle

\tableofcontents
% \addtocontents{toc}{\protect\thispagestyle{empty}}
% \pagenumbering{gobble}
\newpage

\listoffigures

\listoftables

\newpage

\setcounter{figure}{0}
\setcounter{table}{0}

% https://docs.google.com/spreadsheets/d/1xt6MUXLXT1qcxUUfqUIuAjZ7cCSXanxFEFKlJ0jv_x8/edit?usp=sharing

\section{Data Representativeness and pre-processing}

\subsection{Mobility Data Description}

We use a large-scale, longitudinal dataset of anonymized GPS mobility traces provided by Cuebiq. The dataset contains privacy-enhanced location observations for users in the United States who opted in to data sharing for research purposes under a framework compliant with the General Data Protection Regulation (GDPR) and the California Consumer Privacy Act (CCPA). Each record contains an anonymized user identifier, geographic coordinates, timestamp information, and dwell duration.

For the present study, we analyze mobility traces for the state of Colorado from October 2021 to February 2022, spanning the period immediately before and after the Marshall Fire. The analytic sample contains 552,370 users with 53,108,758 stops across the pre-Disaster period of October and November 2021. Relative to Colorado’s January 2021 population, this corresponds to approximately 9.4\% coverage. To preserve privacy, stop locations are subsequently aggregated to the Census Block Group (CBG) level for downstream home-location inference and socio-demographic linkage.

Because the analysis relies on individual mobility traces to infer social interaction opportunities in space, it is important to evaluate whether the sampled users are broadly representative of the study region and whether the preprocessing choices introduce systematic socio-demographic bias. In the following subsection, we therefore describe the representativeness of the mobility data and the ancillary POI datasets used to construct the socio-spatial interaction networks.

\subsection{Population and socio-demographic representativeness}

In addition to assessing the overall size of the mobility sample, we evaluated whether the observed users were systematically over- or under-represented across Census Block Groups (CBGs) with different socio-demographic characteristics. For each CBG, we compared the number of observed users with the total residential population and then examined whether the CBG-level sample rate was associated with median household income, median age, and racial composition. Demographic covariates were obtained from the 2019 American Community Survey (ACS).

\begin{figure}[h]
    \centering
    \includegraphics[width=0.95\linewidth]{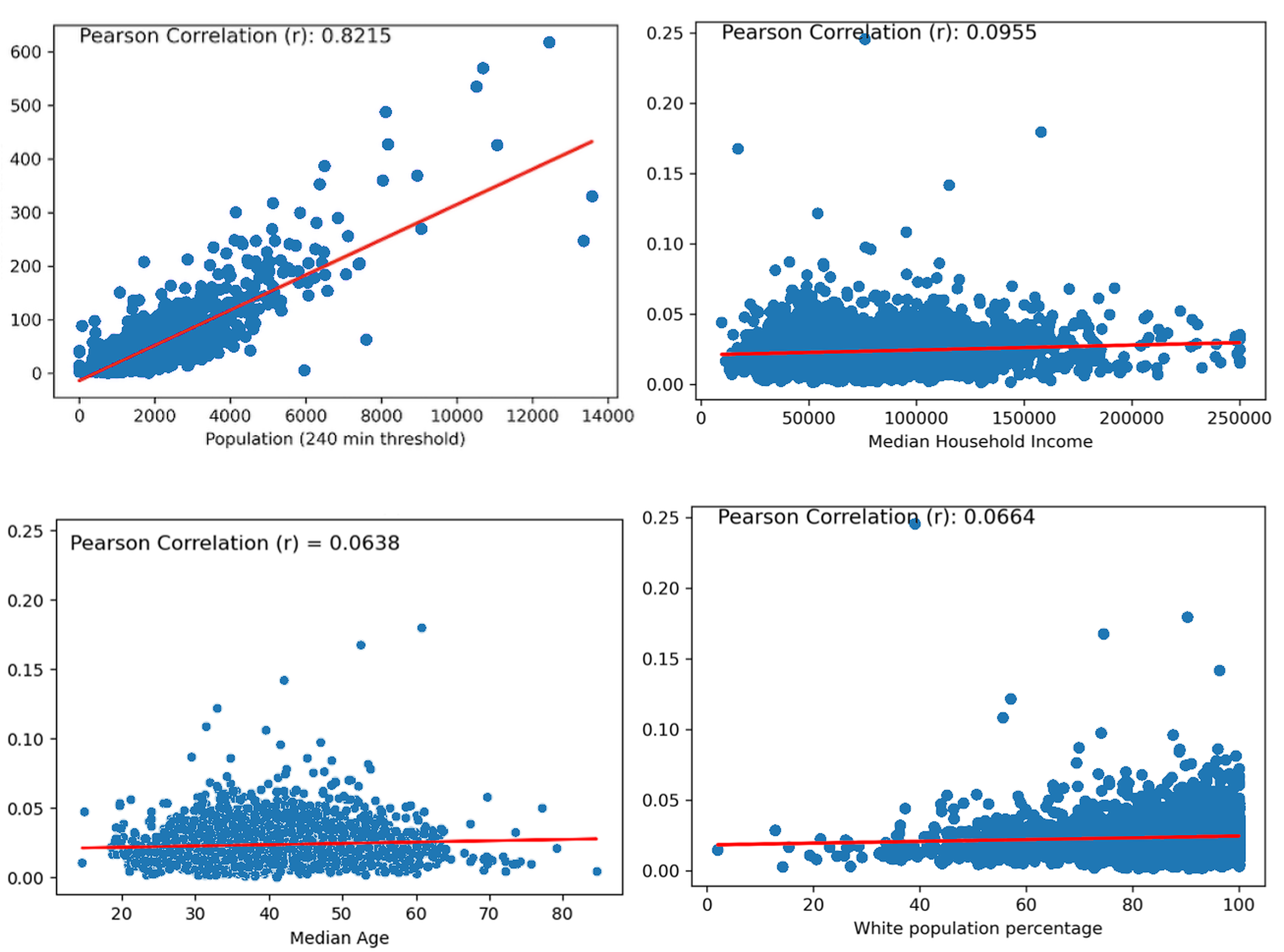}
    \captionsetup{singlelinecheck=off}
    \caption[\textbf{Representativeness of the mobility sample across population and socio-demographic dimensions.}]{
    \textbf{Representativeness of the mobility sample across population and socio-demographic dimensions.}
    CBG-level comparison between the observed mobility sample and demographic characteristics derived from the 2019 American Community Survey. The first panel compares total CBG population with the number of observed users assigned to each CBG, showing that the mobility sample closely tracks the underlying population distribution. The remaining panels compare the CBG-level sample rate with median household income, median age, and White population percentage. Weak correlations in these panels indicate that the sample rate is not strongly associated with specific socio-demographic characteristics.}
    \label{fig:Figures_representativeness_combined}
\end{figure}

    Figure~\ref{fig:Figures_representativeness_combined} shows these CBG-level representativeness checks. The observed number of users is strongly correlated with CBG population size ($r=0.8215$), indicating that the mobility sample broadly tracks the underlying residential population distribution. In contrast, the CBG-level sample rate exhibits only weak correlations with median household income ($r=0.0955$), median age ($r=0.0638$), and White population percentage ($r=0.0664$). These low absolute correlations suggest that the analytic sample is not strongly skewed toward CBGs with particular income, age, or racial compositions. Overall, these checks indicate that the mobility data provide a reasonable basis for downstream spatial network analysis across the Colorado study area.

\subsection{Third place identification from SafeGraph Global Places}
To identify shared social settings, we use the SafeGraph Global Places and Geometry datasets, which provide points of interest (POIs), brand and category information, and both point-based and polygon-based geometries for non-residential places. SafeGraph is a data company that aggregates anonymized location data from numerous applications in order to provide insights about physical places, via the SafeGraph Community. To enhance privacy, SafeGraph excludes census block group information if fewer than two devices visited an establishment in a month from a given census block group. In the present study, we rely on the polygon geometry layer to spatially join user stop locations to POI footprints. We use polygons rather than point representations because they provide a more accurate estimate of the physical extent of a place, improve visit attribution by distinguishing whether a stop falls within a POI rather than merely nearby, and help reduce ambiguity in dense urban settings where multiple POIs may be spatially proximate. In particular, the polygon hierarchy metadata is useful for identifying overlapping or enclosed places and helps avoid attributing a single stop to multiple POIs simultaneously. Additionally, in case of overlapping POI polygons, we match the stop to only one of the polygons, based on which one's centroid is closer to the stop.  

Following the literature on third places, we define third places as socially neutral or community-oriented settings outside the home and workplace that plausibly facilitate repeated, informal social interaction \cite{oldenburg1999great, soja2008thirdspace, klinenberg2018palaces}. These include categories such as cafés, restaurants, bars, parks, libraries, museums, fitness and recreational facilities, and related gathering spaces. \textcolor{black}{Importantly, our identification of third places is theory-driven rather than purely data-driven. While repeated visitation or frequent co-presence alone could be used to identify encounter locations, such patterns may also emerge from routine functional activities including grocery shopping, pharmacies, transit-related mobility, or other utilitarian spaces that generate habitual visitation without necessarily supporting sustained opportunities for social interaction or community formation. In our empirical comparison, theory-defined third places exhibit only moderately greater persistence of recurring visitation relative to other POIs, suggesting that recurring mobility routines are broadly structured across urban space rather than unique to socially oriented environments. However, third places remain systematically more likely to sustain persistent and recurring patterns of visitation, consistent with their theorized role as settings that support discretionary co-presence, informal encounter, and the reinforcement of local social ties. We therefore focus specifically on settings that prior urban sociology and social capital literature identifies as environments conducive to repeated social exposure and interaction opportunity, rather than assuming that all recurring mobility overlap is socially meaningful.} 

\subsection{Home and work CBG inference}

To distinguish discretionary social encounters from mechanically induced overlap in residential or employment settings, we infer each user's home and work Census Block Group (CBG). Home CBG is identified at the monthly scale using the most frequent recurring home location provided in the Cuebiq mobility dataset. Work CBG is inferred analogously from repeated daytime presence patterns using the most frequent recurring work-related CBG available in the same dataset. To preserve individual privacy, Cuebiq spatially transforms these recurring locations to the centroid of the quadrant of their corresponding CBG rather than exposing precise coordinates.

These inferred home and work CBG assignments serve two purposes in the present study. First, they allow mobility records to be linked to CBG-level socio-demographic context. Second, they enable us to exclude POIs located within each individual's home and work CBGs, thereby reducing mechanically induced co-location arising from routine residential proximity or shared employment geography rather than discretionary social encounter.

\subsection{Identifying co-located users}
\label{sec:S_identifying_colocated_users}
\begin{figure}[h]
    \centering
    \includegraphics[width=0.8\linewidth]{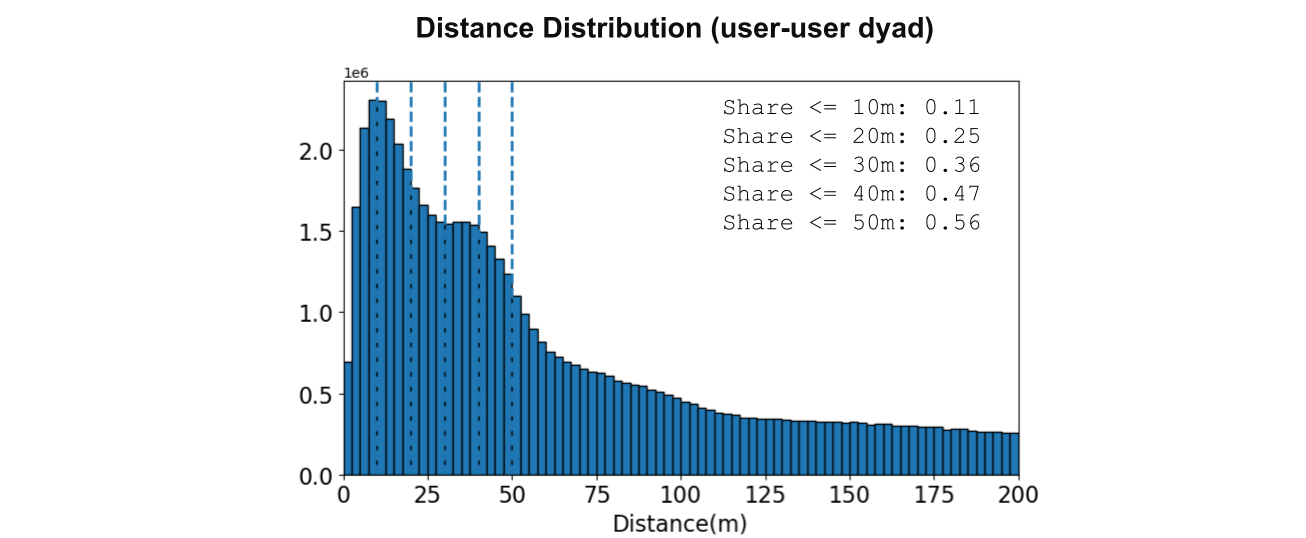}
    \captionsetup{singlelinecheck=off}
    \caption[Distribution of pairwise distances between co-located users at the same third place.]{
    \textbf{Distribution of pairwise distances between co-located users at the same third place.}
    Empirical distribution of user - user distances for pairs matched to the same third-place POI during weekends or weekday evening hours (18:00--23:00). This distribution was used to identify plausible candidate spatial thresholds for co-location and motivated the subsequent sensitivity analysis at 10\,m, 20\.m, 30\,m, 40\,m and 50\,m.}
    \label{fig:Figures - Distance Distribution}
\end{figure}

Using the filtered stop and POI datasets, we infer place-based opportunities for social interaction from repeated co-presence at shared third places. As an initial step, we examine the empirical distribution of pairwise distances between users matched to the same third-place POI in order to identify plausible candidate spatial thresholds for co-location. Figure~\ref{fig:Figures - Distance Distribution} shows that the majority of matched user pairs are concentrated at relatively short distances.  Therefore we use 10\,m, 20\.m, 30\,m, 40\,m and 50\,m as candidate user - user spatial thresholds for subsequent sensitivity analysis.

We then construct co-location graphs by varying both the spatial threshold between user locations and the minimum temporal overlap of shared presence at the same POI \cite{tizzoni2014use, cattuto2010dynamics, stopczynski2014measuring, wang2011human, crandall2010inferring, eagle2009inferring}. . Specifically, we evaluate user - user distance thresholds of 10\,m, 20\.m, 30\,m, 40\,m and 50\,m and temporal overlap thresholds of at least 5, and 15 minutes, under the additional restriction that visits occur during weekends or weekday evening hours (18:00--23:00), when activity is more likely to reflect discretionary social encounter rather than routine daytime circulation. For each parameter combination, we compute the resulting pre- and post-disaster network statistics, including the number of active nodes, total edges, and mean degree. \textcolor{black}{Importantly, the inferred network does not treat all co-presence events as equivalent. Edges are weighted by the frequency of repeated co-presence between the same pair of users at third-place POIs, so single incidental overlaps contribute weakly while persistent encounter opportunities receive greater weight. This weighting strategy reduces the influence of transient spatial coincidence and aligns the network representation with the study’s conceptual focus on repeated, place-based opportunities for social interaction rather than direct observation of friendship ties.}

\begin{figure}[h]
    \centering
    \includegraphics[width=0.80\linewidth]{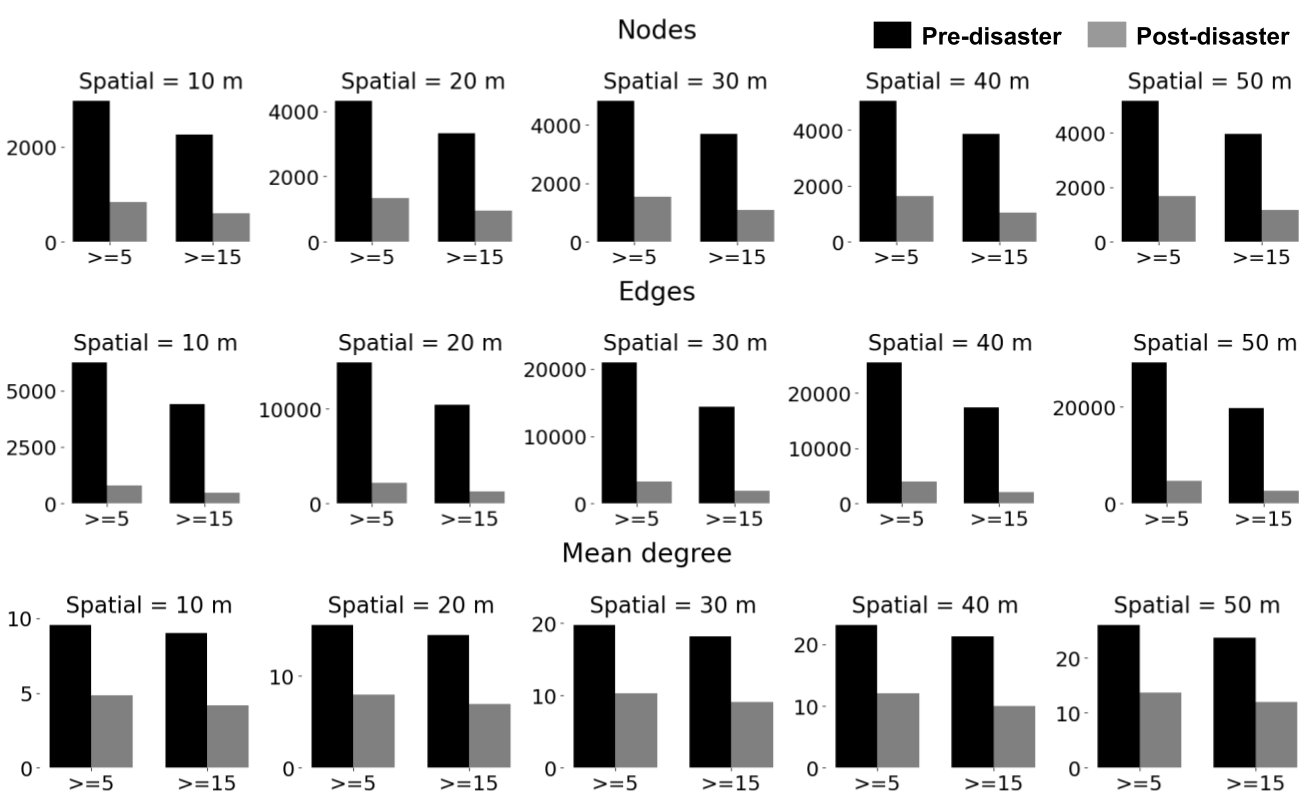}
    \captionsetup{singlelinecheck=off}
    \caption[Network statistics under alternative spatial and temporal co-location thresholds.]{
    \textbf{Network statistics under alternative spatial and temporal co-location thresholds.}
    Sensitivity of inferred pre- and post-disaster co-location networks to alternative user--user spatial thresholds and minimum temporal overlap requirements. Panels report the number of active nodes, total edges, and mean degree under multiple threshold combinations. For each spatial threshold, graphs are constructed using temporal overlap thresholds of 5, and 15 minutes. Across all specifications, more permissive threshold choices produce substantially larger and denser networks, whereas stricter overlap requirements reduce network size and connectivity while preserving the same qualitative pattern of sharp post-disaster contraction. Based on these comparisons, the main analysis adopts a 30\,m spatial threshold and a minimum temporal overlap of 5 minutes, which balances plausible encounter opportunities within the same third-place setting against the risk of inflating incidental co-location.}
    \label{fig:Figures_network_thresholds}
\end{figure}

Figure~\ref{fig:Figures_network_thresholds} shows that the inferred networks are highly sensitive to permissive threshold choices, especially when temporal overlap requirements are weak or absent. Increasing the temporal overlap threshold consistently reduces the number of nodes, edges, under each spatial threshold, while mean degree remains stable, preserving the same qualitative pattern of substantial post-disaster contraction. Based on these comparisons, we proceed with a 30\,m spatial threshold and a minimum temporal overlap of 5 minutes in the main analysis. This specification provides a balance between capturing plausible encounter opportunities within the same third-place setting and avoiding overly permissive matching that may inflate incidental or weak co-location. Alternative threshold combinations are retained as supplementary robustness checks.

For each month $t$, we define a weighted, undirected user--user network $G_t=(V_t,E_t,W_t)$, where nodes represent users and edge weights represent the number of repeated co-location events between a user pair during month $t$. These monthly networks form the basis of the main-text analyses of pre-disaster, post-disaster, and counterfactual network structure. All co-location networks are constructed within the Boulder County study area, such that repeated stops, shared POI visits, and post-disaster network persistence are measured with respect to continued observability within this regional geography.

To assess whether the main findings depend on the specific spatial threshold used to define co-location, we repeat the core analyses on alternative graphs constructed using 10\,m and 50\,m user--user distance thresholds while holding the minimum temporal overlap fixed at 5 minutes. The main analysis uses a 30\,m threshold with a 5-minute overlap requirement, as described above. These alternative specifications test whether the substantive conclusions are robust to a narrower or broader spatial definition of shared presence at the same third place.

Supplementary Figures~S5--S8, and Figures~S10--S39 reproduce the main analyses corresponding to Figures~2 and~3 under these alternative graph definitions. Specifically, Supplementary Figures~S5--S8 replicate the main Figure~2 comparison of observed and counterfactual network structure using 10\,m and 50\,m thresholds, respectively. Supplementary Figures~S10--S39, replicate the main Figure~3 tie-type comparison using the same two alternative thresholds. Across all four replications, the main qualitative conclusions remain unchanged: the observed post-disaster network remains more connected than both counterfactual baselines, and the residual robustness remains concentrated disproportionately in bonding ties relative to bridging ties.

\subsubsection{Robustness to alternative spatial co-location thresholds}

To assess whether the main results depend on the specific spatial threshold used to define co-location, we repeat the core analyses on alternative graphs constructed using 10\,m and 50\,m user--user distance thresholds while holding the minimum temporal overlap fixed at 5 minutes. The main specification in the paper uses a 30\,m spatial threshold and a 5-minute temporal overlap. The alternative graphs therefore provide a direct robustness check for the central empirical findings under stricter and more permissive spatial definitions of co-location.

Supplementary Figures~\ref{fig:Figures_Degree_Distribution_1} to \ref{fig:Figures_centrality_violin_2} replicate the main Figure~2 analysis under the 10\,m and 50\,m graph definitions, respectively. Across both alternative specifications, the observed post-disaster network remains systematically more connected than both the displacement-controlled random-loss null and the behavior-informed counterfactual. Similarly, Supplementary Figures~\ref{fig:Figure_S_tie_robustness_10x5x50} to \ref{fig:Figure_S_tie_robustness_50x15x85} replicate the main Figure~3 tie-type analysis under the same alternative graph definitions. In both cases, the main qualitative conclusion remains unchanged: residual post-disaster robustness is concentrated disproportionately in bonding ties, whereas bridging ties remain fewer, weaker, and more fragile. Together, these replications show that the principal findings of the paper are not an artifact of the specific 30\,m co-location threshold adopted in the main analysis.

To further evaluate whether the inferred co-presence network captures persistent and socially meaningful interaction opportunities rather than primarily incidental spatial overlap, we implemented an additional robustness analysis based on repeated multi-context co-presence. As illustrated in Figure~\ref{fig:copresence_robustness_framework}, the baseline network is constructed by identifying user pairs repeatedly co-located within the same third-place POI under the primary co-location threshold (30\,m spatial proximity and at least 5 minutes of temporal overlap during social hours). This baseline specification is conceptually related to the notion of familiar strangers, where individuals repeatedly encounter one another within shared routine environments without necessarily maintaining explicit social relationships, although it may still capture a small proportion of less socially meaningful ties.

\subsubsection{Validation of co-presence as a proxy for social interaction opportunity}
\begin{figure}[h]
    \centering
    \includegraphics[width=0.9\linewidth]{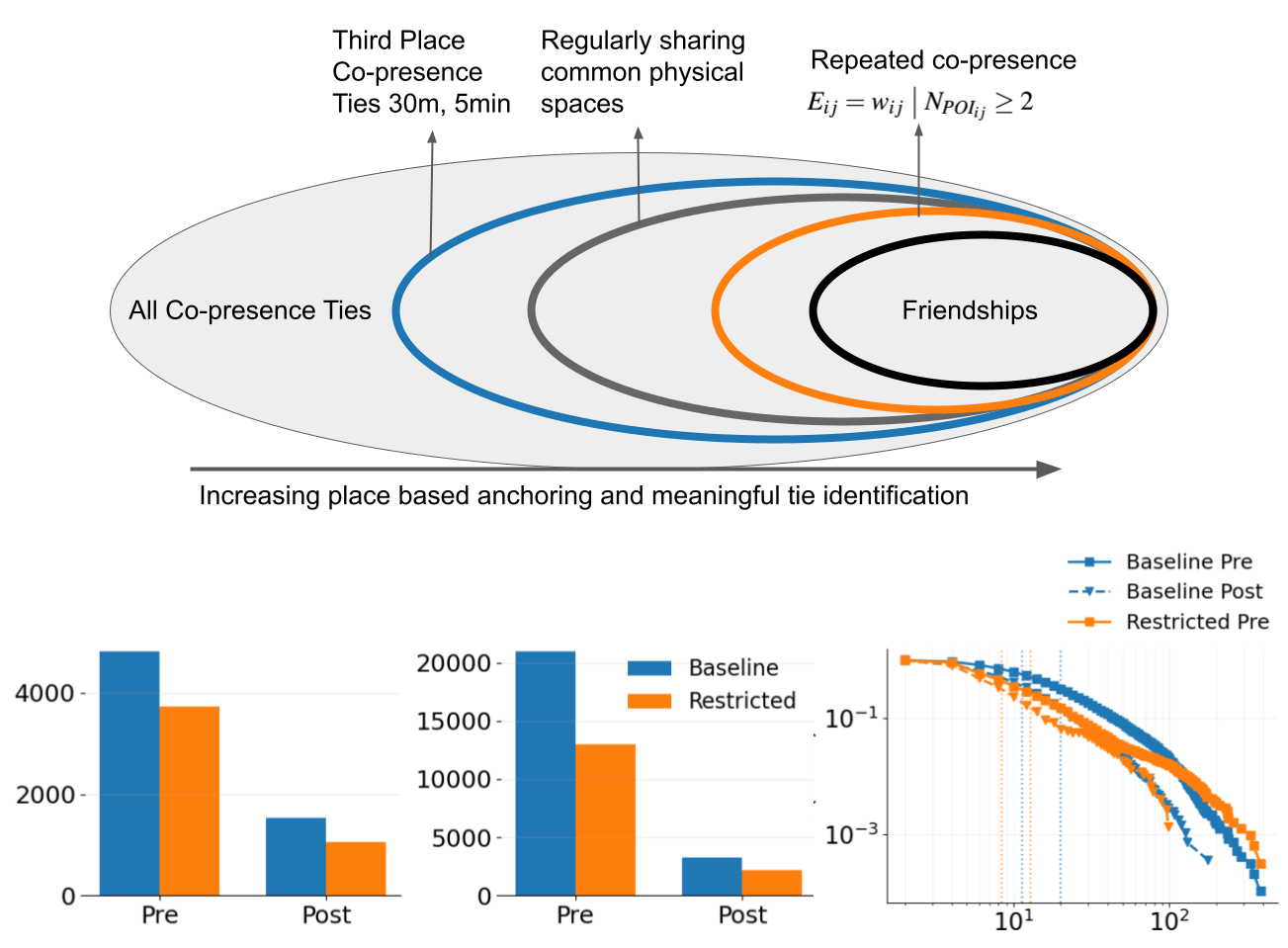}
    \captionsetup{singlelinecheck=off}
    \caption[Robustness framework for identifying repeated place-based co-presence ties.]{
    \textbf{Robustness framework for identifying repeated place-based co-presence ties.}
    Schematic illustration of the conceptual filtering framework used to infer socially embedded co-presence ties from mobility-derived interaction networks. The full set of observed co-presence encounters includes all user pairs repeatedly co-located within the same POI with a temporal overlap of at least 5 minutes. The first subset (blue ellipse) represents co-presence under the baseline co-location threshold: 30\,m spatial proximity within the same third-place POI and at least 5 minutes of temporal overlap during social hours (weekday evenings and weekends). Repeated co-presence within shared places is conceptually related to the notion of ``familiar strangers'', where individuals repeatedly encounter one another in routine social environments without necessarily maintaining explicit social relationships. A stricter robustness specification further restricts the network to dyads repeatedly co-located across at least two distinct POIs whose polygon-derived centroids are at least 100\,m apart, denoted as $E_{ij}=w_{ij}\mid N_{POI_{ij}}\geq2$, where $N_{POI_{ij}}$ represents the number of distinct shared POIs associated with dyad $(i,j)$. This progressively restricts the network toward more spatially anchored and potentially socially meaningful repeated interaction opportunities, while recognizing that inferred co-presence does not directly observe friendship ties themselves. Bottom panels compare baseline and restricted networks across pre- and post-disaster periods. The restricted specification preserves highly similar weighted degree distributions and comparable network structure, indicating that the baseline co-presence framework already captures a large share of persistent and socially embedded recurring interaction opportunities rather than primarily incidental overlap.}
    \label{fig:copresence_robustness_framework}
\end{figure}

\begin{figure}[h]
    \centering
    \includegraphics[width=0.95\linewidth]{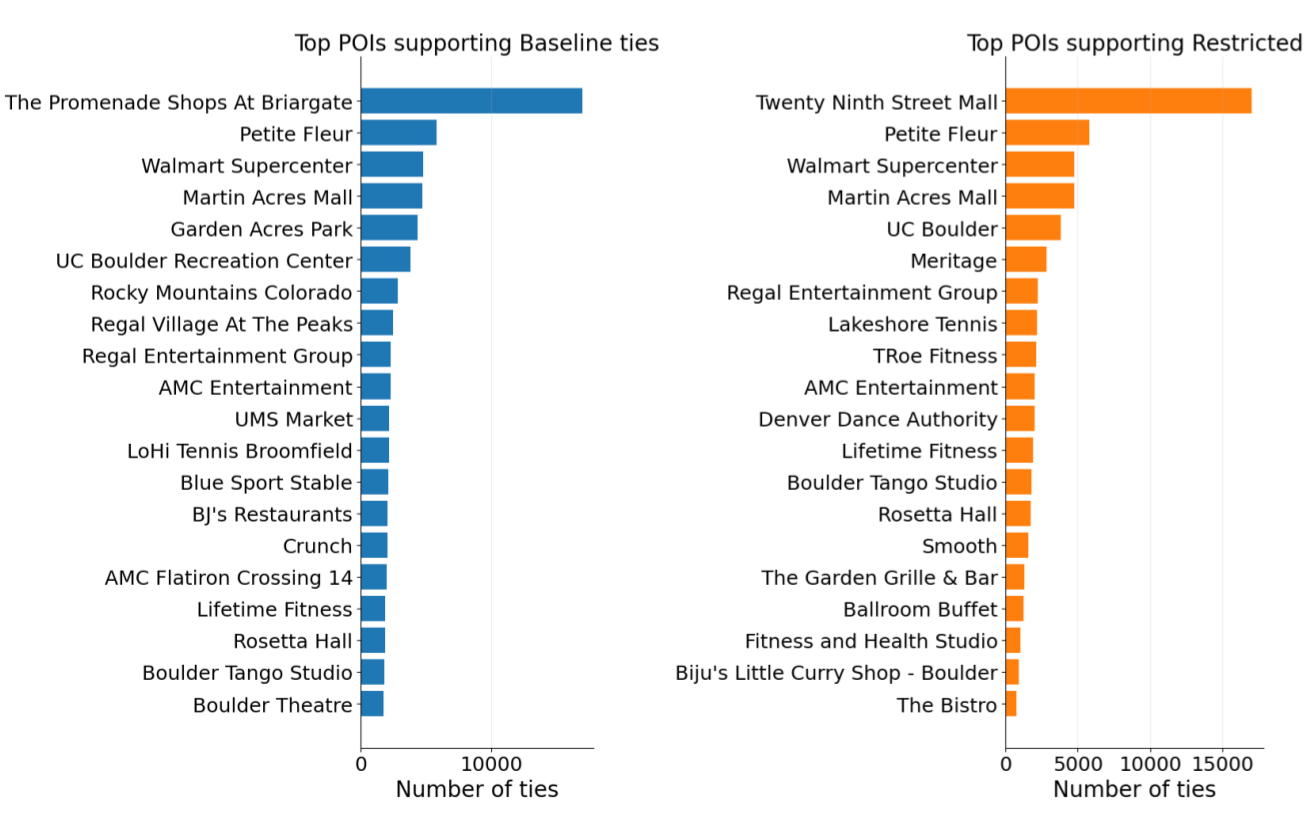}
    \captionsetup{singlelinecheck=off}
    \caption[Top POIs supporting repeated co-presence ties in baseline and restricted networks.]{
    \textbf{Top POIs supporting repeated co-presence ties in baseline and restricted networks.}
    Comparison of the twenty POIs supporting the largest number of repeated co-presence ties under the baseline network specification and the stricter restricted specification requiring repeated co-presence across at least two distinct POIs. Across both specifications, the dominant locations consist primarily of restaurants, recreational facilities, entertainment venues, and mixed-use commercial or community-centered environments commonly associated with discretionary social activity. The similarity of dominant encounter locations across both network definitions indicates that the baseline co-presence framework is not driven primarily by transient or incidental overlap at utilitarian locations, but instead consistently identifies socially active third-place environments supporting persistent repeated encounters.}
    \label{fig:top_pois_baseline_restricted}
\end{figure}

\begin{figure}[h]
    \centering
    \includegraphics[width=0.95\linewidth]{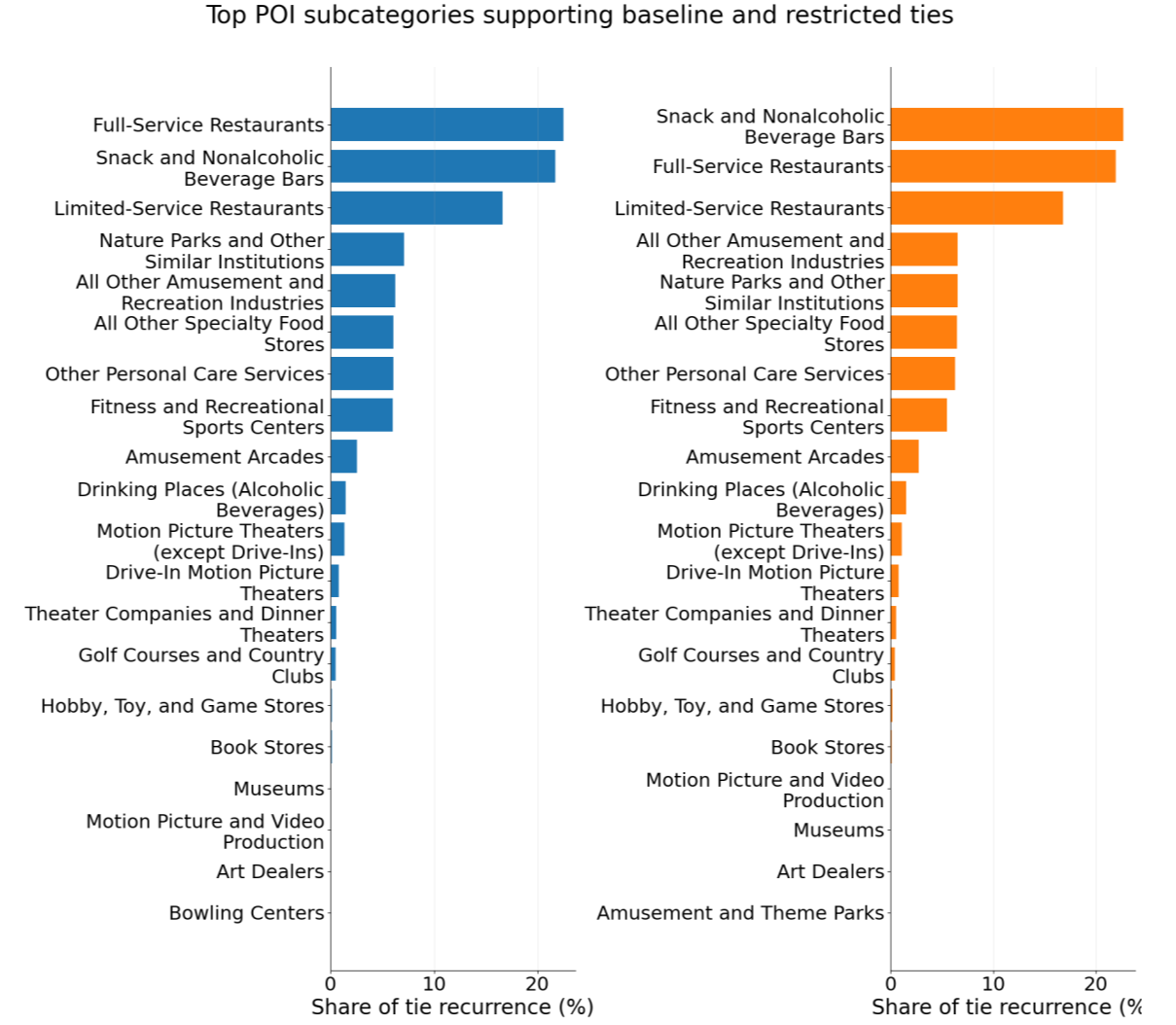}
    \captionsetup{singlelinecheck=off}
    \caption[Top POI subcategories supporting repeated co-presence ties in baseline and restricted networks.]{
    \textbf{Top POI subcategories supporting repeated co-presence ties in baseline and restricted networks.}
    Distribution of repeated co-presence ties across POI subcategories under the baseline and restricted network specifications. Values represent the percentage share of total repeated ties occurring within each POI subcategory. Across both specifications, the dominant categories remain highly consistent and are concentrated in socially oriented environments including full-service restaurants, snack and beverage establishments, recreational facilities, parks, entertainment venues, and fitness-related locations. The stability of the place-category composition under substantially stricter recurrence constraints suggests that the baseline co-presence network already captures persistent socially embedded interaction environments to a great extent.}
    \label{fig:subcategory_baseline_restricted}
\end{figure}

\textcolor{black}{To further restrict this condition, we constructed a substantially stricter network by retaining only dyads repeatedly co-located across at least two distinct POIs whose polygon-derived centroids were separated by at least 100\,m. Formally, the restricted network retains edges satisfying $E_{ij}=w_{ij}\mid N_{POI_{ij}}\geq2$, where $N_{POI_{ij}}$ denotes the number of distinct shared POIs associated with dyad $(i,j)$. This additional filtering step reduces the likelihood that repeated encounters arise solely from repeated overlap within a single establishment, commercial complex, or highly localized activity cluster, and instead emphasizes repeated interaction opportunities occurring across multiple spatially distinct social contexts. Despite the substantially stricter recurrence constraint, the resulting restricted network preserves highly similar structural characteristics relative to the baseline specification (Figure~\ref{fig:copresence_robustness_framework}). In particular, the weighted degree distributions remain broadly similar, although not completely overlapping, across both specifications, suggesting that highly socially embedded individuals remain similarly central even under stricter multi-context recurrence requirements. Furthermore, the dominant POIs and POI subcategories supporting repeated co-presence ties remain highly stable across both network definitions (Figures~\ref{fig:top_pois_baseline_restricted} and~\ref{fig:subcategory_baseline_restricted}). Across both specifications, repeated encounters continue to concentrate primarily within restaurants, recreational facilities, entertainment venues, parks, and other socially oriented third-place environments commonly associated with discretionary social activity. Together, these findings suggest that the baseline co-presence framework already captures a substantial share of persistent and socially embedded recurring interaction opportunities rather than being driven primarily by transient or incidental spatial overlap.}

\section{Measuring displacement and constructing counterfactual models}

This section describes how post-disaster displacement from the observed local social network is operationalized and how the resulting spatial heterogeneity of node loss is used to construct the two counterfactual baselines in the main text: a displacement-controlled random-loss null and a behavior-informed counterfactual based on estimated displacement propensity. Both counterfactual models are generated using 500 bootstrap iterations in order to obtain stable estimates of the resulting network statistics and confidence intervals.

\subsection{Randomized counterfactual}

To test whether the observed contraction in post-disaster connectivity can be explained by displacement counts alone, we construct a displacement-controlled random-loss counterfactual. Starting from the pre-disaster network, we remove nodes uniformly at random within each home CBG while preserving the observed number of displaced users in that CBG. Formally, for each home CBG $k$ and post-disaster period $\tau$, let $\Delta I_{k,\tau}$ denote the observed number of displaced individuals. We then sample the same number of nodes uniformly without replacement from the set of pre-disaster users in CBG $k$ and define the randomized counterfactual as the induced subgraph on the surviving nodes. This procedure is repeated 500 times to generate an ensemble of randomized post-disaster networks and corresponding confidence intervals for the resulting network statistics.

The resulting ensemble provides a null baseline for expected network contraction under spatially heterogeneous but otherwise random node loss. Table~\ref{tab:network_metrics_centrality_R11_CI} reports the mean network metrics under the observed pre-disaster, observed post-disaster, randomized counterfactual, and behavior-informed counterfactual conditions. Figures~\ref{fig:Figures_Degree_Distribution_1}, \ref{fig:Figures_Degree_Distribution_2}, \ref{fig:Figures_centrality_violin_1}, and \ref{fig:Figures_centrality_violin_2} complement these averages by showing the corresponding node-level weighted degree and centrality distributions across alternative co-location thresholds. Together, these results show that the observed post-disaster network remains systematically more connected, clustered, and embedded than the random-loss null, indicating that displacement counts alone are insufficient to explain the surviving connectivity observed after the disaster.

\subsection{Measuring displacement propensity}

To construct a more realistic counterfactual, we estimate each individual's displacement propensity using a logistic regression model in which the dependent variable is the observed displacement indicator. The model takes the form
\[
\Pr(D_i = 1) = \mathrm{logit}^{-1}\!\left(\alpha + \beta_1 C_i^{\mathrm{pre}} + \beta_2 D_i^{\ast} + \beta_3 SDM_i + \beta_4 X_i^{\mathrm{pre}} \right),
\]
where $C_i^{\mathrm{pre}}$ includes pre-disaster network measures such as strength, closeness, and clustering, $D_i^{\ast}$ includes distance-related variables such as distance to the disaster and average distance travelled to third places, $SDM_i$ includes socio-demographic characteristics linked from the user's home CBG, and $X_i^{\mathrm{pre}}$ includes pre-disaster mobility behavior such as exploration rate and category entropy.

The full regression results are reported in Table~\ref{tab:logit_evacuation}. These estimates are used only to construct sampling propensities in the behavior-informed counterfactual and are not interpreted as causal estimates of displacement behavior. Consistent with the main-text results, weaker pre-disaster connectedness and greater travel burden are associated with a higher probability of displacement from the observed local post-disaster network.

\begin{table}[!htbp] \centering
  \caption{Logistic Regression: Evacuation Propensity}
\label{tab:logit_evacuation}
\begin{tabular}{@{\extracolsep{5pt}}lc}
\hline \hline
& \multicolumn{1}{c}{\textit{Dependent variable: Evacuated (yes = 1, no = 0)}} \\
\cline{2-2}
\\[-1.8ex] & \textbf{Model 1} \\
\hline \\[-1.8ex]
Pre strength & -1.2936** ($p=0.0183$) \\
& [-2.3682, -0.219] \\
Pre closeness & 0.4213 ($p=0.1149$) \\
& [-0.1025, 0.945] \\
Pre clustering & 0.7824*** ($p=0.0$) \\
& [0.5573, 1.0075] \\
POI explore rate & 0.9426*** ($p=0.0$) \\
& [0.6945, 1.1907] \\
Category entropy & -0.082 ($p=0.421$) \\
& [-0.2817, 0.1177] \\
Disaster distance & 2.9739** ($p=0.0442$) \\
& [0.0766, 5.8713] \\
Distance traveled & 7.3624*** ($p=0.0$) \\
& [3.9712, 10.7536] \\
Total Pop & -0.0981 ($p=0.7337$) \\
& [-0.6634, 0.4671] \\
Median age & -0.4521** ($p=0.0115$) \\
& [-0.8026, -0.1015] \\
Black (\%) & 2.4102*** ($p=0.0042$) \\
& [0.7616, 4.0587] \\
Median income & 0.3854*** ($p=0.0063$) \\
& [0.1089, 0.6619] \\
Median rent & -0.2856*** ($p=0.0006$) \\
& [-0.4482, -0.1229] \\
Const & -0.4168* ($p=0.0863$) \\
& [-0.893, 0.0594] \\
\hline \\[-1.8ex]
Observations & 6535 \\
Pseudo $R^2$ & 0.041 \\
\hline
\hline \\[-1.8ex]
\textit{Note:} & \multicolumn{1}{r}{$^{*}$p$<$0.1; $^{**}$p$<$0.05; $^{***}$p$<$0.01} \\
\end{tabular}
\end{table}
%%_____________________________________________________________________

\subsection{Behavior-informed counterfactual}

Let $\hat{p}_i$ denote the fitted displacement propensity for individual $i$. Within each home CBG $k$, we define sampling probabilities proportional to $\hat{p}_i$ among all pre-disaster users in that CBG, and then remove exactly $\Delta I_{k,\tau}$ nodes without replacement. The behavior-informed counterfactual network is defined as the induced subgraph on the surviving nodes. This procedure preserves both the observed number of displaced users within each home CBG and systematic heterogeneity in who is more likely to disappear from the observed local post-disaster network. As with the randomized counterfactual, this procedure is repeated 500 times to generate an ensemble of behavior-informed post-disaster networks and corresponding confidence intervals for the resulting network statistics.

\begin{figure}[h]
    \centering
    \includegraphics[width=0.8\linewidth]{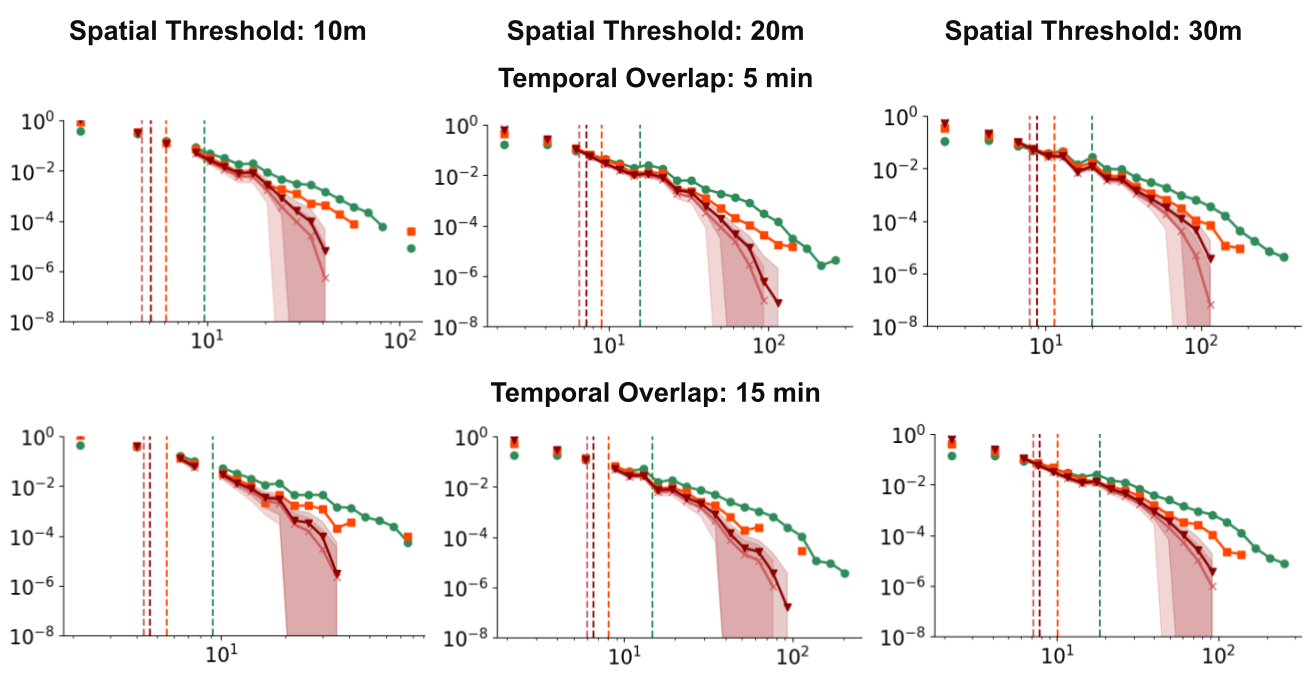}
    \captionsetup{singlelinecheck=off}
    \caption[Weighted degree distributions under alternative co-location thresholds.]{
    \textbf{Weighted degree distributions under alternative co-location thresholds.}
    Weighted degree distributions for the observed pre-disaster network, observed post-disaster network, randomized displacement-controlled counterfactual, and behavior-informed counterfactual under spatial thresholds of 10\,m, 20\,m, and 30\,m, with temporal overlap requirements of 5 and 15 minutes. Across these threshold definitions, the observed post-disaster degree distribution remains systematically above both counterfactual baselines, indicating that the residual robustness of the post-disaster network is not sensitive to the specific co-location threshold used in the main analysis.}
    \label{fig:Figures_Degree_Distribution_1}
\end{figure}

\begin{figure}[h]
    \centering
    \includegraphics[width=0.8\linewidth]{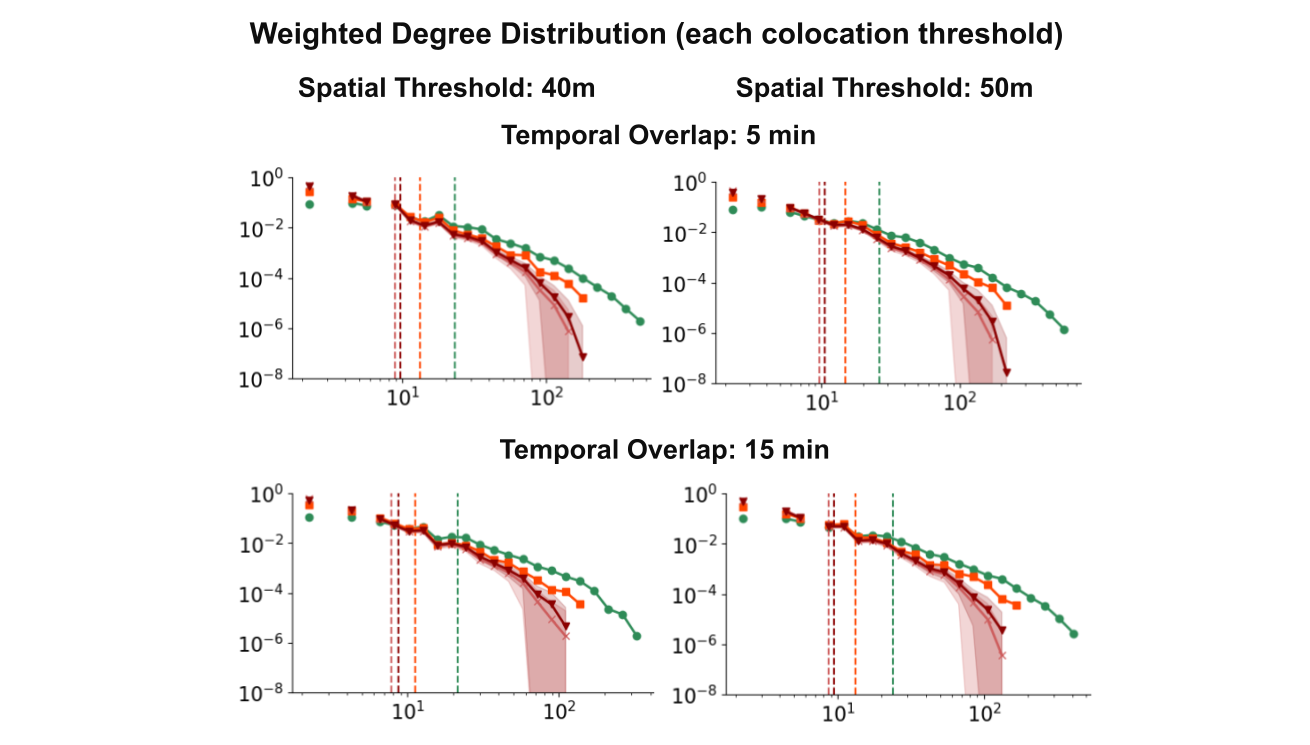}
    \captionsetup{singlelinecheck=off}
    \caption[Weighted degree distributions under broader co-location thresholds.]{
    \textbf{Weighted degree distributions under broader co-location thresholds.}
    Weighted degree distributions for the observed pre-disaster network, observed post-disaster network, randomized displacement-controlled counterfactual, and behavior-informed counterfactual under spatial thresholds of 40\,m and 50\,m, with temporal overlap requirements of 5 and 15 minutes. Even under these more permissive spatial definitions, the observed post-disaster network remains more connected than either counterfactual baseline, confirming the robustness of the main counterfactual results.}
    \label{fig:Figures_Degree_Distribution_2}
\end{figure}

\begin{figure}[h]
    \centering
    \subfloat[Centrality distributions under 10\,m, 20\,m, and 30\,m spatial thresholds with a 5-minute temporal overlap.\label{fig:centrality_violin_a}]{
        \includegraphics[width=0.7\linewidth]{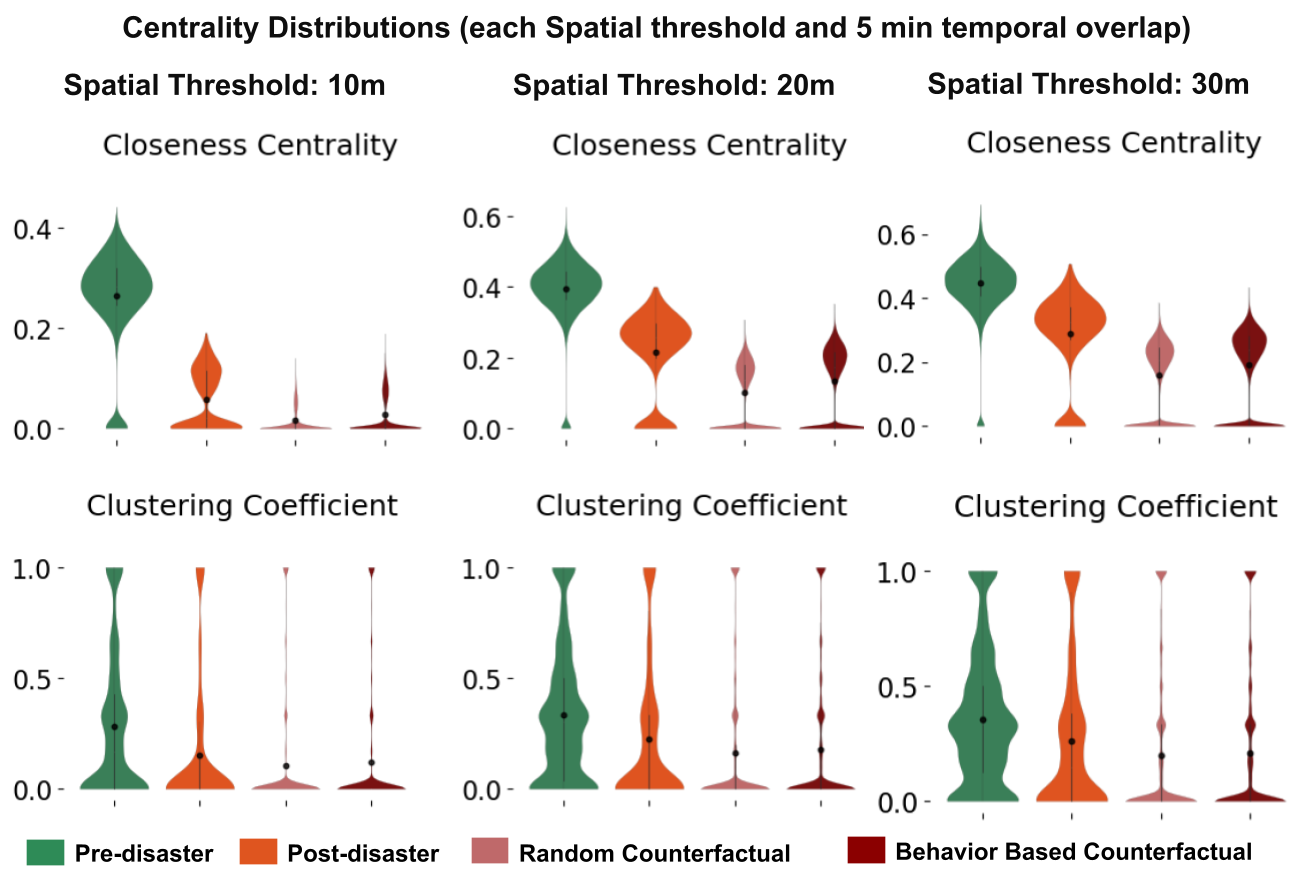}
    }\\[0.8em]
    \subfloat[Centrality distributions under 10\,m, 20\,m, and 30\,m spatial thresholds with a 15-minute temporal overlap.\label{fig:centrality_violin_b}]{
        \includegraphics[width=0.75\linewidth]{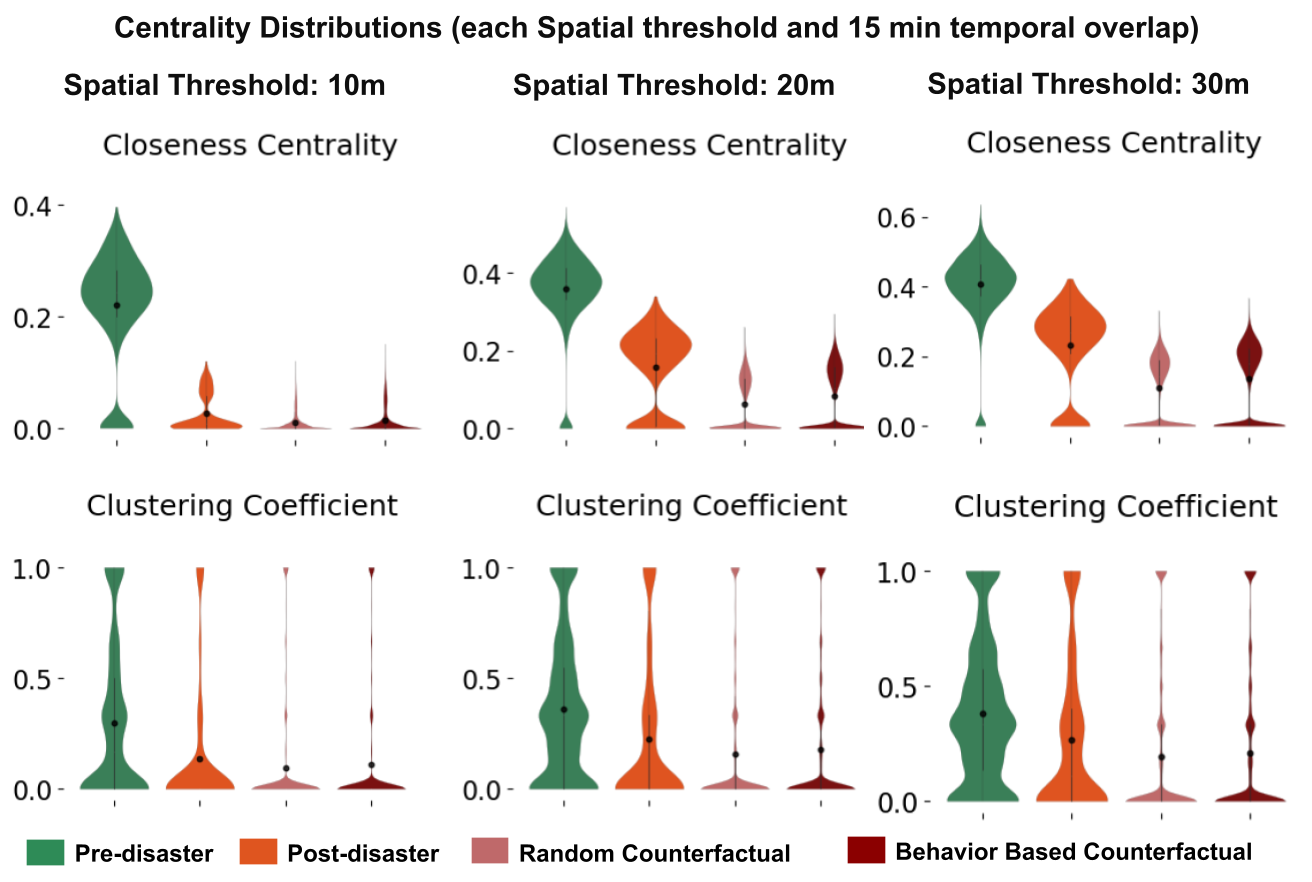}
    }
    \captionsetup{singlelinecheck=off}
    \caption[Node-level centrality distributions across observed and counterfactual networks under alternative co-location thresholds.]{
    \textbf{Node-level centrality distributions across observed and counterfactual networks under alternative co-location thresholds.}
    Violin plots of node-level closeness centrality and clustering coefficient for the observed pre-disaster network, observed post-disaster network, randomized displacement-controlled counterfactual, and behavior-informed counterfactual under alternative co-location graphs with spatial thresholds of 10\,m, 20\,m, and 30\,m. Across both the 5-minute and 15-minute temporal overlap requirements, the observed post-disaster network remains systematically above the counterfactual baselines, indicating that the main centrality-based findings are robust to alternative spatial and temporal threshold definitions.}
    \label{fig:Figures_centrality_violin_1}
\end{figure}

\begin{figure}[h]
    \centering
    \subfloat[Centrality distributions under 40\,m and 50\,m spatial thresholds with a 5-minute temporal overlap.\label{fig:centrality_violin_c}]{
        \includegraphics[width=0.7\linewidth]{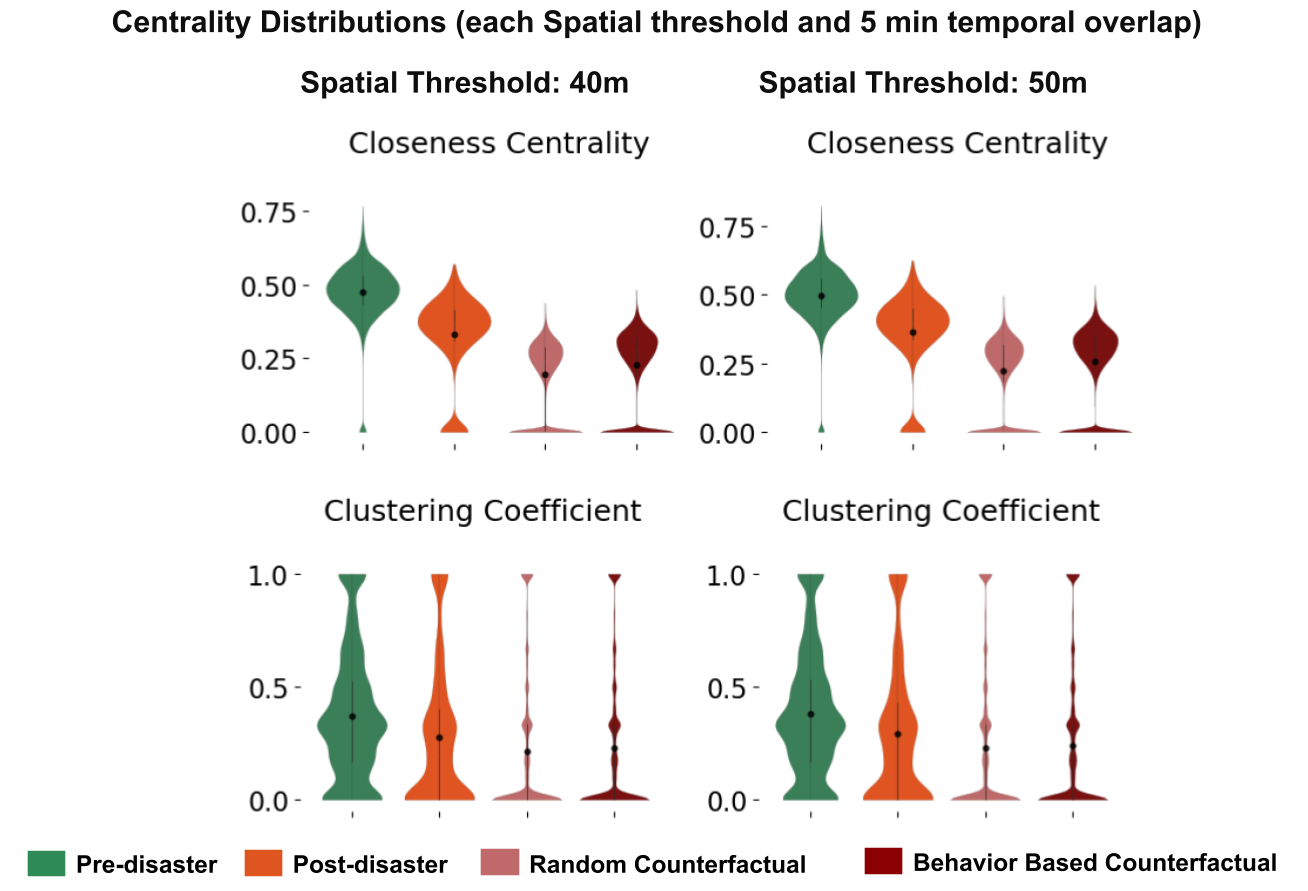}
    }\\[0.8em]
    \subfloat[Centrality distributions under 40\,m and 50\,m spatial thresholds with a 15-minute temporal overlap.\label{fig:centrality_violin_d}]{
        \includegraphics[width=0.75\linewidth]{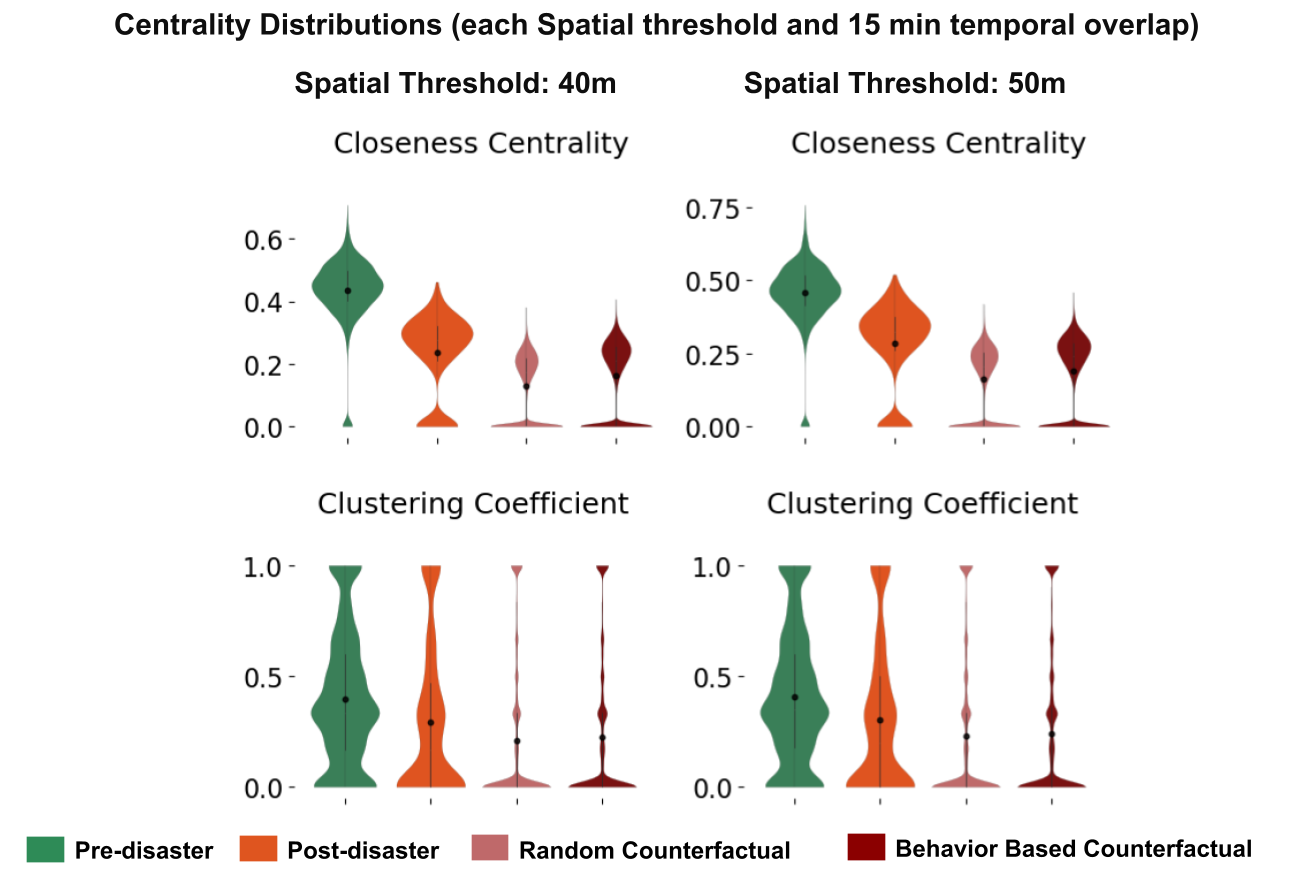}
    }
    \captionsetup{singlelinecheck=off}
    \caption[Node-level centrality distributions across observed and counterfactual networks under broader co-location thresholds.]{
    \textbf{Node-level centrality distributions across observed and counterfactual networks under broader co-location thresholds.}
    Violin plots of node-level closeness centrality and clustering coefficient for the observed pre-disaster network, observed post-disaster network, randomized displacement-controlled counterfactual, and behavior-informed counterfactual under alternative co-location graphs with spatial thresholds of 40\,m and 50\,m. Across both the 5-minute and 15-minute temporal overlap requirements, the observed post-disaster network remains more connected and more clustered than either counterfactual baseline, confirming that the main counterfactual conclusions are robust to broader spatial definitions of co-location.}
    \label{fig:Figures_centrality_violin_2}
\end{figure}

Table~\ref{tab:network_metrics_centrality_R11_CI} summarizes the resulting network-level metrics under the main co-location specification. The behavior-informed counterfactual increases connectivity relative to the randomized null, but still underestimates the observed post-disaster network across mean strength, closeness centrality, and clustering coefficient. Figures~\ref{fig:Figures_Degree_Distribution_1} and \ref{fig:Figures_Degree_Distribution_2} report the weighted degree distributions under alternative spatial and temporal co-location thresholds, while Figures~\ref{fig:Figures_centrality_violin_1} and \ref{fig:Figures_centrality_violin_2} report the corresponding node-level distributions of closeness centrality and clustering coefficient. Across these alternative graph definitions, the observed post-disaster network remains systematically higher than both counterfactual baselines, indicating that the main counterfactual results are robust to alternative spatial and temporal co-location specifications.

\begin{table}[!htbp] \centering
\caption{Network-level metrics across observed and counterfactual conditions with uncertainty intervals \\(Spatial threshold = 30\,m and temporal overlap = 5\,min)}
\label{tab:network_metrics_centrality_R11_CI}
\scriptsize
\setlength{\tabcolsep}{3pt}
\renewcommand{\arraystretch}{1.15}
\begin{tabular}{lcccc}
\hline \hline
Metric & Pre-disaster & Post-disaster & Random survival CF & Behavior-informed CF \\
\hline \\[-1.8ex]
\multicolumn{5}{l}{\textit{Network-level metrics}} \\
Nodes & 4,816 & 1,538 & 1,447 (1,416--1,478) & 1,460 (1,431--1,491) \\
Edges & 20,971 & 3,241 & 2,064 (1,880--2,269) & 2,318 (2,123--2,525) \\
Mean Weighted Degree & 19.94 (19.50--20.40) & 10.33 (9.88--10.82) & 6.43 (5.87--7.05) & 7.32 (6.68--8.01) \\
Closeness centrality & 0.445 (0.443--0.447) & 0.289 (0.284--0.293) & 0.160 (0.137--0.182) & 0.190 (0.166--0.214) \\
Clustering coefficient & 0.355 (0.349--0.361) & 0.261 (0.249--0.272) & 0.196 (0.176--0.220) & 0.210 (0.190--0.231) \\
\hline
\multicolumn{5}{p{0.95\linewidth}}{\scriptsize \textit{Notes:} Observed pre- and post-disaster confidence intervals for mean node-level metrics are node-resampling 95\% percentile intervals using 500 bootstrap samples with replacement. Counterfactual confidence intervals are 95\% percentile intervals across simulation runs. Observed nodes and edges are reported as graph-level point estimates.} \\
\hline
\end{tabular}
\end{table}

\section{Tie classification: bonding, bridging, and unclassified}

To examine whether post-disaster robustness is concentrated in specific forms of social capital, we partition all observed ties into bonding, bridging, and unclassified categories using a sequential rule based on socio-demographic homophily and structural embeddedness. Bonding ties are intended to capture socially similar and locally embedded relationships, whereas bridging ties are intended to capture less redundant ties that survive after filtering the non-bonding network for low local constraint and minimum structural support. All remaining edges are classified as unclassified. 
To test the robustness of the split methodology we check  the percentage of edges in each tie category that connect individuals who share the same home CBG in the pre-disaster and post-disaster networks. We find that in both periods, bonding ties exhibit a higher same-home-CBG share than bridging ties, consistent with the interpretation that bonding ties are more locally embedded and socially homogeneous (Figure \ref{fig:similarity_cbg}).

\subsection{Probabilistic socio-demographic imputation and homophily estimation}

Because individual-level race and income are not directly observed in the mobility data, we estimate dyadic homophily through repeated probabilistic assignment of latent socio-demographic attributes. For each bootstrap iteration $b \in \{1,\dots,B\}$, with $B=100$, race and income are assigned to each user on the basis of the demographic distributions of the user's home CBG. This yields a bootstrap-specific socio-demographic attribute vector $\mathbf{x}_i^{(b)}$ for user $i$, and similarly $\mathbf{x}_j^{(b)}$ for user $j$.

Within each bootstrap iteration, homophily between users $i$ and $j$ is computed using cosine similarity,
\[
H_{ij}^{(b)}=\frac{\mathbf{x}_i^{(b)}\cdot \mathbf{x}_j^{(b)}}{\|\mathbf{x}_i^{(b)}\|\,\|\mathbf{x}_j^{(b)}\|}.
\]
Repeating this procedure across all $B=100$ iterations yields a distribution of homophily values for each dyad. We summarize this distribution using its mean,
\[
\bar{H}_{ij}=\frac{1}{B}\sum_{b=1}^{B}H_{ij}^{(b)}.
\]

Figure \ref{fig:Figures - Similarity (Homophily) Distribution.png} shows the resulting distribution of dyadic homophily values under the main co-location specification. In the main analysis, an edge is classified as bonding if its mean dyadic homophily exceeds the selected pre-disaster homophily threshold. In supplementary robustness analyses, we additionally repeat the tie-splitting procedure under stricter homophily cutoffs to assess whether the main substantive conclusions depend on the precise bonding-threshold choice.

\subsection{Local Constraint}

Among edges not classified as bonding, we quantify structural redundancy using local constraint. We compute local constraint using the pair-specific formulation implemented in \url{networkx.algorithms.structuralholes.local\_constraint}.  For an ordered pair $(u,v)$,
\[
\ell(u,v)=\left(p_{uv}+\sum_{w\in N(v)} p_{uw}p_{wv}\right)^2,
\]
where $p_{uv}$ is the normalized mutual weight of the tie between $u$ and $v$, and $N(v)$ denotes the set of neighbors of node $v$. Because the graph is undirected, we define the symmetric edge-level local constraint of edge $(i,j)$ as
\[
LC_{ij}=\frac{\ell(i,j)+\ell(j,i)}{2}.
\]
Lower values of $LC_{ij}$ indicate that the tie is less redundant within a tightly connected local neighborhood and is therefore more likely to represent brokerage-like structure.

\subsection{Sequential tie classification rule}

Ties are classified sequentially. First, an edge is labeled bonding if its mean dyadic homophily exceeds the selected pre-disaster homophily threshold. Second, among the remaining edges, we retain only those with local constraint below the selected threshold. The resulting low-constraint subgraph is then filtered using a threshold $k > 3$ , so that only ties embedded in a minimally sustained non-redundant structure are retained. Edges that survive both the local-constraint filter and the degree filter are labeled \textit{bridging refined}. All other non-bonding edges are labeled \textit{unclassified}.

To assess the sensitivity of the tie-type decomposition, we repeated the behavior-informed counterfactual comparison across alternative spatial thresholds, temporal overlap requirements, homophily cutoffs, and constraint cutoffs. Figures~\ref{fig:Figure_S_tie_robustness_30x5x50}--\ref{fig:Figure_S_tie_robustness_50x15x85} report the distribution of mean weighted degree across 500 behavior-based counterfactual runs for the bonding, refined bridging, and unclassified subnetworks, with vertical reference lines marking the observed pre-disaster and post-disaster values. Across all specifications except the most restrictive 10\,m spatial co-location threshold, the results replicate the main finding: bonding and unclassified ties remain more connected after the disaster than expected under the behavior-informed counterfactual, whereas refined bridging ties are weaker than expected. The 10\,m specification produces a sparse graph with too few refined bridging ties for stable comparison, so these cases are interpreted cautiously rather than treated as primary robustness evidence.

\subsection{Weighted degree distributions}

For each node $i$, weighted degree (or strength) is defined as
\[
s_i=\sum_j w_{ij},
\]
where $w_{ij}$ is the monthly co-location frequency between users $i$ and $j$. Weighted degree captures the cumulative intensity of repeated interaction opportunities available to an individual. Rather than relying only on mean network summaries, we compare the full weighted degree distributions across observed and counterfactual networks to evaluate whether the observed post-disaster network remains systematically more connected than expected under random or behavior-informed node loss.

\subsection{Centrality measures}

We further compute node-level measures of network embeddedness and local cohesion. Closeness centrality for node $i$ is defined as
\[
CLC_i=\frac{1}{\sum_{j\neq i} d(i,j)},
\]
where $d(i,j)$ denotes the shortest-path distance between nodes $i$ and $j$. When weighted paths are used, edge lengths are defined as the inverse of interaction intensity, $\ell_{ij}=1/w_{ij}$, such that stronger ties correspond to shorter effective distances. Closeness centrality therefore captures how efficiently an individual is embedded within the wider network.

We also compute the local clustering coefficient,
\[
CC_i=\frac{2T_i}{d_i(d_i-1)},
\]
where $T_i$ is the number of triangles involving node $i$ and $d_i$ is the unweighted degree of node $i$. This measure captures triadic closure and the extent to which an individual's neighbors are also connected to one another.

In the supplementary analyses, weighted degree distributions and node-level centrality distributions are compared across observed and counterfactual networks under multiple spatial and temporal co-location thresholds. These distributional comparisons complement the network-level averages reported in the main text and show that the principal counterfactual findings are robust for co-location graphs constructed with spatial thresholds of 20\,m and above. Under these specifications, the observed post-disaster network remains systematically more c 3onnected, clustered, and embedded than either counterfactual baseline. By contrast, the 10\,m specification produces a substantially sparser network with fewer captured ties overall, the bonding ties continue to remain robust in the actual post disaster compared to the counterfactual. However, the bridging sub-network yields un-stable post-disaster versus counterfactual comparisons.

\begin{table}[!htbp]
\centering
\caption{Network-level and mean node-level metrics by tie type and period for the 10 m, 5 min co-location threshold}
\label{tab:network_node_metrics_combined_cf_95ci_r10}
\tiny
\begin{minipage}{\linewidth}
\centering
\begin{tabular}{llccccc}
\hline \hline
Phase & Tie type & Nodes & Edges & Mean degree & Closeness & Clustering \\
\hline
Pre & Bonding & 2410 & 3870 & 3.21 [3.12, 3.30] & 0.18 [0.18, 0.19] & 0.21 [0.20, 0.22] \\
Pre & Unclassified & 1762 & 1950 & 2.21 [2.16, 2.27] & 0.12 [0.11, 0.12] & 0.11 [0.10, 0.11] \\
Pre & Bridging & 148 & 424 & 5.74 [5.38, 6.12] & 0.36 [0.35, 0.37] & 0.46 [0.43, 0.49] \\
Post & Bonding & 430 & 382 & 1.78 [1.69, 1.86] & 0.02 [0.02, 0.03] & 0.10 [0.08, 0.12] \\
Post & Unclassified & 446 & 377 & 1.69 [1.61, 1.78] & 0.02 [0.02, 0.03] & 0.06 [0.05, 0.08] \\
Post & Bridging & 12 & 32 & 5.33 [4.21, 6.65] & 1.88 [1.64, 2.13] & 0.56 [0.45, 0.67] \\
Counterfactual & Bonding & 334 [298, 368] & 285 [241, 329] & 1.71 [1.55, 1.89] & 0.02 [0.01, 0.08] & 0.12 [0.00, 1.00] \\
Counterfactual & Unclassified & 323 [288, 360] & 265 [226, 309] & 1.64 [1.51, 1.80] & 0.03 [0.01, 0.10] & 0.07 [0.00, 1.00] \\
Counterfactual & Bridging & 14 [0, 28] & 30 [0, 66] & 4.10 [0.00, 6.21] & 0.79 [0.29, 2.00] & 0.76 [0.33, 1.00] \\
\hline \hline
\end{tabular}
\vspace{1mm}
\begin{flushleft}
\tiny \textit{Notes:} Each metric cell reports the mean followed by the 95\% confidence interval in brackets. Observed pre- and post-disaster confidence intervals are node-resampling 95\% percentile intervals using 500 bootstrap samples with replacement. Counterfactual confidence intervals are 95\% percentile intervals across 500 simulation runs.
\end{flushleft}
\end{minipage}
\end{table}

\begin{table}[!htbp]
\centering
\caption{Network-level and mean node-level metrics by tie type and period for the 10 m, 15 min co-location threshold}
\label{tab:network_node_metrics_combined_cf_95ci_r10_2}
\tiny
\begin{minipage}{\linewidth}
\centering
\begin{tabular}{llccccc}
\hline \hline
Phase & Tie type & Nodes & Edges & Mean degree & Closeness & Clustering \\
\hline
Pre & Bonding & 1804 & 2770 & 3.07 [2.98, 3.16] & 0.14 [0.14, 0.14] & 0.23 [0.22, 0.24] \\
Pre & Unclassified & 1298 & 1374 & 2.12 [2.06, 2.18] & 0.08 [0.08, 0.08] & 0.11 [0.10, 0.12] \\
Pre & Bridging & 86 & 248 & 5.77 [5.29, 6.22] & 0.42 [0.40, 0.44] & 0.48 [0.44, 0.52] \\
Post & Bonding & 279 & 224 & 1.61 [1.53, 1.69] & 0.02 [0.02, 0.03] & 0.09 [0.07, 0.11] \\
Post & Unclassified & 286 & 225 & 1.58 [1.49, 1.67] & 0.02 [0.02, 0.03] & 0.07 [0.05, 0.09] \\
Post & Bridging & 6 & 16 & 5.64 [4.77, 6.45] & 2.01 [1.80, 2.20] & 0.57 [0.52, 0.63] \\
Counterfactual & Bonding & 217 [188, 246] & 168 [137, 202] & 1.55 [1.38, 1.75] & 0.02 [0.01, 0.07] & 0.11 [0.00, 1.00] \\
Counterfactual & Unclassified & 212 [184, 240] & 162 [132, 193] & 1.52 [1.37, 1.68] & 0.03 [0.01, 0.09] & 0.07 [0.00, 1.00] \\
Counterfactual & Bridging & 8 [0, 19] & 15 [0, 45] & 3.19 [0.00, 5.56] & 1.11 [0.41, 2.22] & 0.79 [0.33, 1.00] \\
\hline \hline
\end{tabular}
\vspace{1mm}
\begin{flushleft}
\tiny \textit{Notes:} Each metric cell reports the mean followed by the 95\% confidence interval in brackets. Observed pre- and post-disaster confidence intervals are node-resampling 95\% percentile intervals using 500 bootstrap samples with replacement. Counterfactual confidence intervals are 95\% percentile intervals across 500 simulation runs.
\end{flushleft}
\end{minipage}
\end{table}

\begin{table}[!htbp]
\centering
\caption{Network-level and mean node-level metrics by tie type and period for the 20 m, 5 min co-location threshold}
\label{tab:network_node_metrics_combined_cf_95ci_r14_1}
\tiny
\begin{minipage}{\linewidth}
\centering
\begin{tabular}{llccccc}
\hline \hline
Phase & Tie type & Nodes & Edges & Mean degree & Closeness & Clustering \\
\hline
Pre & Bonding & 4454 & 15596 & 7.00 [6.82, 7.20] & 0.40 [0.40, 0.41] & 0.29 [0.28, 0.30] \\
Pre & Unclassified & 3404 & 5604 & 3.29 [3.22, 3.36] & 0.27 [0.26, 0.27] & 0.14 [0.14, 0.15] \\
Pre & Bridging & 908 & 4301 & 9.48 [9.05, 9.93] & 0.59 [0.59, 0.60] & 0.27 [0.26, 0.28] \\
Post & Bonding & 1147 & 2048 & 3.57 [3.43, 3.70] & 0.29 [0.29, 0.30] & 0.21 [0.20, 0.22] \\
Post & Unclassified & 1154 & 1832 & 3.18 [3.05, 3.30] & 0.28 [0.27, 0.28] & 0.14 [0.13, 0.15] \\
Post & Bridging & 50 & 133 & 5.27 [4.80, 5.73] & 0.65 [0.62, 0.68] & 0.24 [0.20, 0.27] \\
Counterfactual & Bonding & 957 [908, 1008] & 1378 [1244, 1531] & 2.88 [2.67, 3.11] & 0.20 [0.00, 0.33] & 0.18 [0.00, 1.00] \\
Counterfactual & Unclassified & 931 [878, 986] & 1061 [953, 1159] & 2.28 [2.14, 2.42] & 0.15 [0.00, 0.25] & 0.09 [0.00, 1.00] \\
Counterfactual & Bridging & 124 [69, 172] & 365 [206, 525] & 5.88 [4.87, 7.16] & 0.50 [0.33, 0.68] & 0.39 [0.00, 1.00] \\
\hline \hline
\end{tabular}
\vspace{1mm}
\begin{flushleft}
\tiny \textit{Notes:} Each metric cell reports the mean followed by the 95\% confidence interval in brackets. Observed pre- and post-disaster confidence intervals are node-resampling 95\% percentile intervals using 500 bootstrap samples with replacement. Counterfactual confidence intervals are 95\% percentile intervals across 500 simulation runs.
\end{flushleft}
\end{minipage}
\end{table}

\begin{table}[!htbp]
\centering
\caption{Network-level and mean node-level metrics by tie type and period for the 20 m, 15 min co-location threshold}
\label{tab:network_node_metrics_combined_cf_95ci_r14_2}
\tiny
\begin{minipage}{\linewidth}
\centering
\begin{tabular}{llccccc}
\hline \hline
Phase & Tie type & Nodes & Edges & Mean degree & Closeness & Clustering \\
\hline
Pre & Bonding & 3388 & 10843 & 6.40 [6.21, 6.59] & 0.36 [0.36, 0.36] & 0.32 [0.32, 0.33] \\
Pre & Unclassified & 2508 & 3850 & 3.07 [2.99, 3.16] & 0.23 [0.23, 0.23] & 0.15 [0.14, 0.16] \\
Pre & Bridging & 599 & 2662 & 8.89 [8.44, 9.34] & 0.57 [0.56, 0.57] & 0.30 [0.29, 0.31] \\
Post & Bonding & 694 & 1046 & 3.02 [2.89, 3.16] & 0.18 [0.18, 0.19] & 0.22 [0.20, 0.24] \\
Post & Unclassified & 670 & 968 & 2.89 [2.73, 3.04] & 0.19 [0.19, 0.20] & 0.19 [0.17, 0.20] \\
Post & Bridging & 8 & 17 & 4.00 [3.59, 4.53] & 0.82 [0.75, 0.89] & 0.38 [0.27, 0.48] \\
Counterfactual & Bonding & 563 [523, 609] & 706 [614, 806] & 2.50 [2.26, 2.76] & 0.12 [0.00, 0.26] & 0.19 [0.00, 1.00] \\
Counterfactual & Unclassified & 559 [519, 599] & 603 [531, 678] & 2.16 [1.99, 2.33] & 0.11 [0.00, 0.24] & 0.11 [0.00, 1.00] \\
Counterfactual & Bridging & 64 [33, 100] & 171 [84, 275] & 5.38 [4.31, 6.99] & 0.46 [0.15, 0.74] & 0.48 [0.00, 1.00] \\
\hline \hline
\end{tabular}
\vspace{1mm}
\begin{flushleft}
\tiny \textit{Notes:} Each metric cell reports the mean followed by the 95\% confidence interval in brackets. Observed pre- and post-disaster confidence intervals are node-resampling 95\% percentile intervals using 500 bootstrap samples with replacement. Counterfactual confidence intervals are 95\% percentile intervals across 500 simulation runs.
\end{flushleft}
\end{minipage}
\end{table}

\begin{table}[!htbp]
\centering
\caption{Network-level and mean node-level metrics by tie type and period for the 30 m, 5 min co-location threshold}
\label{tab:network_node_metrics_combined_cf_95ci_r11}
\tiny
\begin{minipage}{\linewidth}
\centering
\begin{tabular}{llccccc}
\hline \hline
Phase & Tie type & Nodes & Edges & Mean degree & Closeness & Clustering \\
\hline
Pre & Bonding & 4226 & 12962 & 6.13 [6.00, 6.28] & 0.37 [0.37, 0.38] & 0.28 [0.27, 0.28] \\
Pre & Unclassified & 3220 & 4826 & 3.00 [2.93, 3.06] & 0.23 [0.23, 0.24] & 0.13 [0.12, 0.13] \\
Pre & Bridging & 774 & 3184 & 8.22 [7.90, 8.60] & 0.53 [0.53, 0.53] & 0.28 [0.26, 0.29] \\
Post & Bonding & 1022 & 1589 & 3.11 [3.00, 3.24] & 0.24 [0.23, 0.24] & 0.18 [0.17, 0.19] \\
Post & Unclassified & 1074 & 1616 & 3.01 [2.89, 3.12] & 0.25 [0.24, 0.25] & 0.15 [0.13, 0.16] \\
Post & Bridging & 18 & 36 & 3.94 [3.61, 4.29] & 0.92 [0.82, 1.03] & 0.29 [0.23, 0.35] \\
Counterfactual & Bonding & 864 [813, 912] & 1134 [1010, 1254] & 2.62 [2.42, 2.82] & 0.16 [0.00, 0.29] & 0.17 [0.00, 1.00] \\
Counterfactual & Unclassified & 843 [796, 892] & 918 [835, 1005] & 2.18 [2.05, 2.31] & 0.13 [0.00, 0.23] & 0.08 [0.00, 1.00] \\
Counterfactual & Bridging & 94 [58, 135] & 284 [166, 427] & 6.07 [4.86, 7.57] & 0.47 [0.21, 0.67] & 0.49 [0.00, 1.00] \\
\hline \hline
\end{tabular}
\vspace{1mm}
\begin{flushleft}
\tiny \textit{Notes:} Each metric cell reports the mean followed by the 95\% confidence interval in brackets. Observed pre- and post-disaster confidence intervals are node-resampling 95\% percentile intervals using 500 bootstrap samples with replacement. Counterfactual confidence intervals are 95\% percentile intervals across 500 simulation runs.
\end{flushleft}
\end{minipage}
\end{table}

\begin{table}[!htbp]
\centering
\caption{Network-level and mean node-level metrics by tie type and period for the 30 m, 15 min co-location threshold}
\label{tab:network_node_metrics_combined_cf_95ci_r11_2}
\tiny
\begin{minipage}{\linewidth}
\centering
\begin{tabular}{llccccc}
\hline \hline
Phase & Tie type & Nodes & Edges & Mean degree & Closeness & Clustering \\
\hline
Pre & Bonding & 3224 & 9003 & 5.58 [5.44, 5.74] & 0.33 [0.32, 0.33] & 0.30 [0.29, 0.30] \\
Pre & Unclassified & 2368 & 3389 & 2.86 [2.79, 2.94] & 0.21 [0.21, 0.21] & 0.14 [0.13, 0.15] \\
Pre & Bridging & 532 & 2016 & 7.57 [7.20, 7.92] & 0.50 [0.49, 0.51] & 0.31 [0.30, 0.33] \\
Post & Bonding & 718 & 972 & 2.71 [2.60, 2.84] & 0.16 [0.16, 0.17] & 0.18 [0.17, 0.20] \\
Post & Unclassified & 694 & 942 & 2.71 [2.58, 2.87] & 0.18 [0.18, 0.19] & 0.14 [0.13, 0.15] \\
Post & Bridging & 0 & 0 & 0.00 &  &  \\
Counterfactual & Bonding & 564 [527, 605] & 646 [566, 733] & 2.29 [2.08, 2.52] & 0.08 [0.00, 0.20] & 0.17 [0.00, 1.00] \\
Counterfactual & Unclassified & 551 [511, 591] & 583 [519, 653] & 2.11 [1.96, 2.27] & 0.10 [0.00, 0.22] & 0.10 [0.00, 1.00] \\
Counterfactual & Bridging & 49 [23, 82] & 131 [55, 224] & 5.36 [4.02, 7.22] & 0.47 [0.15, 0.92] & 0.58 [0.00, 1.00] \\
\hline \hline
\end{tabular}
\vspace{1mm}
\begin{flushleft}
\tiny \textit{Notes:} Each metric cell reports the mean followed by the 95\% confidence interval in brackets. Observed pre- and post-disaster confidence intervals are node-resampling 95\% percentile intervals using 500 bootstrap samples with replacement. Counterfactual confidence intervals are 95\% percentile intervals across 500 simulation runs.
\end{flushleft}
\end{minipage}
\end{table}

\begin{table}[!htbp]
\centering
\caption{Network-level and mean node-level metrics by tie type and period for the 40 m, 5 min co-location threshold}
\label{tab:network_node_metrics_combined_cf_95ci_r13_1}
\tiny
\begin{minipage}{\linewidth}
\centering
\begin{tabular}{llccccc}
\hline \hline
Phase & Tie type & Nodes & Edges & Mean degree & Closeness & Clustering \\
\hline
Pre & Bonding & 3682 & 9276 & 5.04 [4.90, 5.16] & 0.32 [0.32, 0.33] & 0.26 [0.26, 0.27] \\
Pre & Unclassified & 2816 & 3809 & 2.71 [2.64, 2.76] & 0.20 [0.20, 0.20] & 0.12 [0.11, 0.12] \\
Pre & Bridging & 536 & 1848 & 6.89 [6.58, 7.20] & 0.47 [0.47, 0.48] & 0.27 [0.26, 0.29] \\
Post & Bonding & 835 & 1096 & 2.63 [2.51, 2.73] & 0.15 [0.15, 0.16] & 0.16 [0.14, 0.17] \\
Post & Unclassified & 848 & 1048 & 2.47 [2.37, 2.57] & 0.17 [0.16, 0.17] & 0.12 [0.11, 0.14] \\
Post & Bridging & 12 & 28 & 4.96 [4.19, 5.78] & 1.36 [1.28, 1.45] & 0.45 [0.38, 0.53] \\
Counterfactual & Bonding & 695 [655, 737] & 798 [709, 890] & 2.30 [2.12, 2.49] & 0.10 [0.00, 0.22] & 0.14 [0.00, 1.00] \\
Counterfactual & Unclassified & 664 [617, 709] & 656 [588, 736] & 1.98 [1.85, 2.13] & 0.08 [0.00, 0.19] & 0.07 [0.00, 1.00] \\
Counterfactual & Bridging & 60 [32, 91] & 166 [78, 262] & 5.55 [4.36, 7.33] & 0.44 [0.16, 0.71] & 0.48 [0.00, 1.00] \\
\hline \hline
\end{tabular}
\vspace{1mm}
\begin{flushleft}
\tiny \textit{Notes:} Each metric cell reports the mean followed by the 95\% confidence interval in brackets. Observed pre- and post-disaster confidence intervals are node-resampling 95\% percentile intervals using 500 bootstrap samples with replacement. Counterfactual confidence intervals are 95\% percentile intervals across 500 simulation runs.
\end{flushleft}
\end{minipage}
\end{table}

\begin{table}[!htbp]
\centering
\caption{Network-level and mean node-level metrics by tie type and period for the 40 m, 15 min co-location threshold}
\label{tab:network_node_metrics_combined_cf_95ci_r13_2}
\tiny
\begin{minipage}{\linewidth}
\centering
\begin{tabular}{llccccc}
\hline \hline
Phase & Tie type & Nodes & Edges & Mean degree & Closeness & Clustering \\
\hline
Pre & Bonding & 2824 & 6556 & 4.64 [4.51, 4.80] & 0.28 [0.28, 0.28] & 0.28 [0.27, 0.29] \\
Pre & Unclassified & 2090 & 2696 & 2.58 [2.52, 2.66] & 0.17 [0.17, 0.18] & 0.13 [0.12, 0.14] \\
Pre & Bridging & 342 & 1168 & 6.84 [6.51, 7.26] & 0.46 [0.45, 0.47] & 0.30 [0.28, 0.32] \\
Post & Bonding & 554 & 641 & 2.31 [2.20, 2.43] & 0.09 [0.09, 0.09] & 0.18 [0.17, 0.20] \\
Post & Unclassified & 580 & 646 & 2.23 [2.13, 2.34] & 0.10 [0.10, 0.11] & 0.11 [0.09, 0.12] \\
Post & Bridging & 5 & 10 & 4.00 [3.50, 4.60] & 1.38 [1.30, 1.46] & 0.26 [0.15, 0.37] \\
Counterfactual & Bonding & 440 [403, 476] & 443 [376, 509] & 2.01 [1.82, 2.21] & 0.05 [0.00, 0.17] & 0.13 [0.00, 1.00] \\
Counterfactual & Unclassified & 453 [415, 492] & 437 [382, 491] & 1.93 [1.79, 2.08] & 0.06 [0.00, 0.17] & 0.09 [0.00, 1.00] \\
Counterfactual & Bridging & 31 [11, 56] & 81 [23, 151] & 5.12 [3.55, 7.18] & 0.54 [0.20, 1.24] & 0.57 [0.00, 1.00] \\
\hline \hline
\end{tabular}
\vspace{1mm}
\begin{flushleft}
\tiny \textit{Notes:} Each metric cell reports the mean followed by the 95\% confidence interval in brackets. Observed pre- and post-disaster confidence intervals are node-resampling 95\% percentile intervals using 500 bootstrap samples with replacement. Counterfactual confidence intervals are 95\% percentile intervals across 500 simulation runs.
\end{flushleft}
\end{minipage}
\end{table}

\begin{table}[!htbp]
\centering
\caption{Network-level and mean node-level metrics by tie type and period for the 50 m, 5 min co-location threshold}
\label{tab:network_node_metrics_combined_cf_95ci_r12_1}
\tiny
\begin{minipage}{\linewidth}
\centering
\begin{tabular}{llccccc}
\hline \hline
Phase & Tie type & Nodes & Edges & Mean degree & Closeness & Clustering \\
\hline
Pre & Bonding & 4624 & 17968 & 7.77 [7.58, 7.99] & 0.43 [0.43, 0.43] & 0.31 [0.30, 0.31] \\
Pre & Unclassified & 3553 & 6160 & 3.47 [3.40, 3.54] & 0.28 [0.28, 0.28] & 0.15 [0.14, 0.15] \\
Pre & Bridging & 958 & 5111 & 10.66 [10.13, 11.20] & 0.62 [0.62, 0.63] & 0.28 [0.27, 0.29] \\
Post & Bonding & 1181 & 2328 & 3.94 [3.78, 4.12] & 0.33 [0.32, 0.33] & 0.22 [0.21, 0.24] \\
Post & Unclassified & 1235 & 2145 & 3.47 [3.32, 3.61] & 0.31 [0.30, 0.31] & 0.16 [0.15, 0.17] \\
Post & Bridging & 70 & 200 & 5.66 [5.16, 6.27] & 0.64 [0.62, 0.67] & 0.22 [0.19, 0.25] \\
Counterfactual & Bonding & 996 [940, 1045] & 1542 [1382, 1709] & 3.09 [2.85, 3.37] & 0.24 [0.00, 0.38] & 0.20 [0.00, 1.00] \\
Counterfactual & Unclassified & 999 [950, 1050] & 1168 [1065, 1271] & 2.34 [2.20, 2.49] & 0.16 [0.00, 0.26] & 0.09 [0.00, 1.00] \\
Counterfactual & Bridging & 145 [93, 193] & 457 [281, 625] & 6.32 [5.22, 7.69] & 0.54 [0.35, 0.73] & 0.36 [0.00, 1.00] \\
\hline \hline
\end{tabular}
\vspace{1mm}
\begin{flushleft}
\tiny \textit{Notes:} Each metric cell reports the mean followed by the 95\% confidence interval in brackets. Observed pre- and post-disaster confidence intervals are node-resampling 95\% percentile intervals using 500 bootstrap samples with replacement. Counterfactual confidence intervals are 95\% percentile intervals across 500 simulation runs.
\end{flushleft}
\end{minipage}
\end{table}

\begin{table}[!htbp]
\centering
\caption{Network-level and mean node-level metrics by tie type and period for the 50 m, 15 min co-location threshold}
\label{tab:network_node_metrics_combined_cf_95ci_r12_2}
\tiny
\begin{minipage}{\linewidth}
\centering
\begin{tabular}{llccccc}
\hline \hline
Phase & Tie type & Nodes & Edges & Mean degree & Closeness & Clustering \\
\hline
Pre & Bonding & 3498 & 12290 & 7.03 [6.83, 7.23] & 0.38 [0.38, 0.39] & 0.32 [0.32, 0.33] \\
Pre & Unclassified & 2601 & 4232 & 3.25 [3.17, 3.34] & 0.25 [0.25, 0.26] & 0.16 [0.15, 0.17] \\
Pre & Bridging & 668 & 3221 & 9.65 [9.19, 10.20] & 0.59 [0.59, 0.60] & 0.29 [0.28, 0.30] \\
Post & Bonding & 791 & 1338 & 3.38 [3.23, 3.56] & 0.24 [0.24, 0.25] & 0.23 [0.21, 0.24] \\
Post & Unclassified & 830 & 1267 & 3.05 [2.92, 3.20] & 0.23 [0.23, 0.24] & 0.17 [0.15, 0.18] \\
Post & Bridging & 32 & 80 & 4.95 [4.42, 5.51] & 0.77 [0.70, 0.84] & 0.19 [0.15, 0.23] \\
Counterfactual & Bonding & 647 [602, 688] & 869 [763, 974] & 2.69 [2.46, 2.93] & 0.15 [0.00, 0.29] & 0.20 [0.00, 1.00] \\
Counterfactual & Unclassified & 644 [599, 690] & 733 [651, 812] & 2.28 [2.10, 2.43] & 0.13 [0.00, 0.25] & 0.11 [0.00, 1.00] \\
Counterfactual & Bridging & 77 [43, 115] & 217 [111, 330] & 5.58 [4.52, 6.78] & 0.50 [0.19, 0.74] & 0.40 [0.00, 1.00] \\
\hline \hline
\end{tabular}
\vspace{1mm}
\begin{flushleft}
\tiny \textit{Notes:} Each metric cell reports the mean followed by the 95\% confidence interval in brackets. Observed pre- and post-disaster confidence intervals are node-resampling 95\% percentile intervals using 500 bootstrap samples with replacement. Counterfactual confidence intervals are 95\% percentile intervals across 500 simulation runs.
\end{flushleft}
\end{minipage}
\end{table}

\section{POI-level residual interaction models}

To examine whether excess post-disaster interaction becomes concentrated in particular third places, we estimate POI-level models in which the dependent variable is the standardized residual change in interaction intensity relative to the behavior-informed counterfactual,
\[
Z_{\log,i}=\frac{\log(\mathrm{post}_i)-\mu\left(\log(\mathrm{rand}_i)\right)}{\sigma\left(\log(\mathrm{rand}_i)\right)}.
\]
We fit models separately for all ties, bonding ties, bridging ties, and unclassified ties. The main text focuses on the bonding-tie model because the tie-type results show that the residual robustness is concentrated disproportionately in bonding ties. The full regression outputs are reported in Table~\ref{tab:zlog_regression_all_panels}.

Table~\ref{tab:zlog_regression_all_panels} shows that positive residual deviations are strongest and most consistent in the bonding-tie model, whereas the bridging-tie model exhibits substantially weaker and often statistically insignificant category effects. This extended table therefore supports the main-text interpretation that post-disaster excess interaction becomes concentrated in specific third places through the reinforcement of already embedded, socially similar ties.

\subsection{Interactive POI visitation deviation map}

We provide an interactive supplementary map to visualize how observed post-disaster visitation to third-place POIs differs from counterfactual-predicted post-disaster visitation. Each polygon represents a third-place POI and includes aggregated tooltip information on pre-disaster visits, observed post-disaster visits, POI category, and standardized deviation from the counterfactual prediction. The map is colored by $z_{\log}$, a log-adjusted standardized deviation score. Negative values indicate that observed post-disaster visitation was lower than predicted by the counterfactual model, while positive values indicate that observed post-disaster visitation was higher than predicted. The log adjustment accounts for baseline differences in POI visitation intensity, reducing the influence of highly visited locations when comparing deviations across POIs. The visualization is based only on aggregated POI-level measures and does not include individual-level mobility traces. This map can be found at \href{https://github.com/VaidehiRaipat/Social-networks/tree/main/Maps}{Third Place POI Map}{GitHub repository}.

%%_____________________________________________________________________
%%_____________________________________________________________________
\begin{figure}[h]
    \centering
    \includegraphics[width=0.8\linewidth]{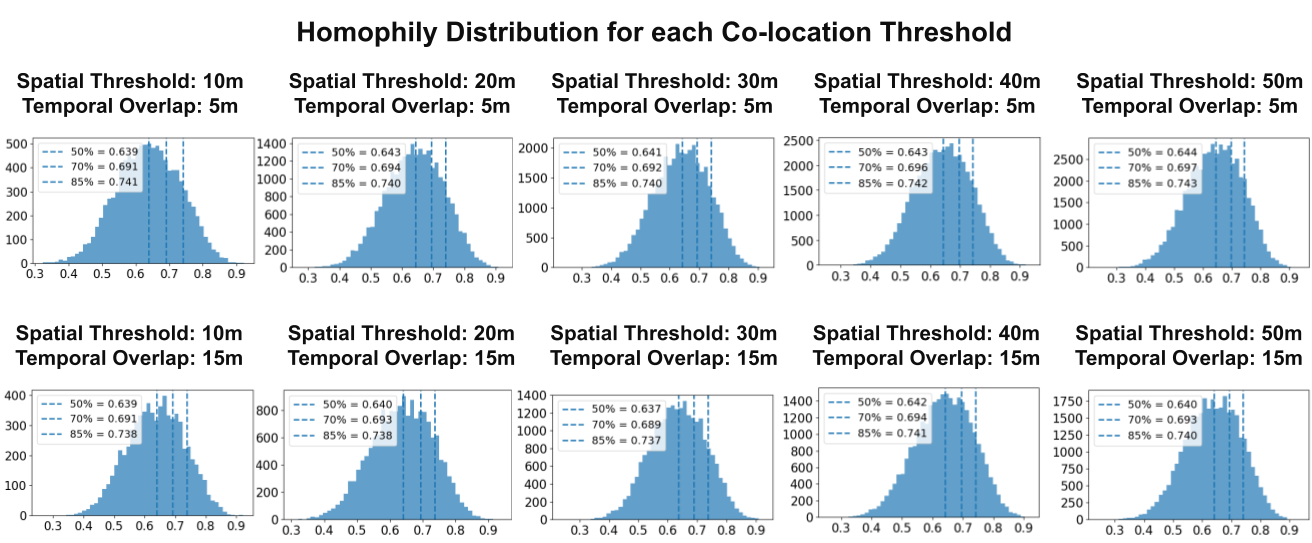}
    \captionsetup{singlelinecheck=off}
    \caption[Distribution of estimated dyadic homophily across co-location thresholds.]{
    \textbf{Distribution of estimated dyadic homophily across co-location thresholds.}
    Distribution of mean dyadic race--income similarity scores obtained from repeated probabilistic assignment of latent socio-demographic attributes based on users' home CBG demographic distributions, shown for each co-location graph definition. Panels compare alternative spatial thresholds (10\,m, 20\,m, 30\,m, 40\,m, and 50\,m) under 5-minute and 15-minute temporal overlap requirements. For each threshold combination, vertical dashed lines indicate the 50th, 70th, and 85th percentiles of the pre-disaster homophily distribution, which are used in the supplementary robustness analyses as alternative cutoffs for classifying bonding ties. The distribution remains broadly stable across co-location thresholds, indicating that the homophily-based tie classification is not driven by a single graph specification.}
    \label{fig:Figure_S_similarity}
\end{figure}

% Supplementary robustness figures: tie classification sensitivity

\begin{figure}[h]
    \centering
    \includegraphics[width=0.8\linewidth]{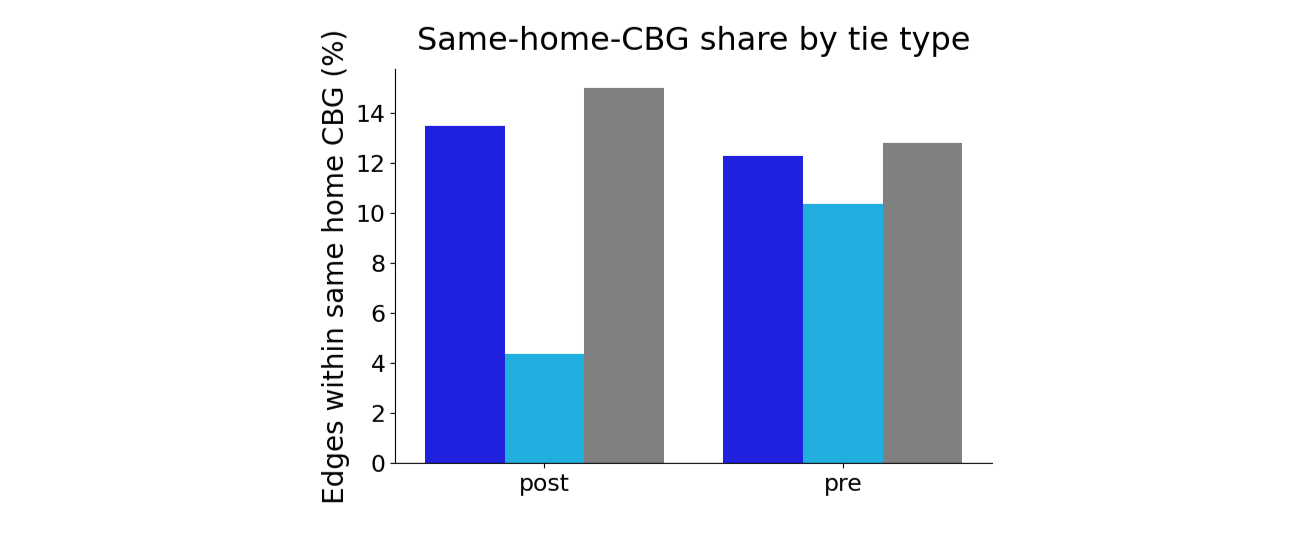}
    \captionsetup{singlelinecheck=off}
    \caption[Share of ties connecting individuals from the same home Census Block Group (CBG), by tie type, before and after the disaster.]{
    \textbf{Share of ties connecting individuals from the same home Census Block Group (CBG), by tie type, before and after the disaster.} The figure shows the percentage of edges in each tie category that connect individuals who share the same home CBG in the pre-disaster and post-disaster networks. In both periods, bonding ties exhibit a higher same-home-CBG share than bridging ties, consistent with the interpretation that bonding ties are more locally embedded and socially homogeneous.}
    \label{fig:similarity_cbg}
\end{figure}

\begin{figure}[h]
    \centering
    \includegraphics[width=0.95\linewidth]{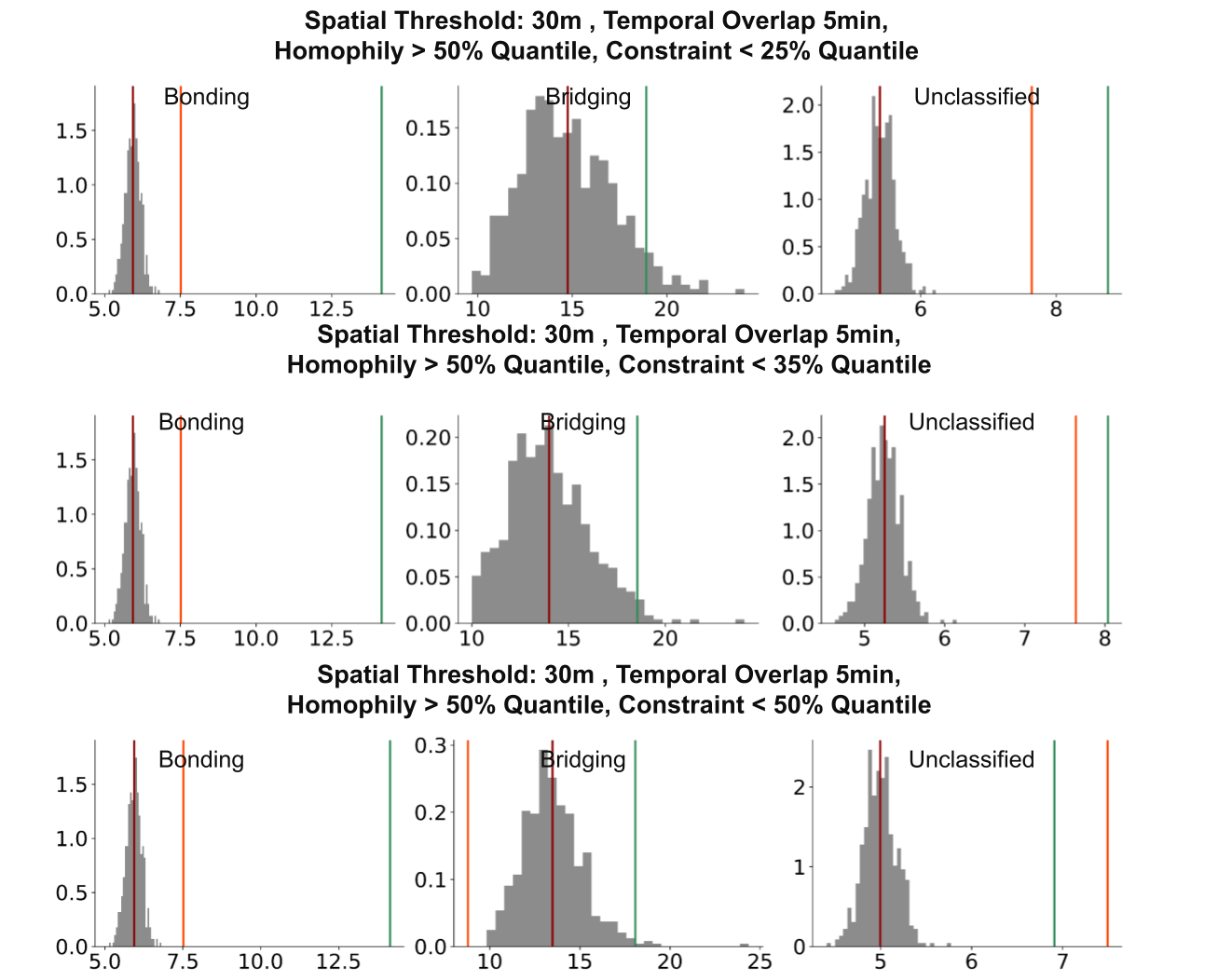}
    \captionsetup{singlelinecheck=off}
    \caption[Robustness of mean weighted degree under a 30\,m spatial threshold, 5-minute temporal overlap, and 50th-percentile homophily cutoff.]{
    \textbf{Robustness of mean weighted degree under a 30\,m spatial threshold, 5-minute temporal overlap, and 50th-percentile homophily cutoff.}
    The figure shows mean weighted degree distributions across 500 behavior-based counterfactual runs for the bonding, refined bridging, and unclassified subnetworks under the specified co-location and homophily thresholds. Histograms represent the counterfactual distributions, while vertical reference lines mark the observed pre-disaster and post-disaster values. This main specification replicates the central result: bonding and unclassified subnetworks remain more connected after the disaster than expected under the behavior-informed counterfactual, whereas refined bridging connectivity is weaker than expected.}
    \label{fig:Figure_S_tie_robustness_30x5x50}
\end{figure}

\begin{figure}[h]
    \centering
    \includegraphics[width=0.95\linewidth]{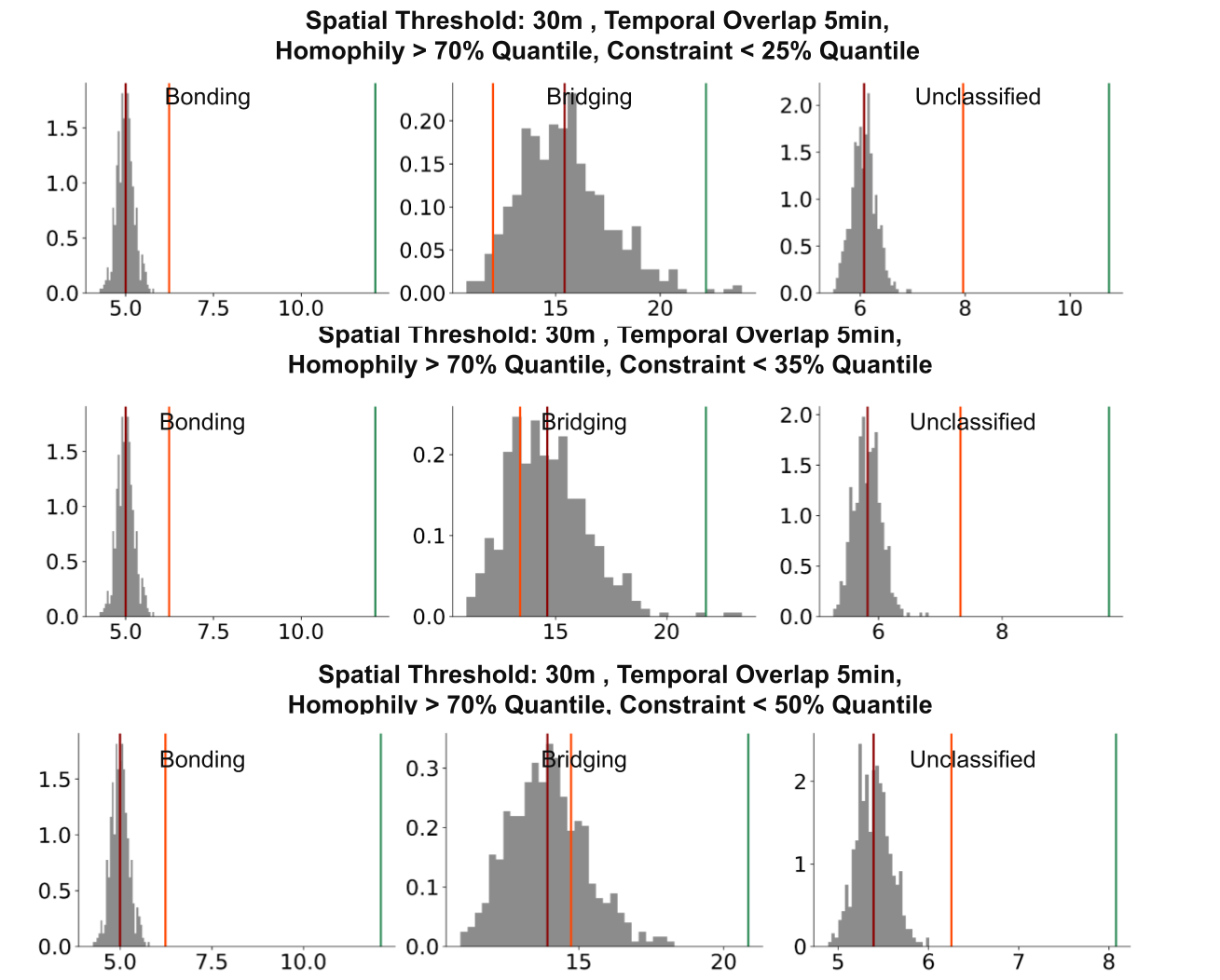}
    \captionsetup{singlelinecheck=off}
    \caption[Robustness of mean weighted degree under a 30\,m spatial threshold, 5-minute temporal overlap, and 70th-percentile homophily cutoff.]{
    \textbf{Robustness of mean weighted degree under a 30\,m spatial threshold, 5-minute temporal overlap, and 70th-percentile homophily cutoff.}
    The figure shows mean weighted degree distributions across 500 behavior-based counterfactual runs for the bonding, refined bridging, and unclassified subnetworks under the specified co-location and homophily thresholds. Histograms represent the counterfactual distributions, while vertical reference lines mark the observed pre-disaster and post-disaster values. Despite the stricter homophily cutoff, this specification replicates the main result: bonding and unclassified subnetworks remain more connected after the disaster than expected under the behavior-informed counterfactual, whereas refined bridging connectivity is weaker than expected.}
    \label{fig:Figure_S_tie_robustness_30x5x70}
\end{figure}

\begin{figure}[h]
    \centering
    \includegraphics[width=0.95\linewidth]{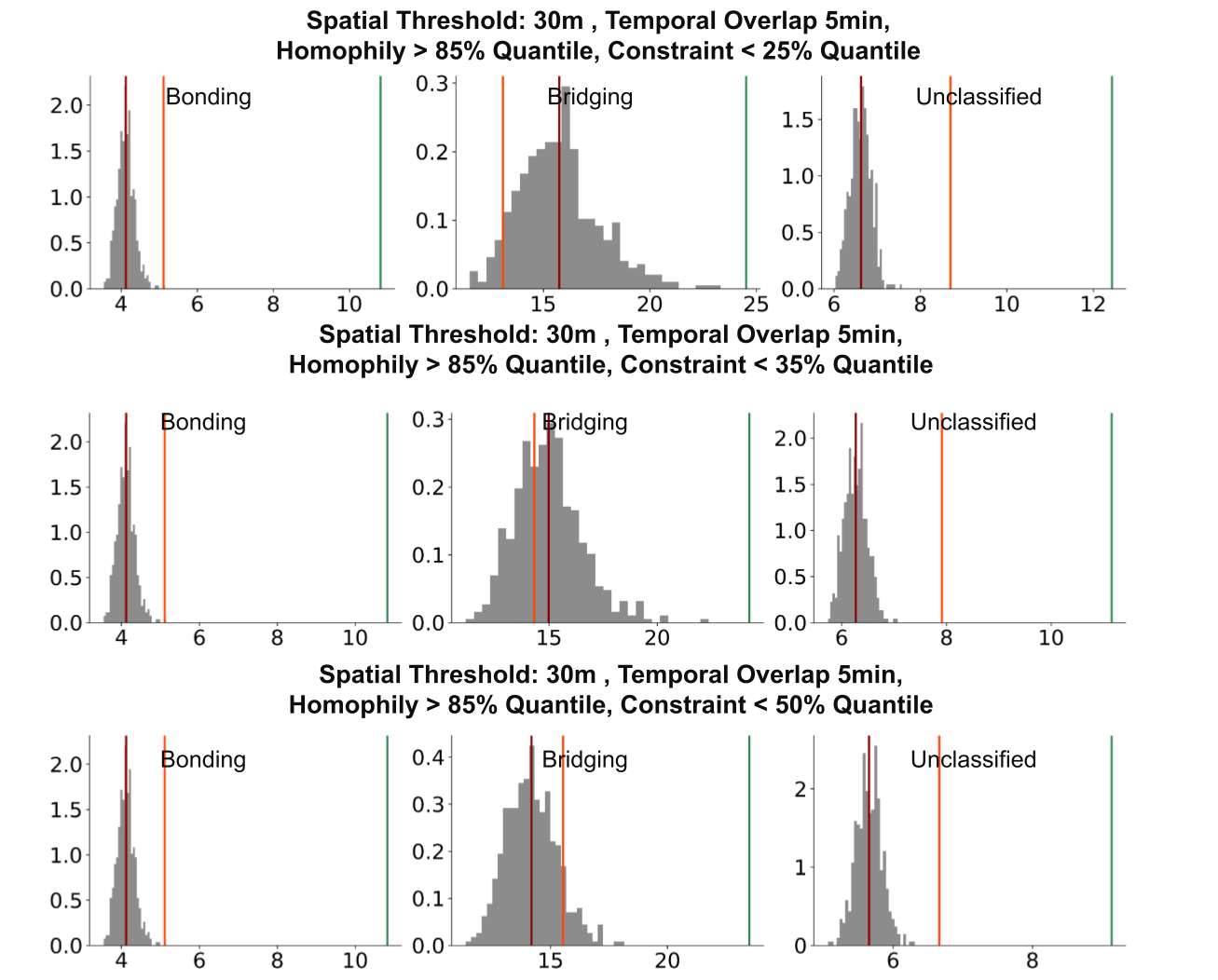}
    \captionsetup{singlelinecheck=off}
    \caption[Robustness of mean weighted degree under a 30\,m spatial threshold, 5-minute temporal overlap, and 85th-percentile homophily cutoff.]{
    \textbf{Robustness of mean weighted degree under a 30\,m spatial threshold, 5-minute temporal overlap, and 85th-percentile homophily cutoff.}
    The figure shows mean weighted degree distributions across 500 behavior-based counterfactual runs for the bonding, refined bridging, and unclassified subnetworks under the specified co-location and homophily thresholds. Histograms represent the counterfactual distributions, while vertical reference lines mark the observed pre-disaster and post-disaster values. Even under this highly selective definition of bonding, the main pattern is reproduced: bonding and unclassified subnetworks remain more connected after the disaster than expected under the behavior-informed counterfactual, whereas refined bridging connectivity is weaker than expected.}
    \label{fig:Figure_S_tie_robustness_30x5x85}
\end{figure}

\begin{figure}[h]
    \centering
    \includegraphics[width=0.95\linewidth]{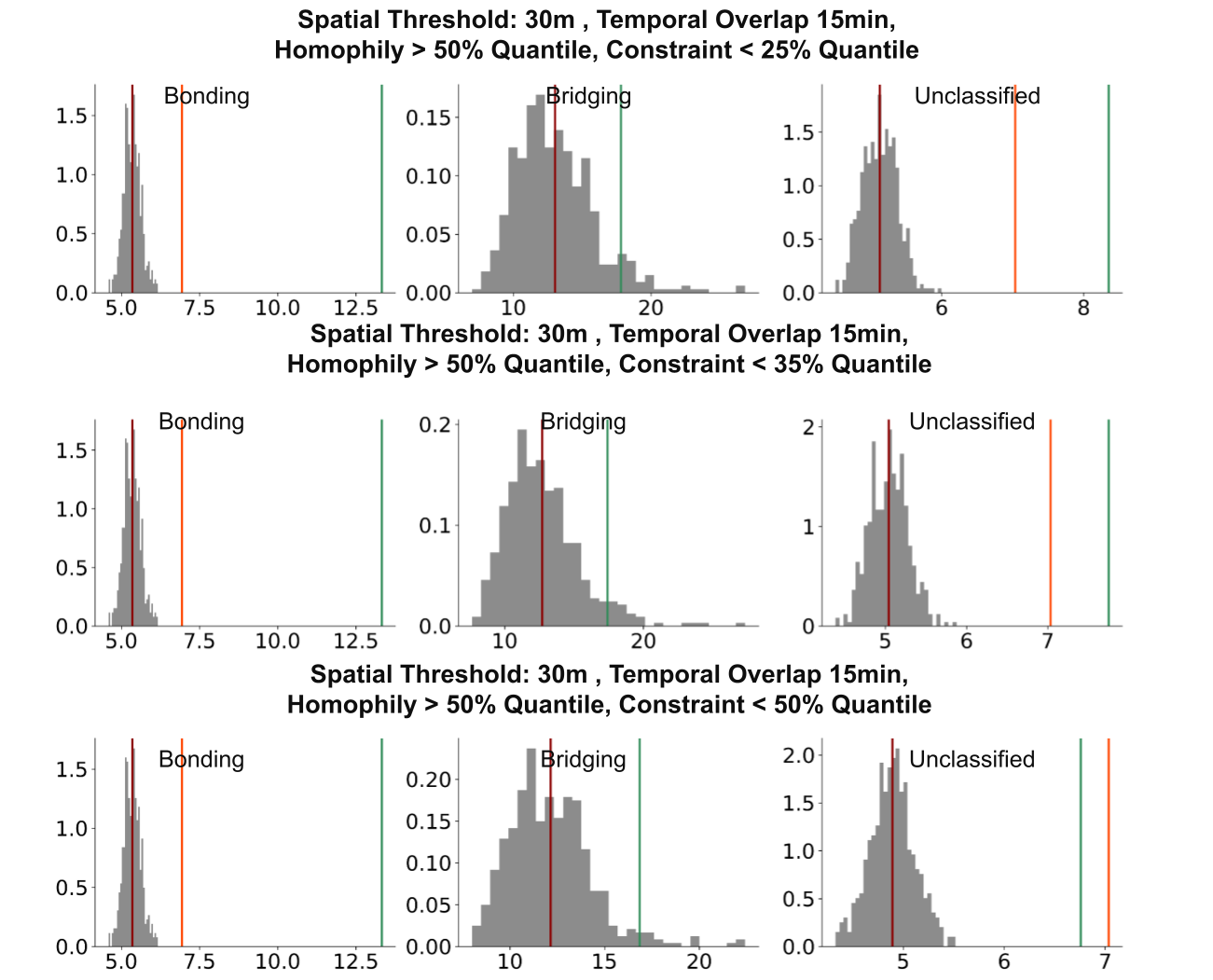}
    \captionsetup{singlelinecheck=off}
    \caption[Robustness of mean weighted degree under a 30\,m spatial threshold, 15-minute temporal overlap, and 50th-percentile homophily cutoff.]{
    \textbf{Robustness of mean weighted degree under a 30\,m spatial threshold, 15-minute temporal overlap, and 50th-percentile homophily cutoff.}
    The figure shows mean weighted degree distributions across 500 behavior-based counterfactual runs for the bonding, refined bridging, and unclassified subnetworks under the specified co-location and homophily thresholds. Histograms represent the counterfactual distributions, while vertical reference lines mark the observed pre-disaster and post-disaster values. Although the 15-minute overlap requirement produces a sparser network than the main specification, the result remains robust: bonding and unclassified subnetworks remain more connected after the disaster than expected, whereas refined bridging connectivity is weaker than expected.}
    \label{fig:Figure_S_tie_robustness_30x15x50}
\end{figure}

\begin{figure}[h]
    \centering
    \includegraphics[width=0.95\linewidth]{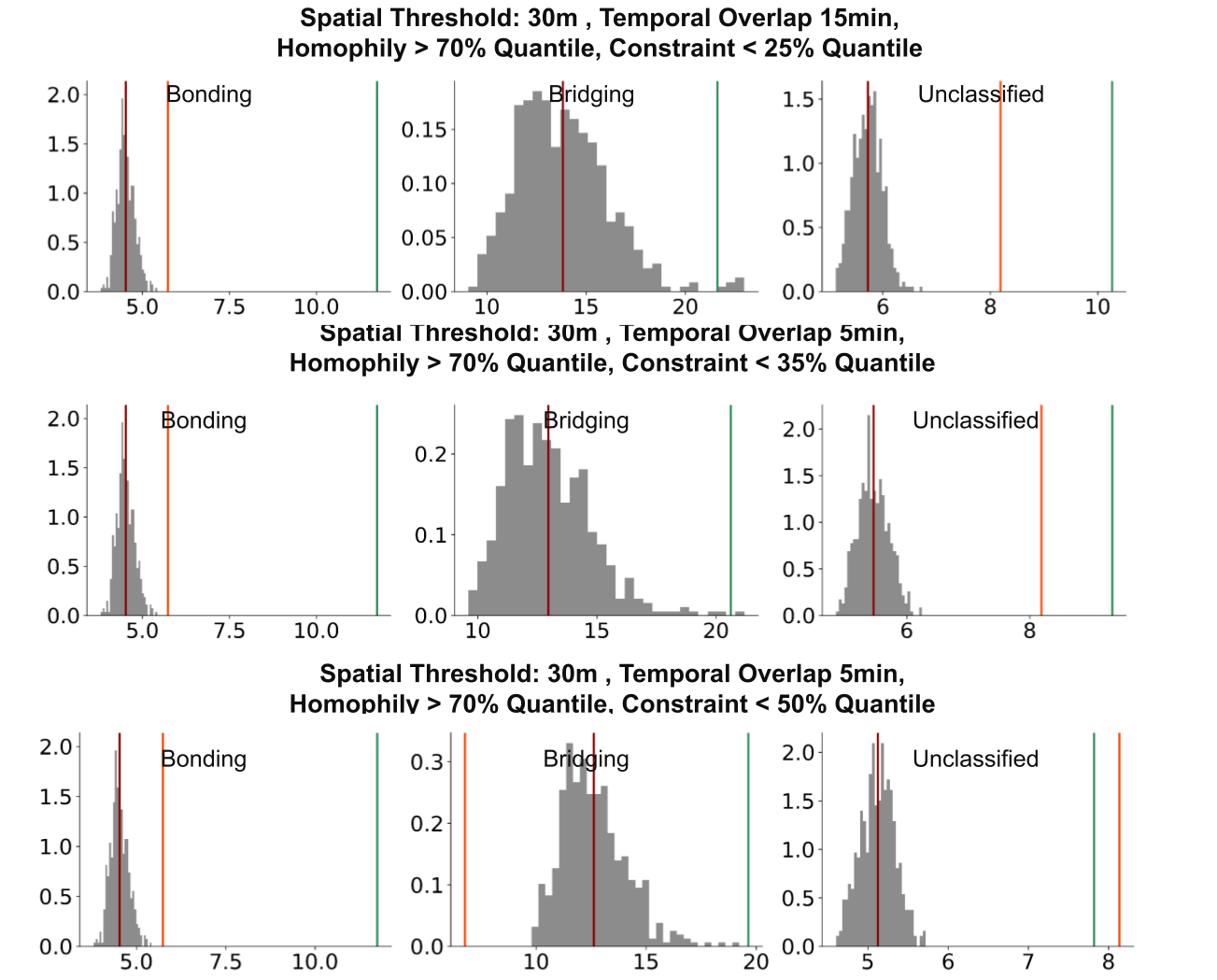}
    \captionsetup{singlelinecheck=off}
    \caption[Robustness of mean weighted degree under a 30\,m spatial threshold, 15-minute temporal overlap, and 70th-percentile homophily cutoff.]{
    \textbf{Robustness of mean weighted degree under a 30\,m spatial threshold, 15-minute temporal overlap, and 70th-percentile homophily cutoff.}
    The figure shows mean weighted degree distributions across 500 behavior-based counterfactual runs for the bonding, refined bridging, and unclassified subnetworks under the specified co-location and homophily thresholds. Histograms represent the counterfactual distributions, while vertical reference lines mark the observed pre-disaster and post-disaster values. With both a stricter temporal overlap rule and a stricter homophily cutoff, the robustness pattern persists: bonding and unclassified subnetworks remain more connected than expected under the behavior-informed counterfactual, whereas refined bridging connectivity is weaker than expected.}
    \label{fig:Figure_S_tie_robustness_30x15x70}
\end{figure}

\begin{figure}[h]
    \centering
    \includegraphics[width=0.95\linewidth]{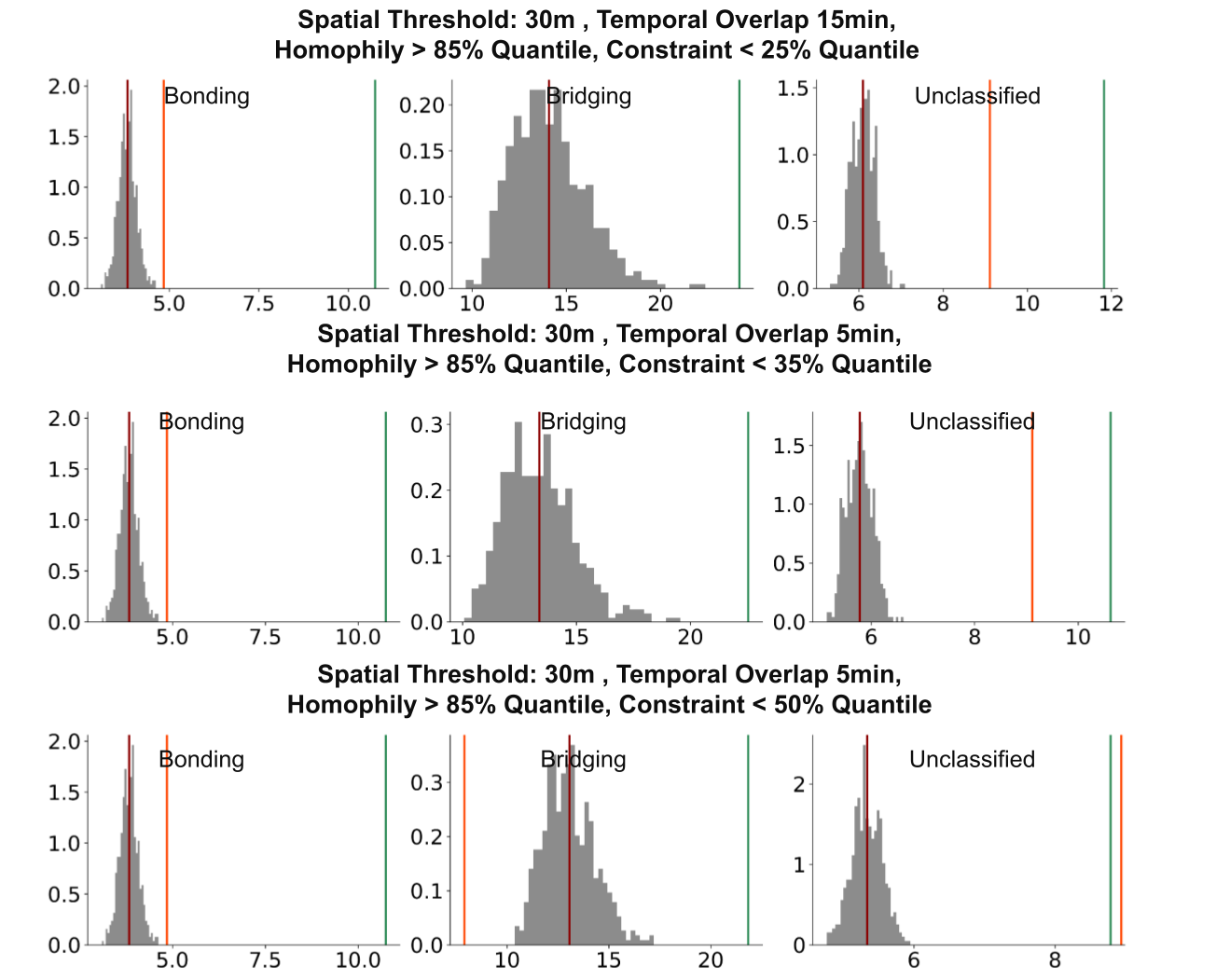}
    \captionsetup{singlelinecheck=off}
    \caption[Robustness of mean weighted degree under a 30\,m spatial threshold, 15-minute temporal overlap, and 85th-percentile homophily cutoff.]{
    \textbf{Robustness of mean weighted degree under a 30\,m spatial threshold, 15-minute temporal overlap, and 85th-percentile homophily cutoff.}
    The figure shows mean weighted degree distributions across 500 behavior-based counterfactual runs for the bonding, refined bridging, and unclassified subnetworks under the specified co-location and homophily thresholds. Histograms represent the counterfactual distributions, while vertical reference lines mark the observed pre-disaster and post-disaster values. Even under this conservative specification, the main result is reproduced: bonding and unclassified subnetworks remain more connected than expected, whereas refined bridging connectivity is weaker than expected.}
    \label{fig:Figure_S_tie_robustness_30x15x85}
\end{figure}

\begin{figure}[h]
    \centering
    \includegraphics[width=0.95\linewidth]{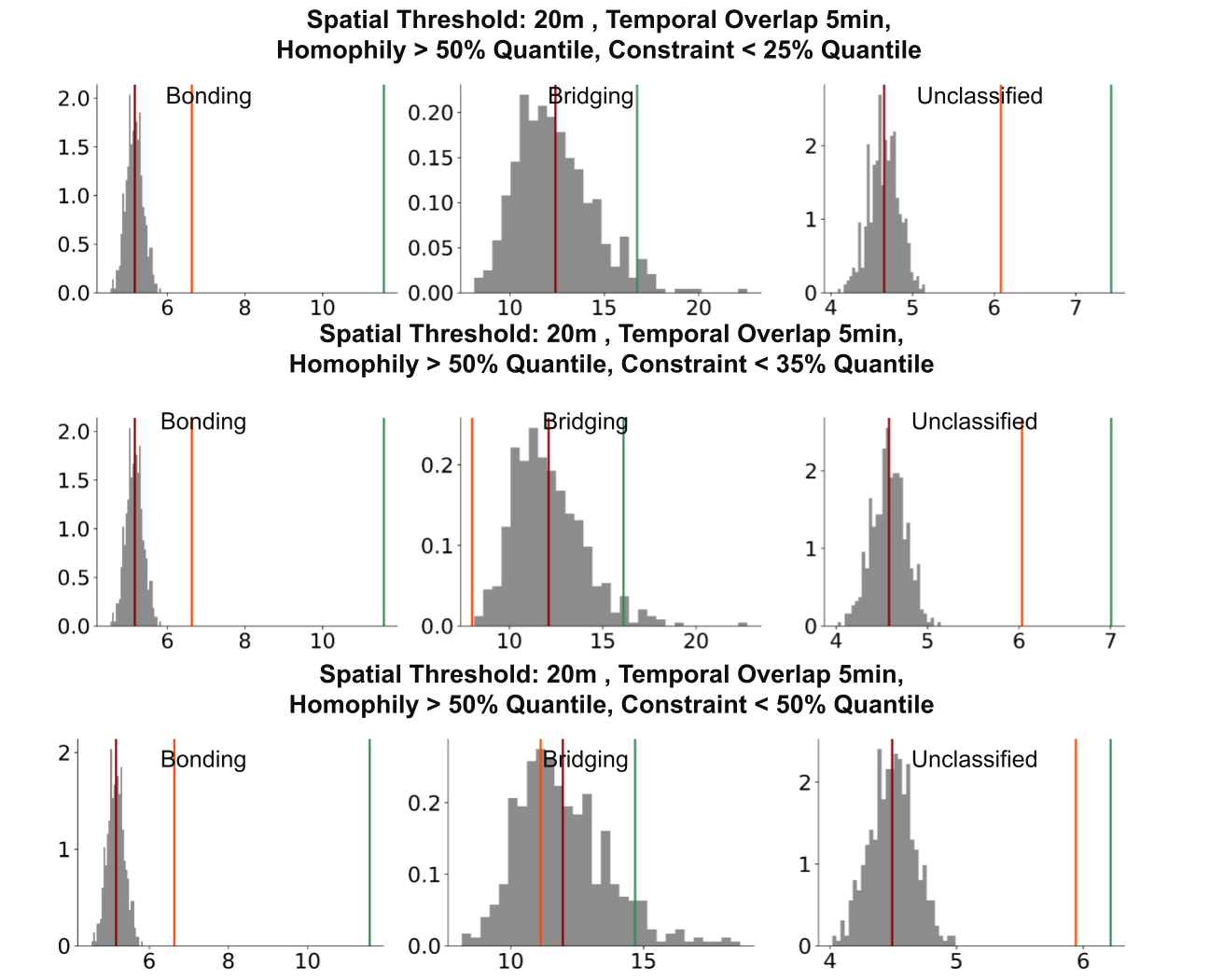}
    \captionsetup{singlelinecheck=off}
    \caption[Robustness of mean weighted degree under a 20\,m spatial threshold, 5-minute temporal overlap, and 50th-percentile homophily cutoff.]{
    \textbf{Robustness of mean weighted degree under a 20\,m spatial threshold, 5-minute temporal overlap, and 50th-percentile homophily cutoff.}
    The figure shows mean weighted degree distributions across 500 behavior-based counterfactual runs for the bonding, refined bridging, and unclassified subnetworks under the specified co-location and homophily thresholds. Histograms represent the counterfactual distributions, while vertical reference lines mark the observed pre-disaster and post-disaster values. Under this more restrictive spatial threshold, the main result remains stable: bonding and unclassified subnetworks remain more connected than expected under the behavior-informed counterfactual, while refined bridging connectivity is weaker than expected.}
    \label{fig:Figure_S_tie_robustness_20x5x50}
\end{figure}

\begin{figure}[h]
    \centering
    \includegraphics[width=0.95\linewidth]{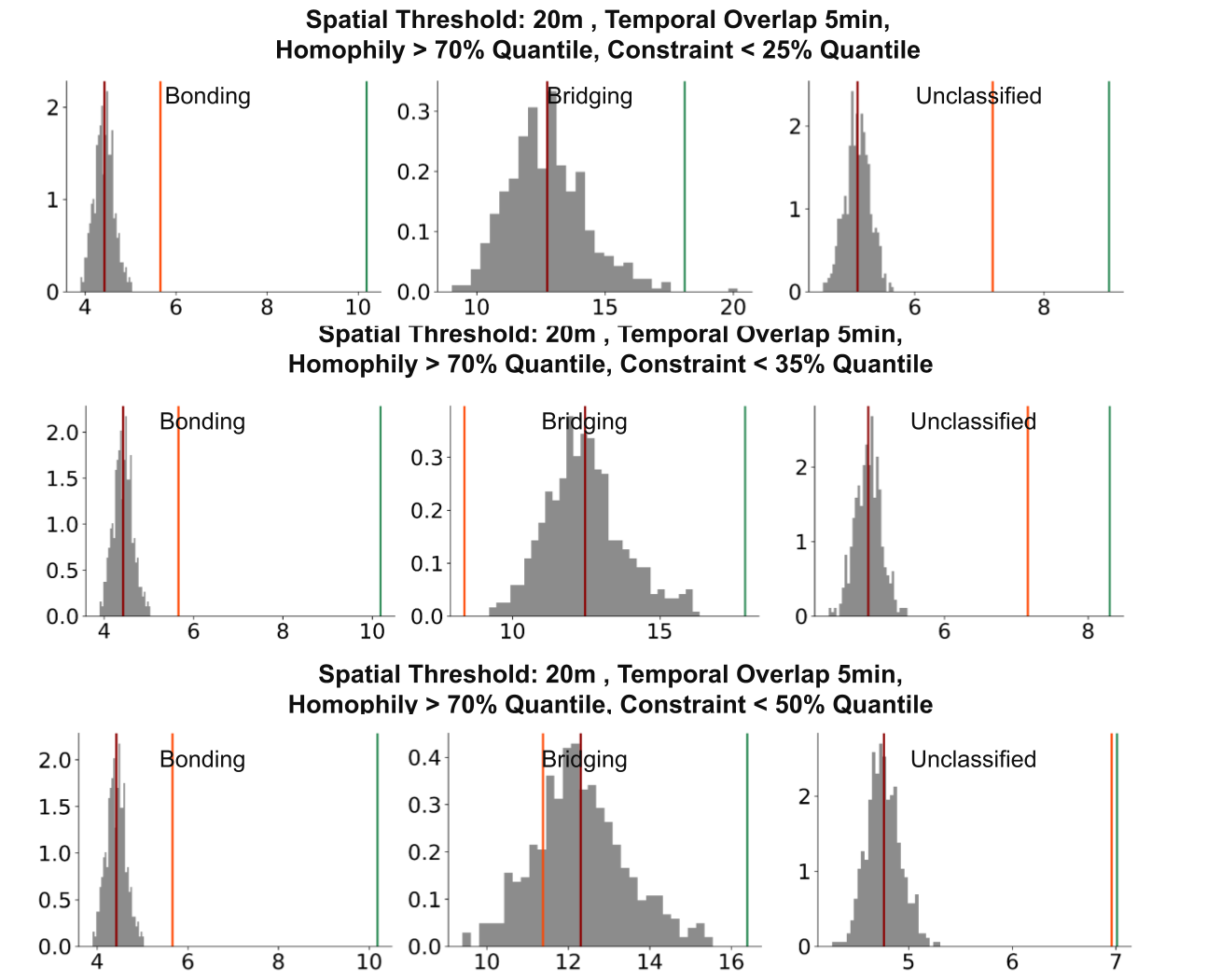}
    \captionsetup{singlelinecheck=off}
    \caption[Robustness of mean weighted degree under a 20\,m spatial threshold, 5-minute temporal overlap, and 70th-percentile homophily cutoff.]{
    \textbf{Robustness of mean weighted degree under a 20\,m spatial threshold, 5-minute temporal overlap, and 70th-percentile homophily cutoff.}
    The figure shows mean weighted degree distributions across 500 behavior-based counterfactual runs for the bonding, refined bridging, and unclassified subnetworks under the specified co-location and homophily thresholds. Histograms represent the counterfactual distributions, while vertical reference lines mark the observed pre-disaster and post-disaster values. The stricter homophily cutoff reduces the bonding set, but the main qualitative result remains unchanged: bonding and unclassified subnetworks remain more connected than expected, whereas refined bridging connectivity is weaker than expected.}
    \label{fig:Figure_S_tie_robustness_20x5x70}
\end{figure}

\begin{figure}[h]
    \centering
    \includegraphics[width=0.95\linewidth]{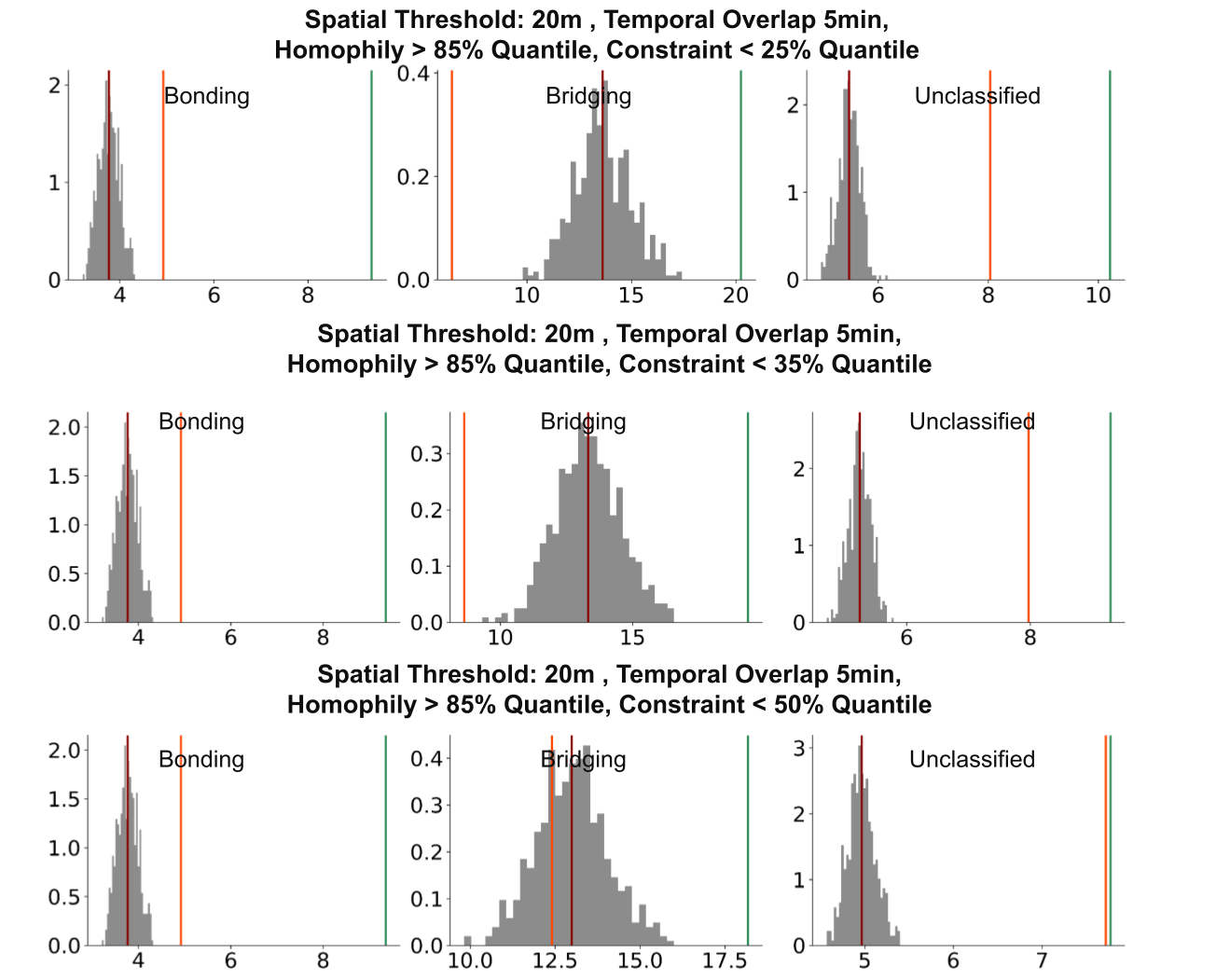}
    \captionsetup{singlelinecheck=off}
    \caption[Robustness of mean weighted degree under a 20\,m spatial threshold, 5-minute temporal overlap, and 85th-percentile homophily cutoff.]{
    \textbf{Robustness of mean weighted degree under a 20\,m spatial threshold, 5-minute temporal overlap, and 85th-percentile homophily cutoff.}
    The figure shows mean weighted degree distributions across 500 behavior-based counterfactual runs for the bonding, refined bridging, and unclassified subnetworks under the specified co-location and homophily thresholds. Histograms represent the counterfactual distributions, while vertical reference lines mark the observed pre-disaster and post-disaster values. Even with a highly selective bonding definition, the robustness pattern is reproduced: bonding and unclassified subnetworks remain more connected than expected, while refined bridging connectivity is weaker than expected.}
    \label{fig:Figure_S_tie_robustness_20x5x85}
\end{figure}

\begin{figure}[h]
    \centering
    \includegraphics[width=0.95\linewidth]{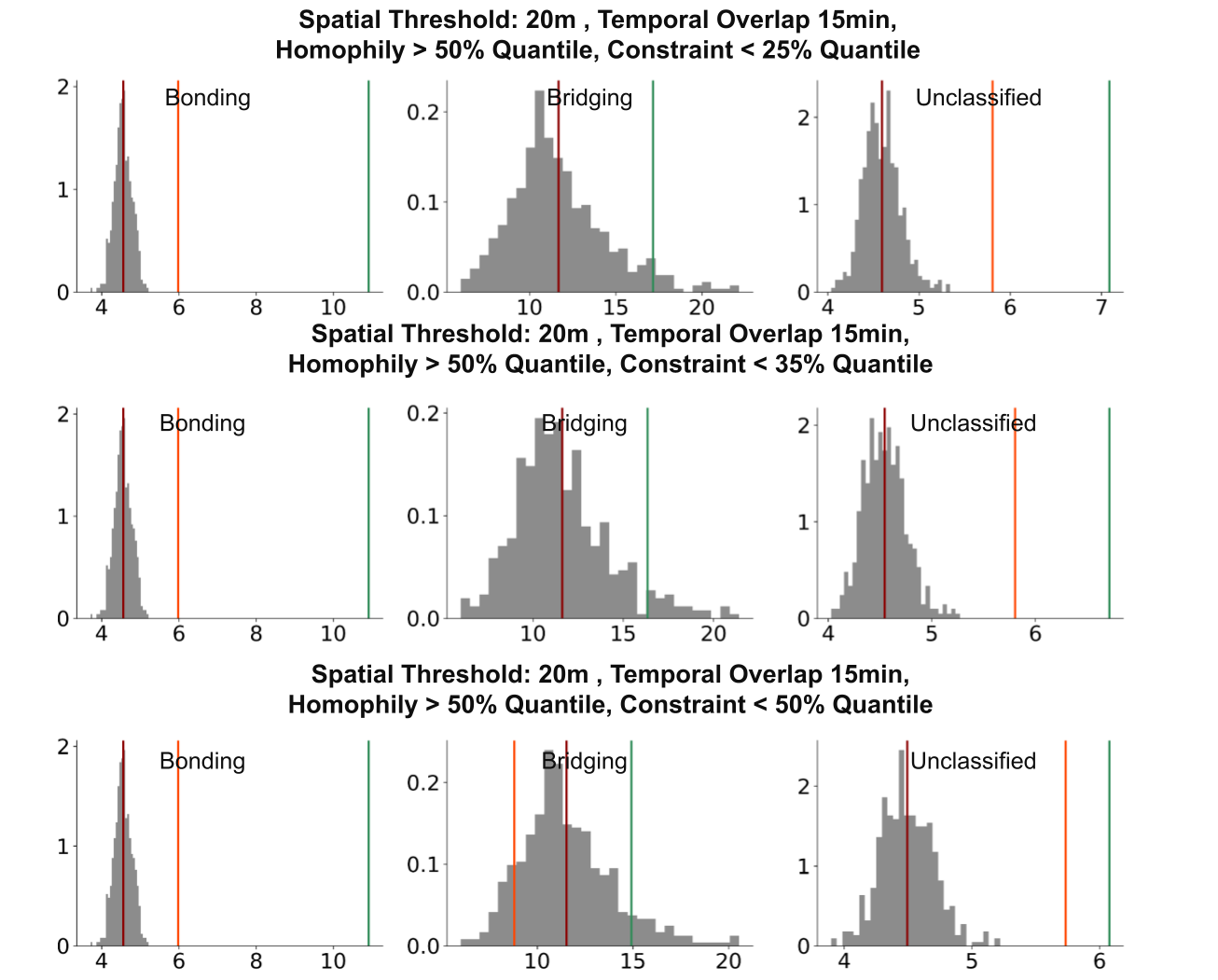}
    \captionsetup{singlelinecheck=off}
    \caption[Robustness of mean weighted degree under a 20\,m spatial threshold, 15-minute temporal overlap, and 50th-percentile homophily cutoff.]{
    \textbf{Robustness of mean weighted degree under a 20\,m spatial threshold, 15-minute temporal overlap, and 50th-percentile homophily cutoff.}
    The figure shows mean weighted degree distributions across 500 behavior-based counterfactual runs for the bonding, refined bridging, and unclassified subnetworks under the specified co-location and homophily thresholds. Histograms represent the counterfactual distributions, while vertical reference lines mark the observed pre-disaster and post-disaster values. Although this specification is restrictive in both space and time, it replicates the main result: bonding and unclassified subnetworks remain more connected after the disaster than expected, whereas refined bridging connectivity is weaker than expected.}
    \label{fig:Figure_S_tie_robustness_20x15x50}
\end{figure}

\begin{figure}[h]
    \centering
    \includegraphics[width=0.95\linewidth]{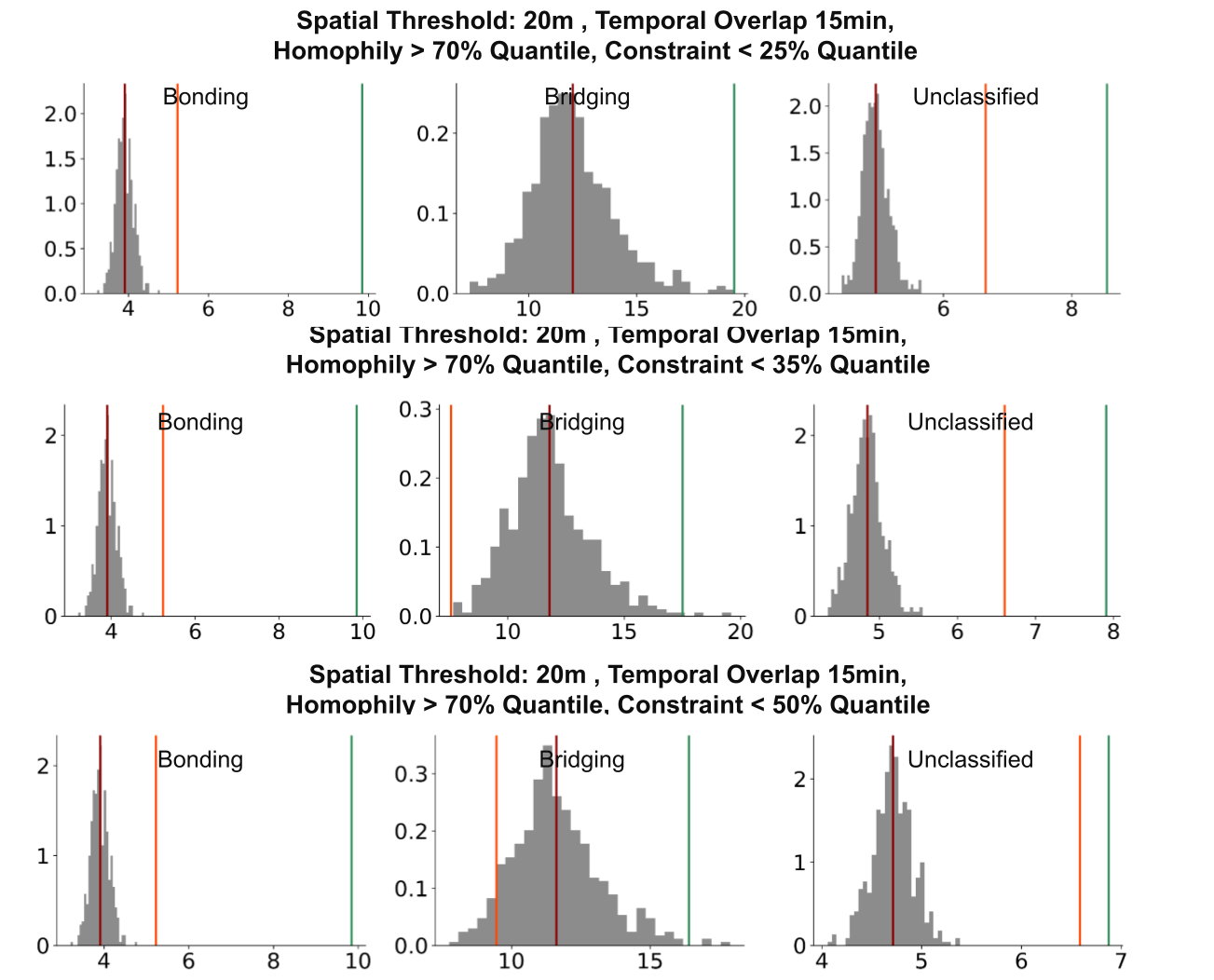}
    \captionsetup{singlelinecheck=off}
    \caption[Robustness of mean weighted degree under a 20\,m spatial threshold, 15-minute temporal overlap, and 70th-percentile homophily cutoff.]{
    \textbf{Robustness of mean weighted degree under a 20\,m spatial threshold, 15-minute temporal overlap, and 70th-percentile homophily cutoff.}
    The figure shows mean weighted degree distributions across 500 behavior-based counterfactual runs for the bonding, refined bridging, and unclassified subnetworks under the specified co-location and homophily thresholds. Histograms represent the counterfactual distributions, while vertical reference lines mark the observed pre-disaster and post-disaster values. The same broad pattern remains visible under this stricter specification: bonding and unclassified subnetworks remain more connected than expected, while refined bridging connectivity is weaker than expected.}
    \label{fig:Figure_S_tie_robustness_20x15x70}
\end{figure}

\begin{figure}[h]
    \centering
    \includegraphics[width=0.95\linewidth]{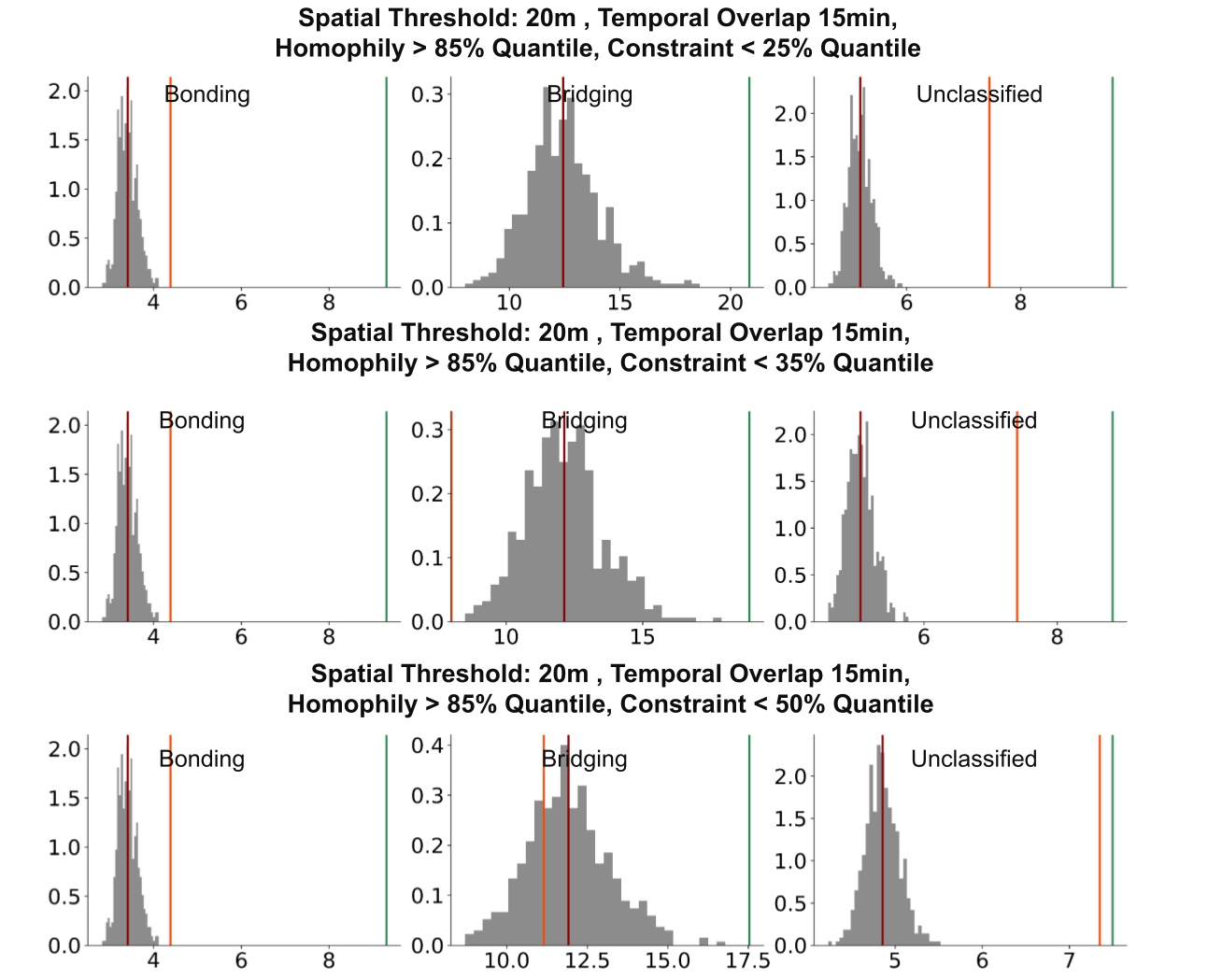}
    \captionsetup{singlelinecheck=off}
    \caption[Robustness of mean weighted degree under a 20\,m spatial threshold, 15-minute temporal overlap, and 85th-percentile homophily cutoff.]{
    \textbf{Robustness of mean weighted degree under a 20\,m spatial threshold, 15-minute temporal overlap, and 85th-percentile homophily cutoff.}
    The figure shows mean weighted degree distributions across 500 behavior-based counterfactual runs for the bonding, refined bridging, and unclassified subnetworks under the specified co-location and homophily thresholds. Histograms represent the counterfactual distributions, while vertical reference lines mark the observed pre-disaster and post-disaster values. Even under this conservative specification, the main result holds: bonding and unclassified subnetworks remain more connected than expected, whereas refined bridging connectivity is weaker than expected.}
    \label{fig:Figure_S_tie_robustness_20x15x85}
\end{figure}

\begin{figure}[h]
    \centering
    \includegraphics[width=0.95\linewidth]{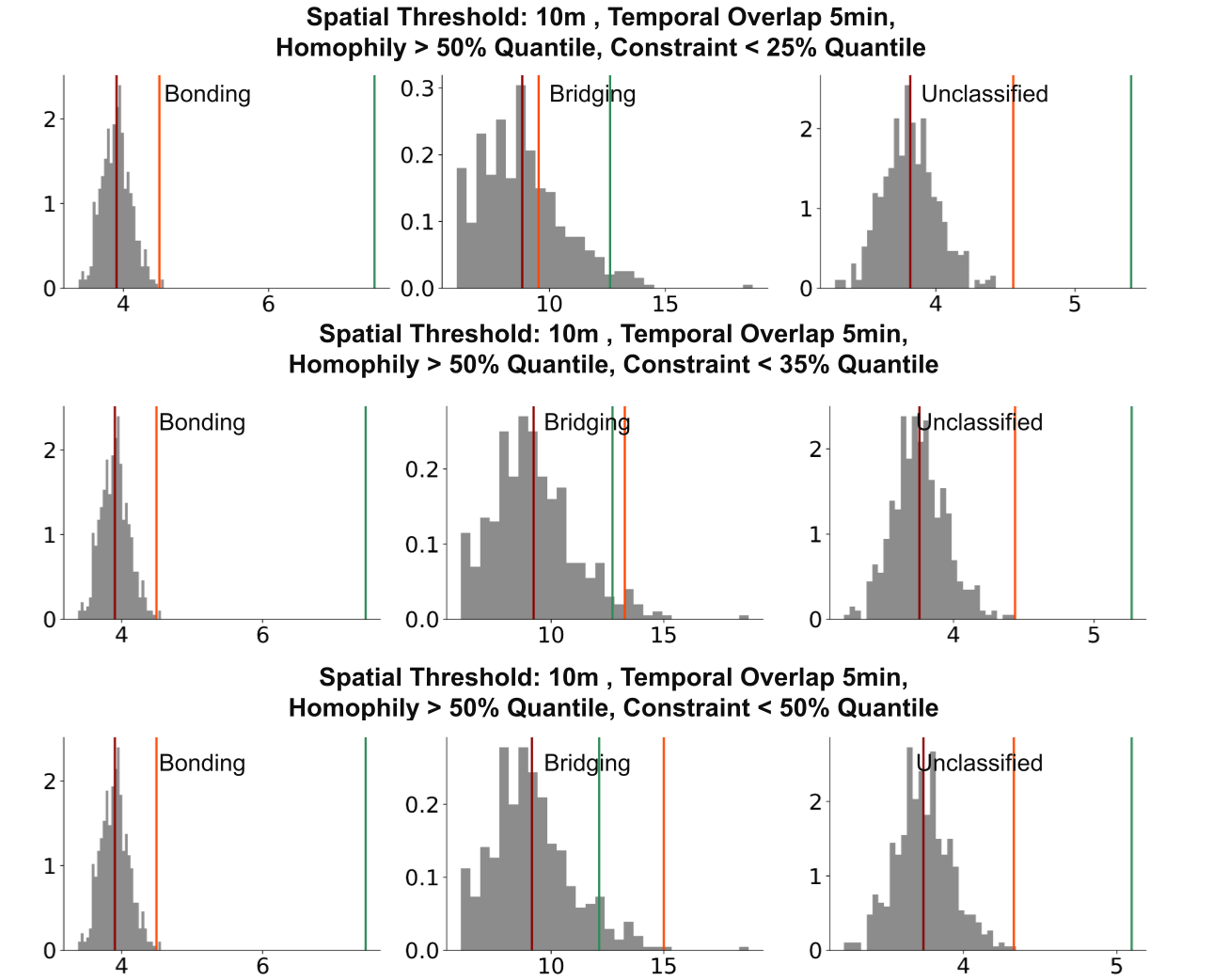}
    \captionsetup{singlelinecheck=off}
    \caption[Mean weighted degree under a restrictive 10\,m spatial threshold, 5-minute temporal overlap, and 50th-percentile homophily cutoff.]{
    \textbf{Mean weighted degree under a restrictive 10\,m spatial threshold, 5-minute temporal overlap, and 50th-percentile homophily cutoff.}
    The figure shows mean weighted degree distributions across 500 behavior-based counterfactual runs for the bonding, refined bridging, and unclassified subnetworks under a highly restrictive 10\,m spatial co-location threshold. Histograms represent the counterfactual distributions, while vertical reference lines mark the observed pre-disaster and post-disaster values. Because the 10\,m threshold produces a sparse graph with too few refined bridging ties for stable comparison, these results are interpreted cautiously and are not treated as primary robustness evidence.}
    \label{fig:Figure_S_tie_robustness_10x5x50}
\end{figure}

\begin{figure}[h]
    \centering
    \includegraphics[width=0.95\linewidth]{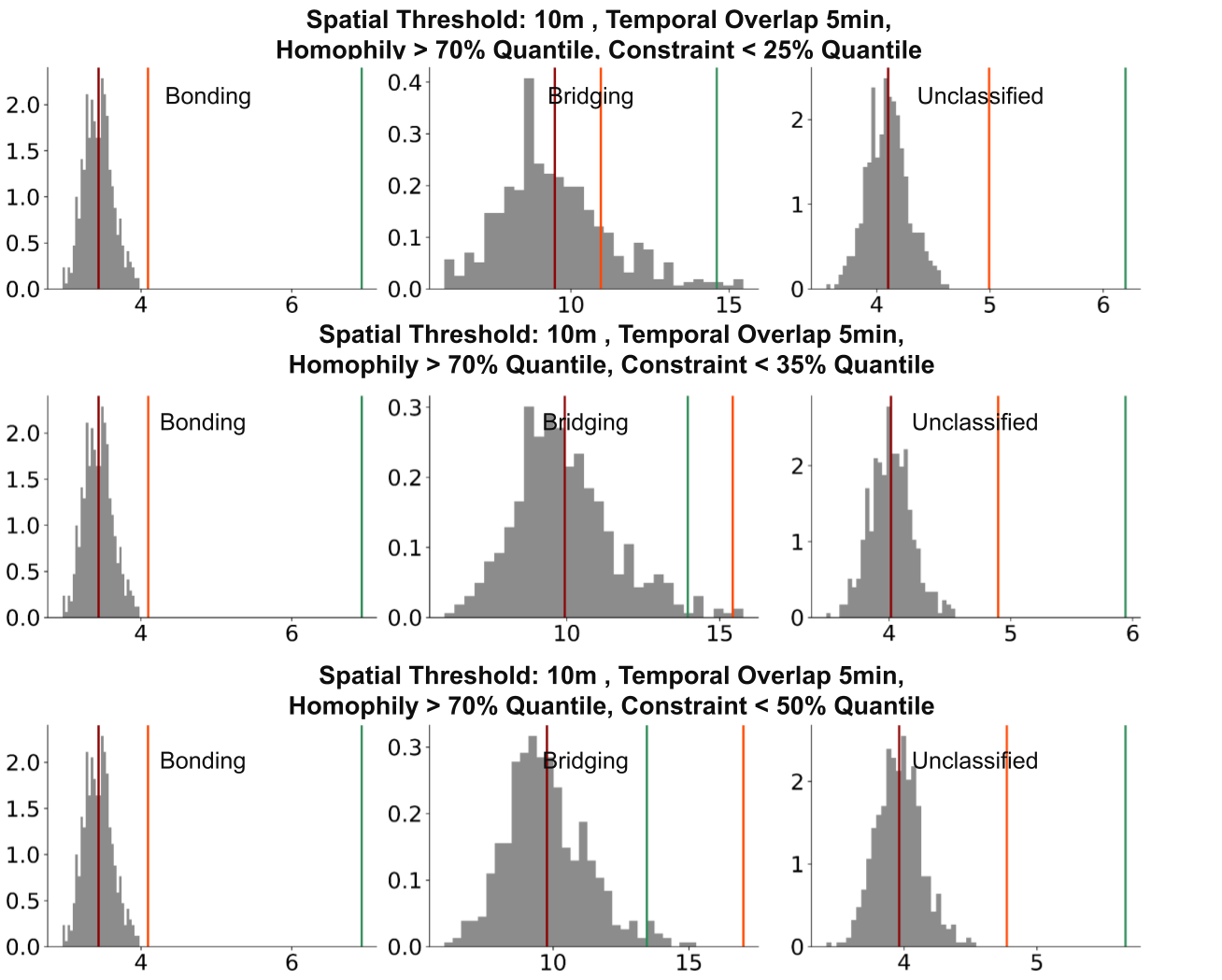}
    \captionsetup{singlelinecheck=off}
    \caption[Mean weighted degree under a restrictive 10\,m spatial threshold, 5-minute temporal overlap, and 70th-percentile homophily cutoff.]{
    \textbf{Mean weighted degree under a restrictive 10\,m spatial threshold, 5-minute temporal overlap, and 70th-percentile homophily cutoff.}
    The figure shows mean weighted degree distributions across 500 behavior-based counterfactual runs for the bonding, refined bridging, and unclassified subnetworks under a highly restrictive 10\,m spatial co-location threshold. Histograms represent the counterfactual distributions, while vertical reference lines mark the observed pre-disaster and post-disaster values. The 10\,m graph is sparse, especially for refined bridging ties, so the bridging comparison is less stable than in broader spatial specifications. These results are therefore interpreted cautiously rather than used as primary robustness evidence.}
    \label{fig:Figure_S_tie_robustness_10x5x70}
\end{figure}

\begin{figure}[h]
    \centering
    \includegraphics[width=0.95\linewidth]{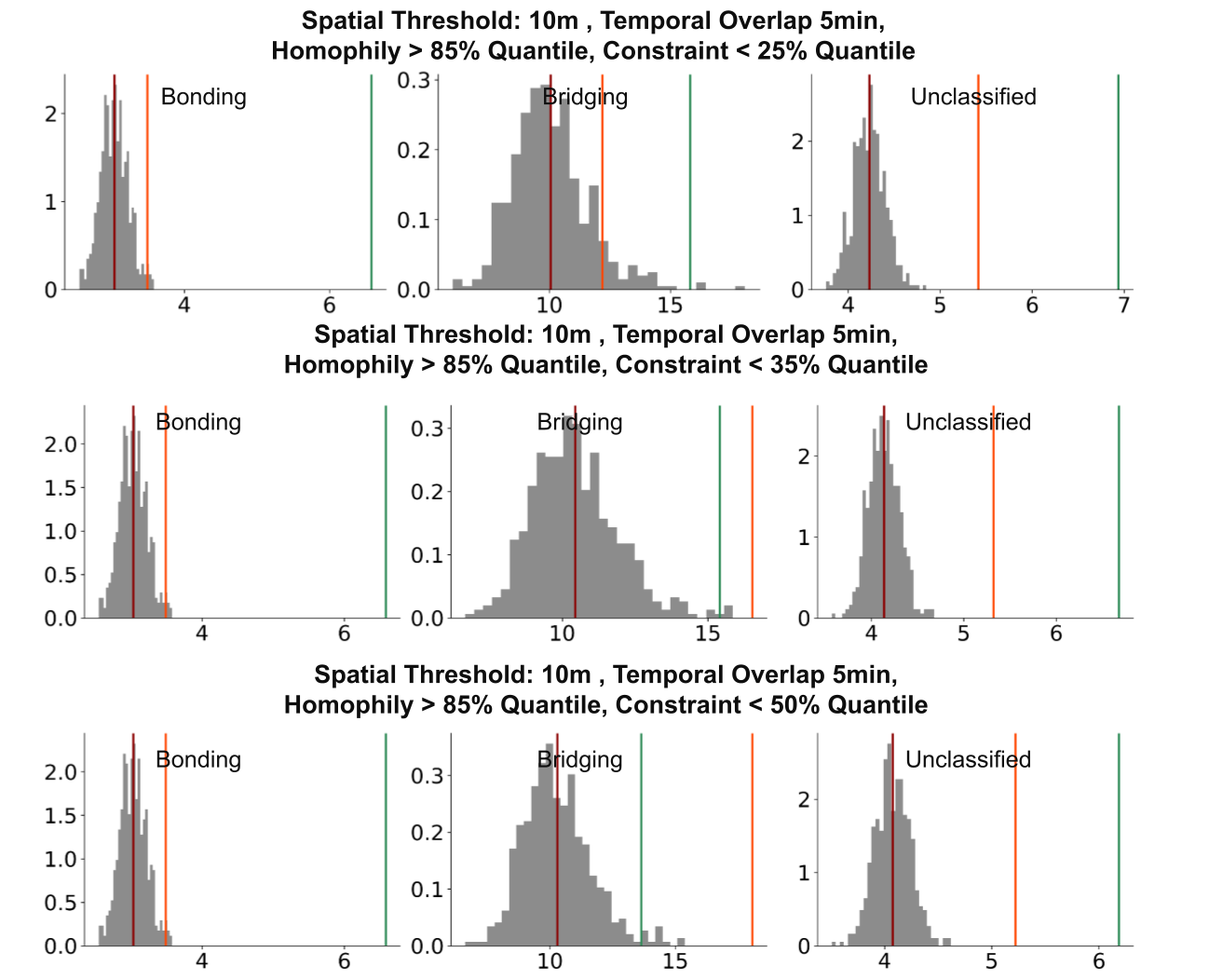}
    \captionsetup{singlelinecheck=off}
    \caption[Mean weighted degree under a restrictive 10\,m spatial threshold, 5-minute temporal overlap, and 85th-percentile homophily cutoff.]{
    \textbf{Mean weighted degree under a restrictive 10\,m spatial threshold, 5-minute temporal overlap, and 85th-percentile homophily cutoff.}
    The figure shows mean weighted degree distributions across 500 behavior-based counterfactual runs for the bonding, refined bridging, and unclassified subnetworks under a highly restrictive 10\,m spatial co-location threshold. Histograms represent the counterfactual distributions, while vertical reference lines mark the observed pre-disaster and post-disaster values. This conservative specification produces a sparse observed network and very limited refined bridging structure, making the bridging comparison unstable. These results are interpreted cautiously and are not treated as primary evidence for robustness.}
    \label{fig:Figure_S_tie_robustness_10x5x85}
\end{figure}

\begin{figure}[h]
    \centering
    \includegraphics[width=0.95\linewidth]{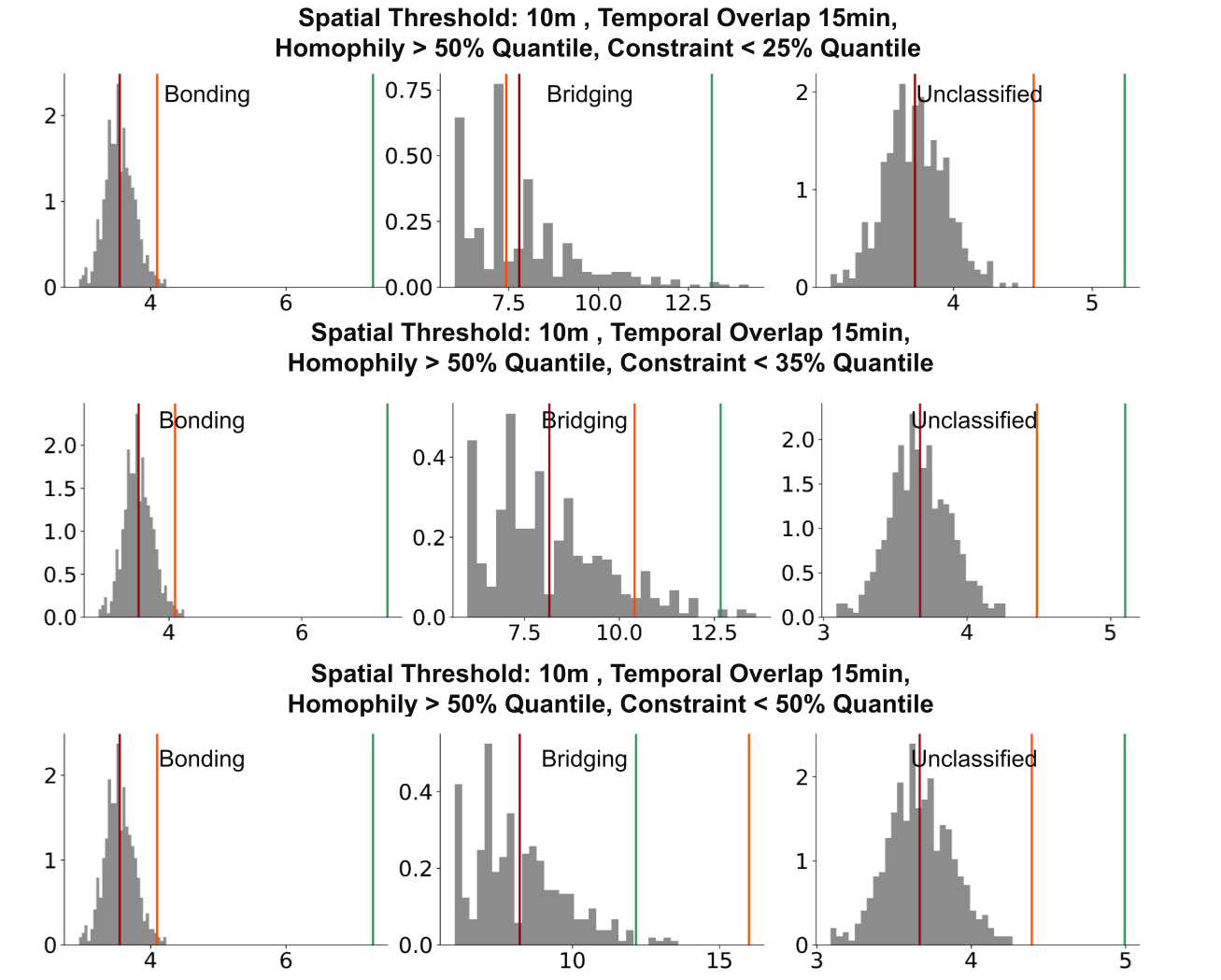}
    \captionsetup{singlelinecheck=off}
    \caption[Mean weighted degree under a restrictive 10\,m spatial threshold, 15-minute temporal overlap, and 50th-percentile homophily cutoff.]{
    \textbf{Mean weighted degree under a restrictive 10\,m spatial threshold, 15-minute temporal overlap, and 50th-percentile homophily cutoff.}
    The figure shows mean weighted degree distributions across 500 behavior-based counterfactual runs for the bonding, refined bridging, and unclassified subnetworks under the most restrictive spatial and temporal co-location definition considered here. Histograms represent the counterfactual distributions, while vertical reference lines mark the observed pre-disaster and post-disaster values. This specification yields a sparse graph with too few refined bridging ties for stable comparison, so the results are interpreted cautiously rather than treated as primary robustness evidence.}
    \label{fig:Figure_S_tie_robustness_10x15x50}
\end{figure}

\begin{figure}[h]
    \centering
    \includegraphics[width=0.95\linewidth]{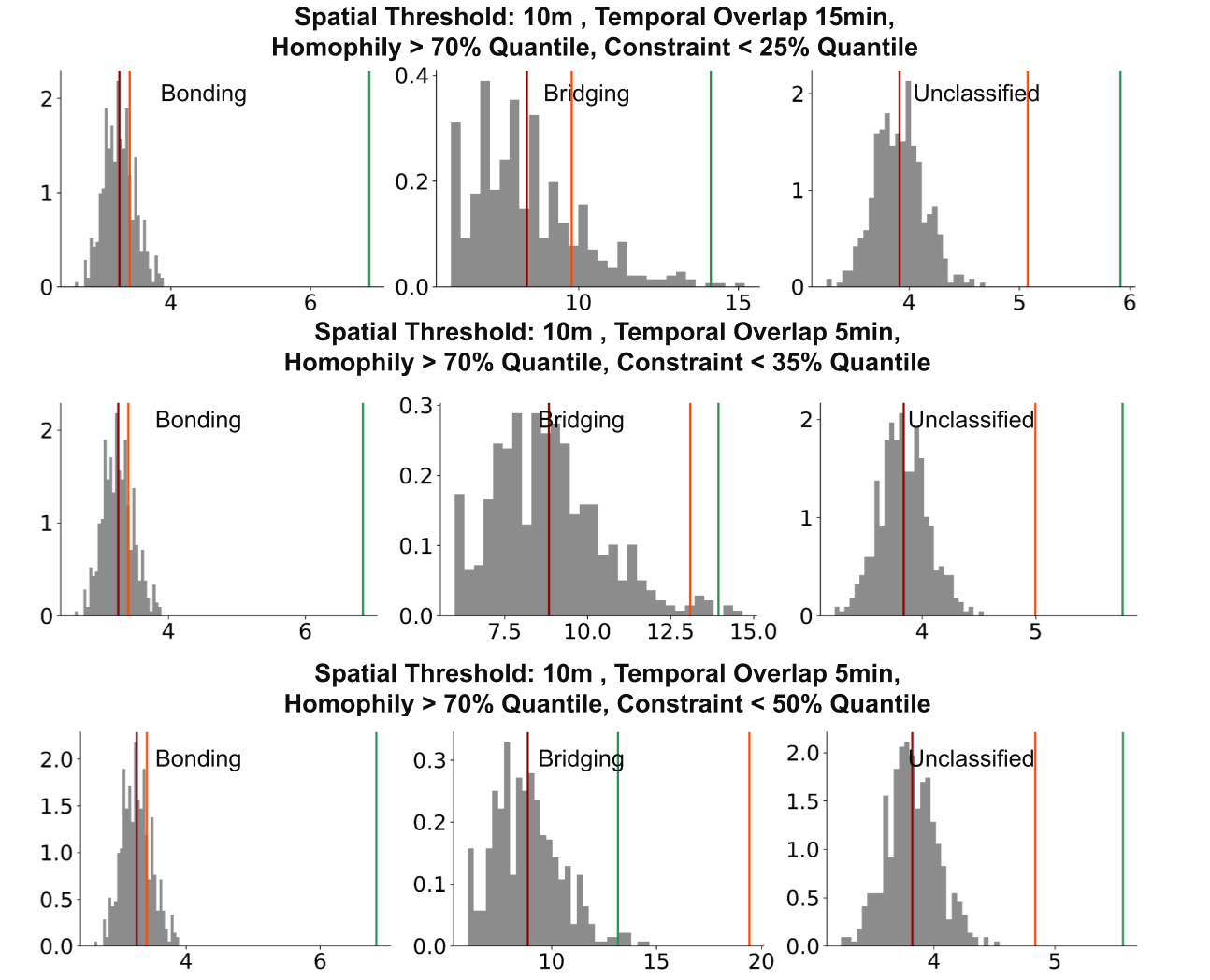}
    \captionsetup{singlelinecheck=off}
    \caption[Mean weighted degree under a restrictive 10\,m spatial threshold, 15-minute temporal overlap, and 70th-percentile homophily cutoff.]{
    \textbf{Mean weighted degree under a restrictive 10\,m spatial threshold, 15-minute temporal overlap, and 70th-percentile homophily cutoff.}
    The figure shows mean weighted degree distributions across 500 behavior-based counterfactual runs for the bonding, refined bridging, and unclassified subnetworks under the most restrictive spatial and temporal co-location definition considered here. Histograms represent the counterfactual distributions, while vertical reference lines mark the observed pre-disaster and post-disaster values. Because this specification produces a very sparse graph, especially for refined bridging ties, bridging comparisons are less stable. These results are interpreted cautiously and are not treated as primary evidence for robustness.}
    \label{fig:Figure_S_tie_robustness_10x15x70}
\end{figure}

\begin{figure}[h]
    \centering
    \includegraphics[width=0.95\linewidth]{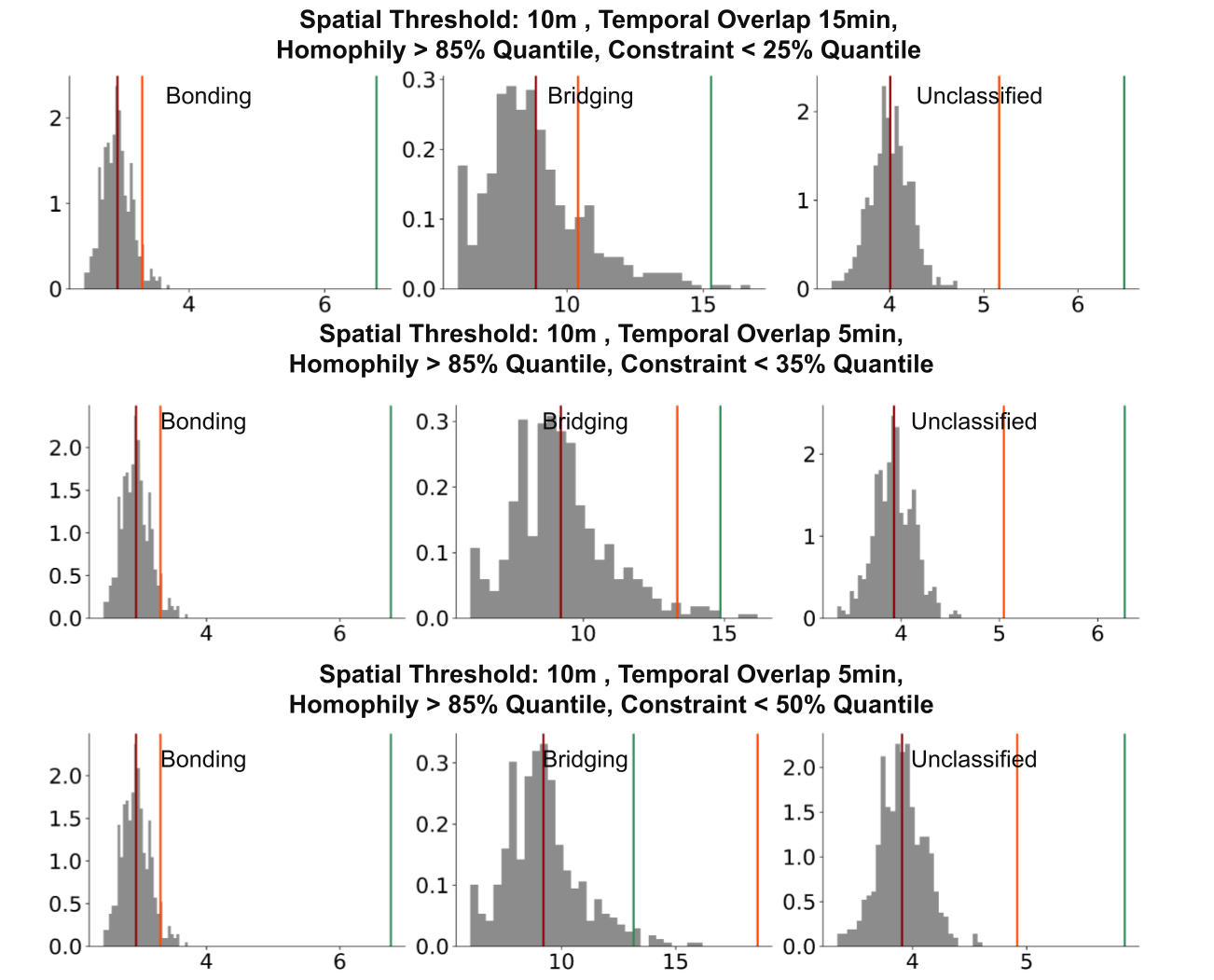}
    \captionsetup{singlelinecheck=off}
    \caption[Mean weighted degree under a restrictive 10\,m spatial threshold, 15-minute temporal overlap, and 85th-percentile homophily cutoff.]{
    \textbf{Mean weighted degree under a restrictive 10\,m spatial threshold, 15-minute temporal overlap, and 85th-percentile homophily cutoff.}
    The figure shows mean weighted degree distributions across 500 behavior-based counterfactual runs for the bonding, refined bridging, and unclassified subnetworks under the most restrictive spatial, temporal, and homophily specification considered here. Histograms represent the counterfactual distributions, while vertical reference lines mark the observed pre-disaster and post-disaster values. The resulting graph is sparse and contains too few refined bridging ties for stable comparison. We therefore interpret this case cautiously and do not treat it as primary robustness evidence.}
    \label{fig:Figure_S_tie_robustness_10x15x85}
\end{figure}

\begin{figure}[h]
    \centering
    \includegraphics[width=0.95\linewidth]{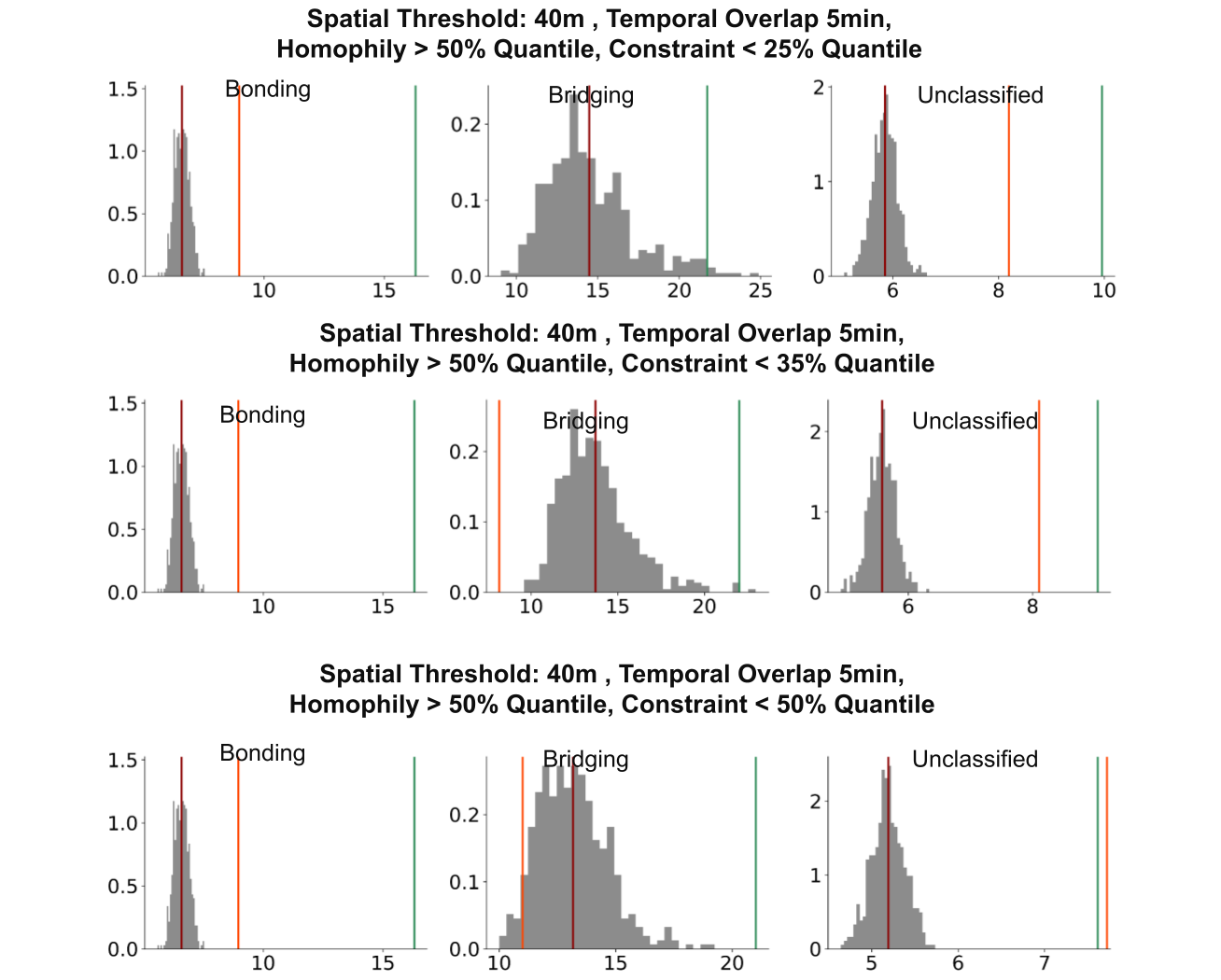}
    \captionsetup{singlelinecheck=off}
    \caption[Robustness of mean weighted degree under a 40\,m spatial threshold, 5-minute temporal overlap, and 50th-percentile homophily cutoff.]{
    \textbf{Robustness of mean weighted degree under a 40\,m spatial threshold, 5-minute temporal overlap, and 50th-percentile homophily cutoff.}
    The figure shows mean weighted degree distributions across 500 behavior-based counterfactual runs for the bonding, refined bridging, and unclassified subnetworks under the specified co-location and homophily thresholds. Histograms represent the counterfactual distributions, while vertical reference lines mark the observed pre-disaster and post-disaster values. Despite the larger and denser graph produced under this broader spatial threshold, the main result is reproduced: bonding and unclassified subnetworks remain more connected than expected, whereas refined bridging connectivity is weaker than expected.}
    \label{fig:Figure_S_tie_robustness_40x5x50}
\end{figure}

\begin{figure}[h]
    \centering
    \includegraphics[width=0.95\linewidth]{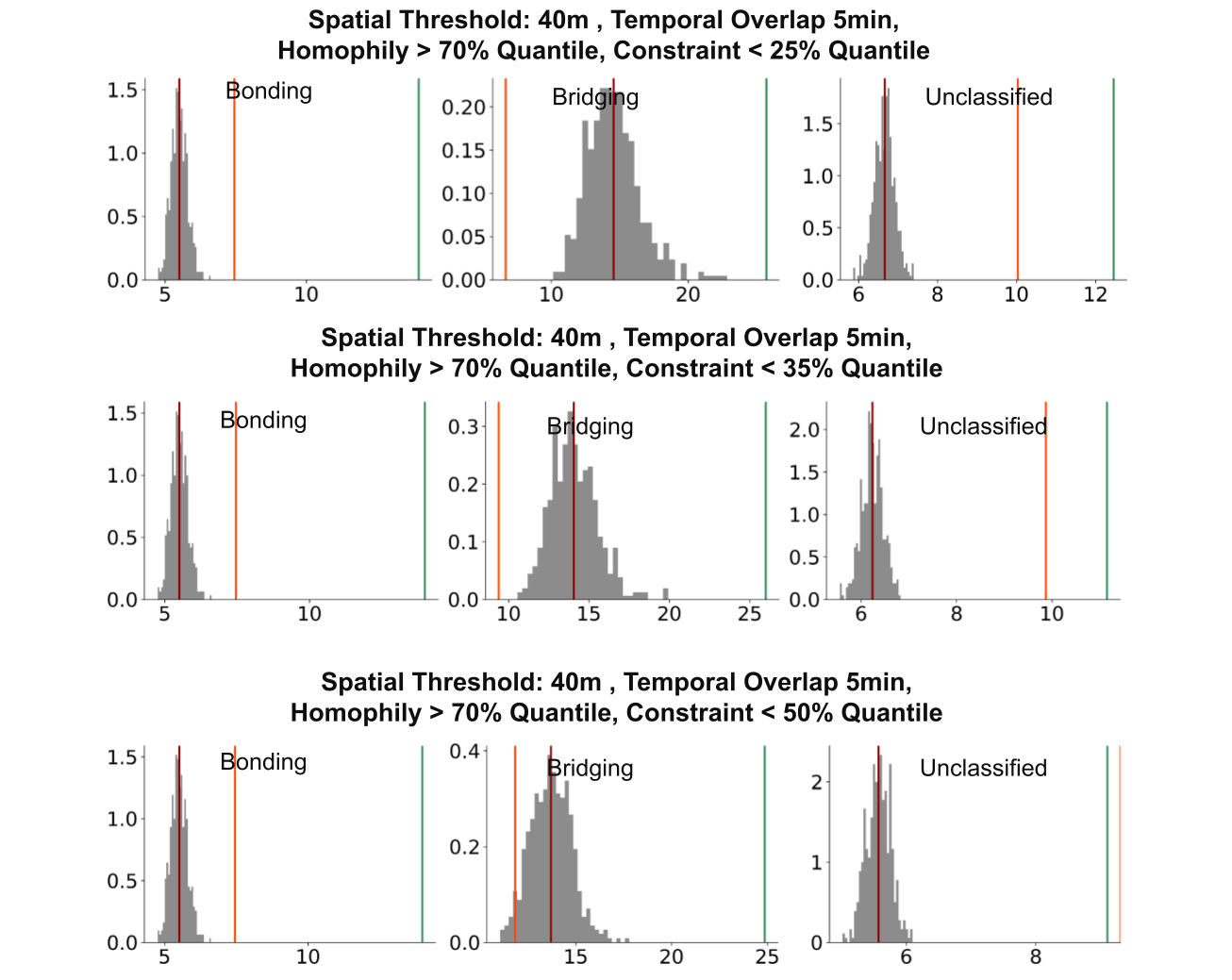}
    \captionsetup{singlelinecheck=off}
    \caption[Robustness of mean weighted degree under a 40\,m spatial threshold, 5-minute temporal overlap, and 70th-percentile homophily cutoff.]{
    \textbf{Robustness of mean weighted degree under a 40\,m spatial threshold, 5-minute temporal overlap, and 70th-percentile homophily cutoff.}
    The figure shows mean weighted degree distributions across 500 behavior-based counterfactual runs for the bonding, refined bridging, and unclassified subnetworks under the specified co-location and homophily thresholds. Histograms represent the counterfactual distributions, while vertical reference lines mark the observed pre-disaster and post-disaster values. The main qualitative finding remains unchanged under the broader spatial threshold and stricter homophily cutoff: bonding and unclassified subnetworks remain more connected after the disaster than expected, whereas refined bridging connectivity is weaker than expected.}
    \label{fig:Figure_S_tie_robustness_40x5x70}
\end{figure}

\begin{figure}[h]
    \centering
    \includegraphics[width=0.95\linewidth]{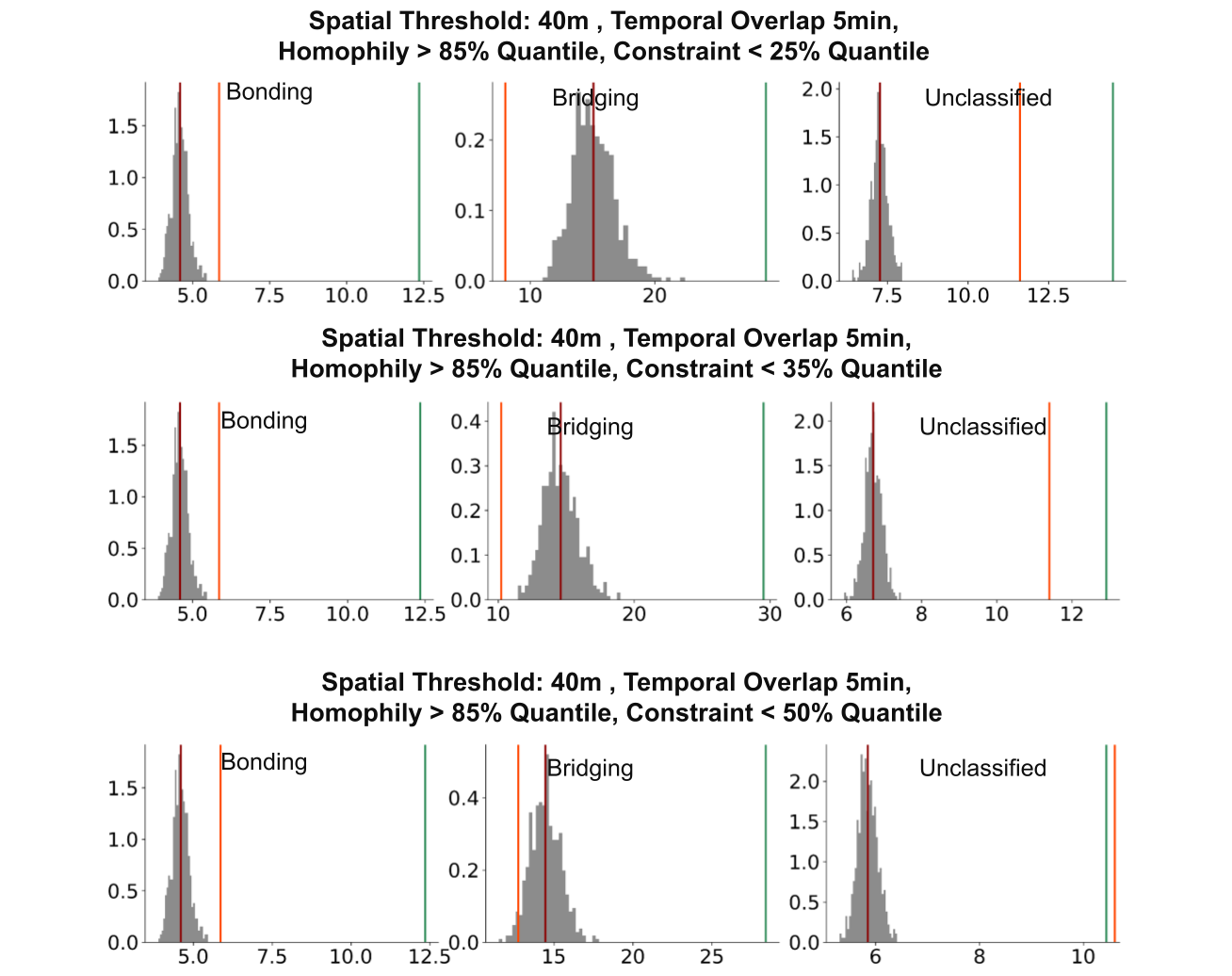}
    \captionsetup{singlelinecheck=off}
    \caption[Robustness of mean weighted degree under a 40\,m spatial threshold, 5-minute temporal overlap, and 85th-percentile homophily cutoff.]{
    \textbf{Robustness of mean weighted degree under a 40\,m spatial threshold, 5-minute temporal overlap, and 85th-percentile homophily cutoff.}
    The figure shows mean weighted degree distributions across 500 behavior-based counterfactual runs for the bonding, refined bridging, and unclassified subnetworks under the specified co-location and homophily thresholds. Histograms represent the counterfactual distributions, while vertical reference lines mark the observed pre-disaster and post-disaster values. Even under this conservative classification of bonding ties, the observed post-disaster pattern remains robust: bonding and unclassified subnetworks are more connected than expected, while refined bridging connectivity is weaker than expected.}
    \label{fig:Figure_S_tie_robustness_40x5x85}
\end{figure}

\begin{figure}[h]
    \centering
    \includegraphics[width=0.95\linewidth]{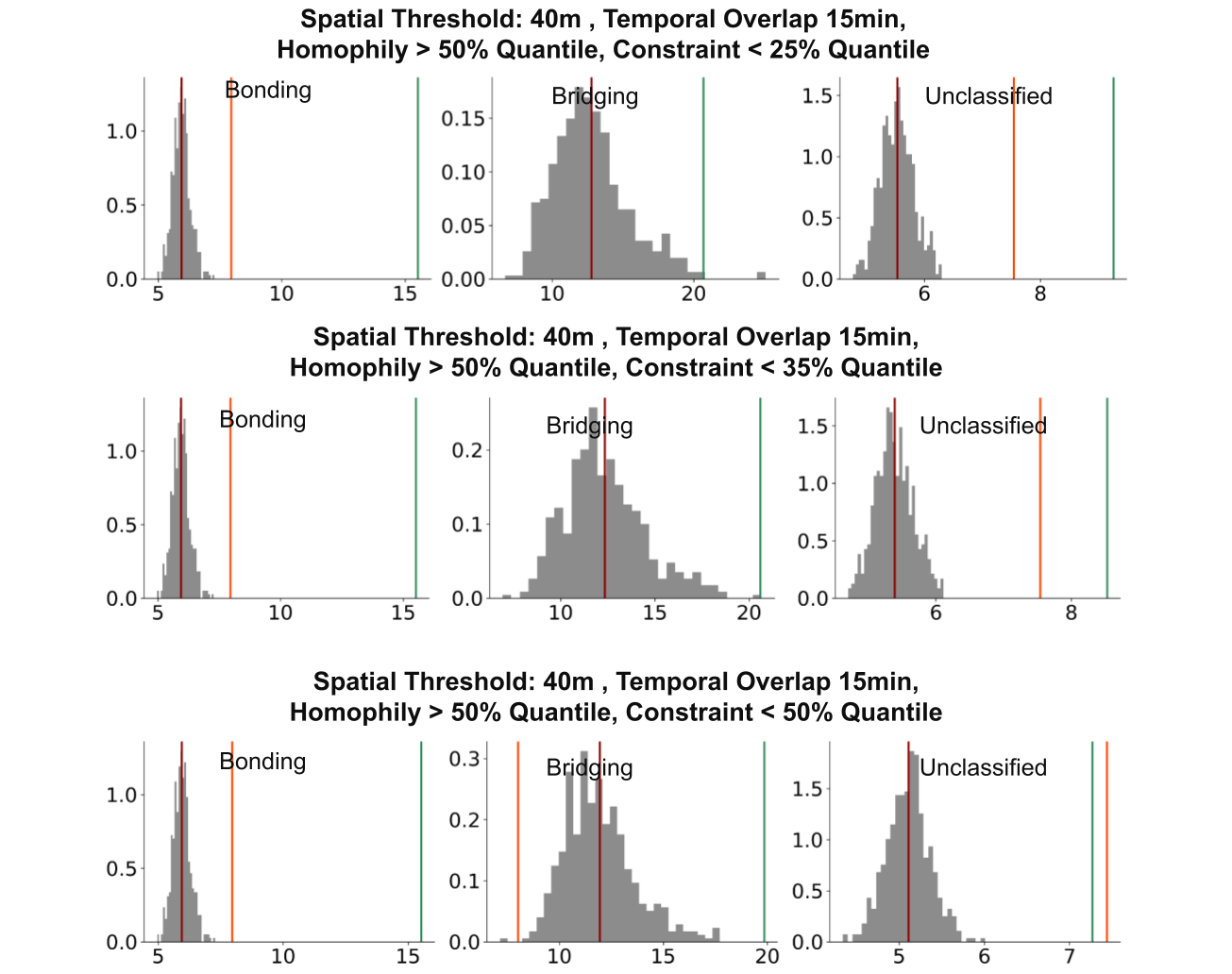}
    \captionsetup{singlelinecheck=off}
    \caption[Robustness of mean weighted degree under a 40\,m spatial threshold, 15-minute temporal overlap, and 50th-percentile homophily cutoff.]{
    \textbf{Robustness of mean weighted degree under a 40\,m spatial threshold, 15-minute temporal overlap, and 50th-percentile homophily cutoff.}
    The figure shows mean weighted degree distributions across 500 behavior-based counterfactual runs for the bonding, refined bridging, and unclassified subnetworks under the specified co-location and homophily thresholds. Histograms represent the counterfactual distributions, while vertical reference lines mark the observed pre-disaster and post-disaster values. The main tie-type pattern remains visible under the broader spatial threshold and stricter temporal overlap rule: bonding and unclassified subnetworks remain more connected than expected, whereas refined bridging connectivity is weaker than expected.}
    \label{fig:Figure_S_tie_robustness_40x15x50}
\end{figure}

\begin{figure}[h]
    \centering
    \includegraphics[width=0.95\linewidth]{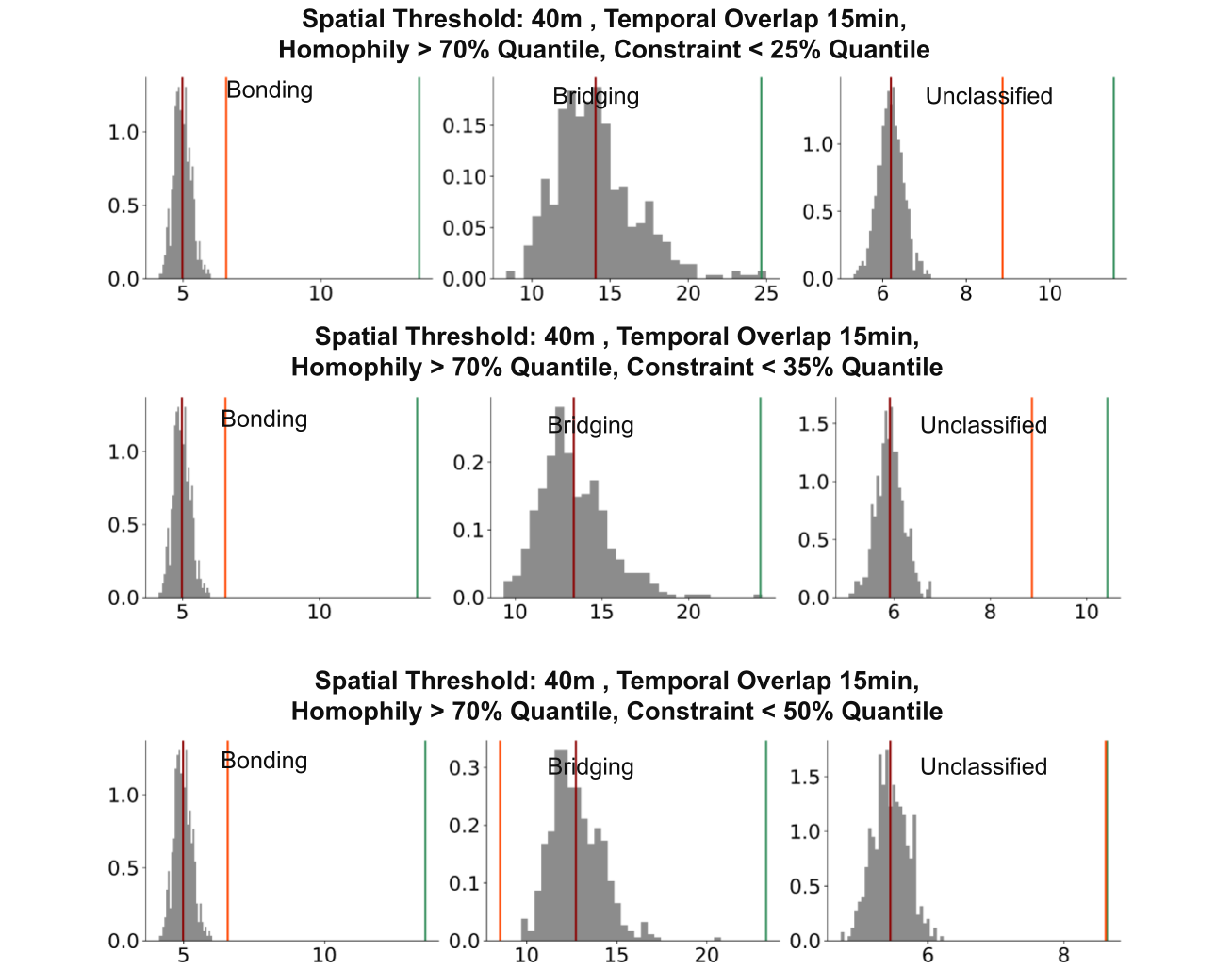}
    \captionsetup{singlelinecheck=off}
    \caption[Robustness of mean weighted degree under a 40\,m spatial threshold, 15-minute temporal overlap, and 70th-percentile homophily cutoff.]{
    \textbf{Robustness of mean weighted degree under a 40\,m spatial threshold, 15-minute temporal overlap, and 70th-percentile homophily cutoff.}
    The figure shows mean weighted degree distributions across 500 behavior-based counterfactual runs for the bonding, refined bridging, and unclassified subnetworks under the specified co-location and homophily thresholds. Histograms represent the counterfactual distributions, while vertical reference lines mark the observed pre-disaster and post-disaster values. The same qualitative structure persists under this stricter classification: bonding and unclassified subnetworks remain more connected than expected, while refined bridging connectivity is weaker than expected.}
    \label{fig:Figure_S_tie_robustness_40x15x70}
\end{figure}

\begin{figure}[h]
    \centering
    \includegraphics[width=0.95\linewidth]{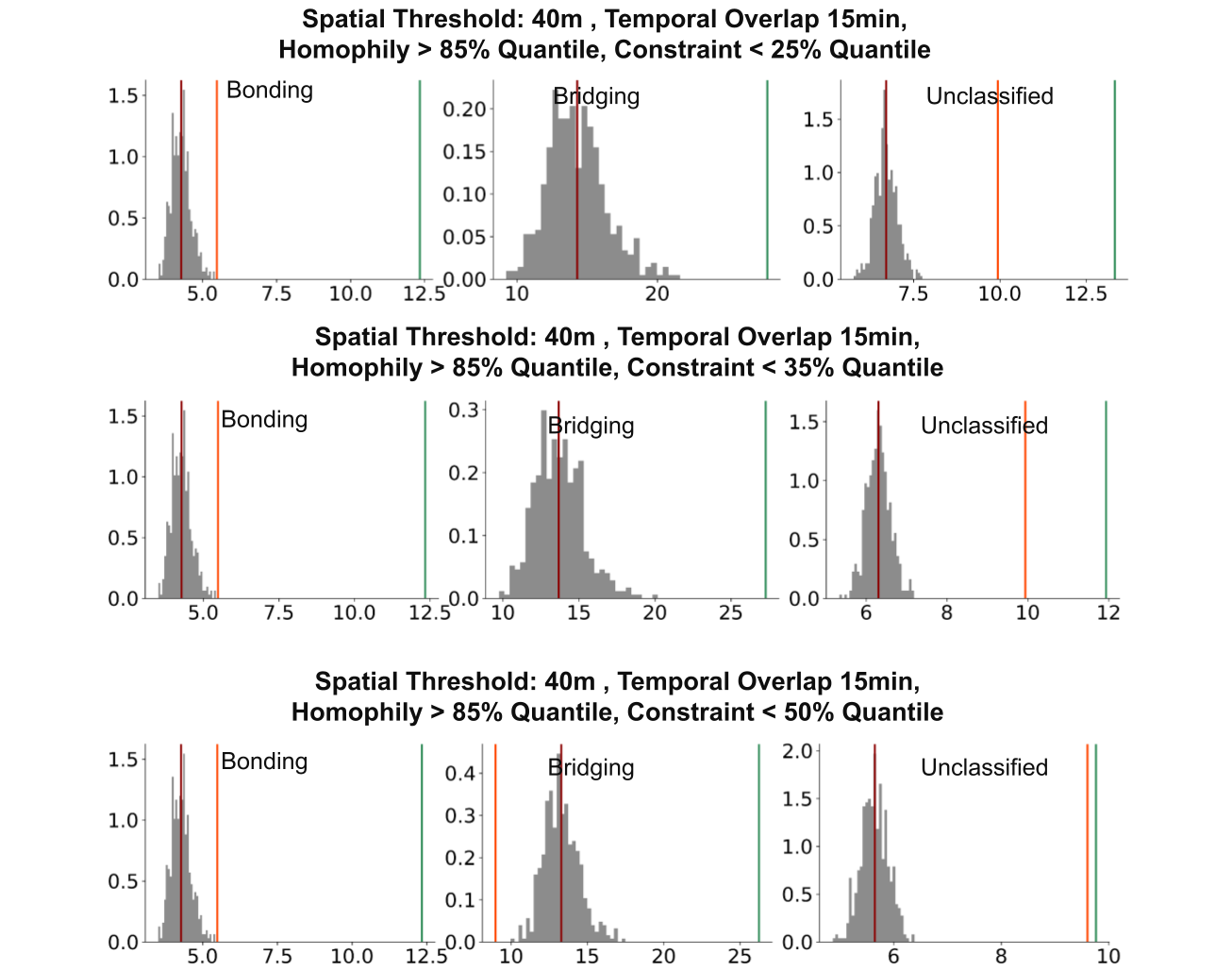}
    \captionsetup{singlelinecheck=off}
    \caption[Robustness of mean weighted degree under a 40\,m spatial threshold, 15-minute temporal overlap, and 85th-percentile homophily cutoff.]{
    \textbf{Robustness of mean weighted degree under a 40\,m spatial threshold, 15-minute temporal overlap, and 85th-percentile homophily cutoff.}
    The figure shows mean weighted degree distributions across 500 behavior-based counterfactual runs for the bonding, refined bridging, and unclassified subnetworks under the specified co-location and homophily thresholds. Histograms represent the counterfactual distributions, while vertical reference lines mark the observed pre-disaster and post-disaster values. Even under this conservative specification, the result is replicated: bonding and unclassified subnetworks remain more connected after the disaster than expected, whereas refined bridging connectivity is weaker than expected.}
    \label{fig:Figure_S_tie_robustness_40x15x85}
\end{figure}

\begin{figure}[h]
    \centering
    \includegraphics[width=0.95\linewidth]{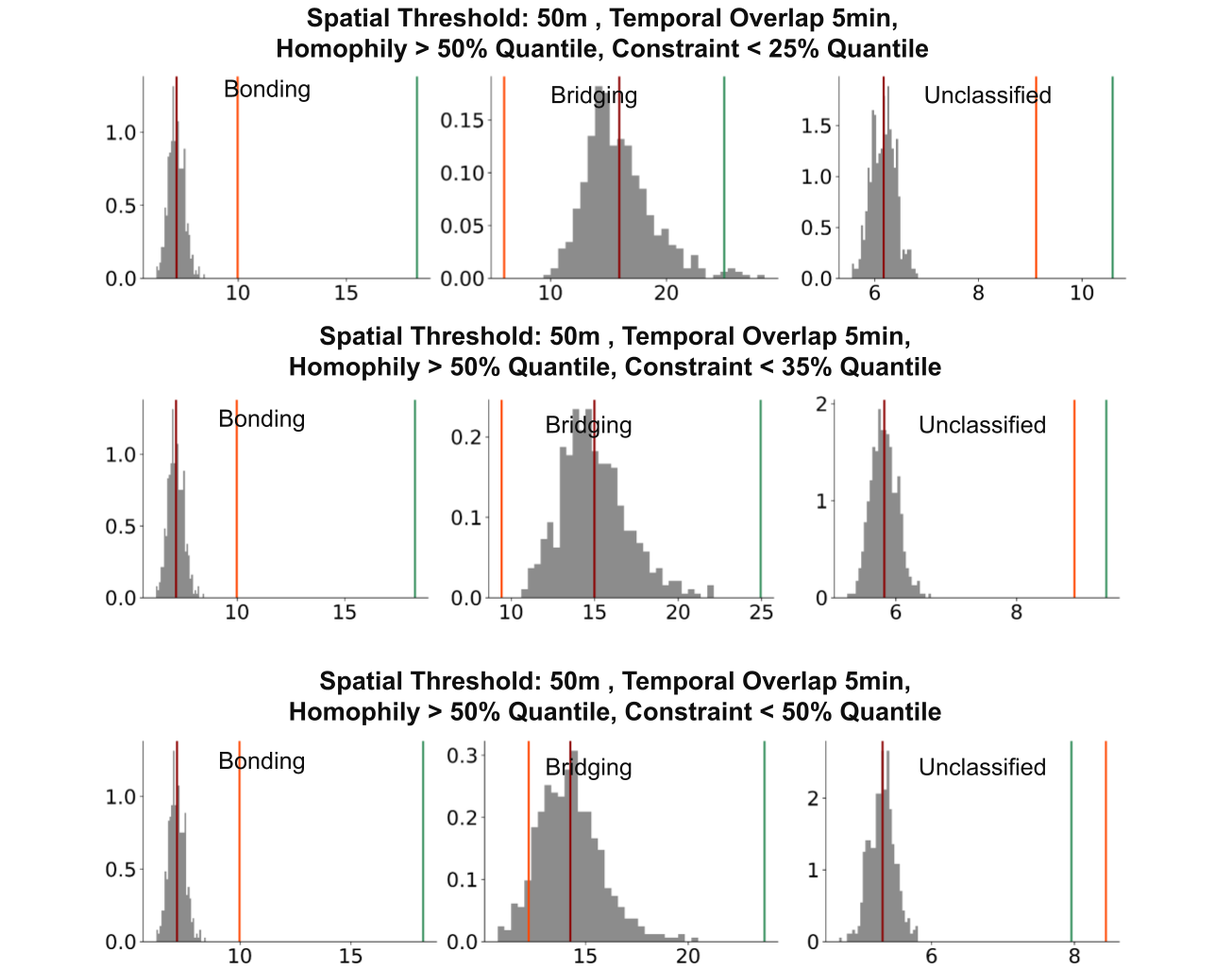}
    \captionsetup{singlelinecheck=off}
    \caption[Robustness of mean weighted degree under a 50\,m spatial threshold, 5-minute temporal overlap, and 50th-percentile homophily cutoff.]{
    \textbf{Robustness of mean weighted degree under a 50\,m spatial threshold, 5-minute temporal overlap, and 50th-percentile homophily cutoff.}
    The figure shows mean weighted degree distributions across 500 behavior-based counterfactual runs for the bonding, refined bridging, and unclassified subnetworks under the specified co-location and homophily thresholds. Histograms represent the counterfactual distributions, while vertical reference lines mark the observed pre-disaster and post-disaster values. Despite the increased density of the graph under the broadest spatial threshold considered here, the main result is reproduced: bonding and unclassified subnetworks remain more connected than expected, whereas refined bridging connectivity is weaker than expected.}
    \label{fig:Figure_S_tie_robustness_50x5x50}
\end{figure}

\begin{figure}[h]
    \centering
    \includegraphics[width=0.95\linewidth]{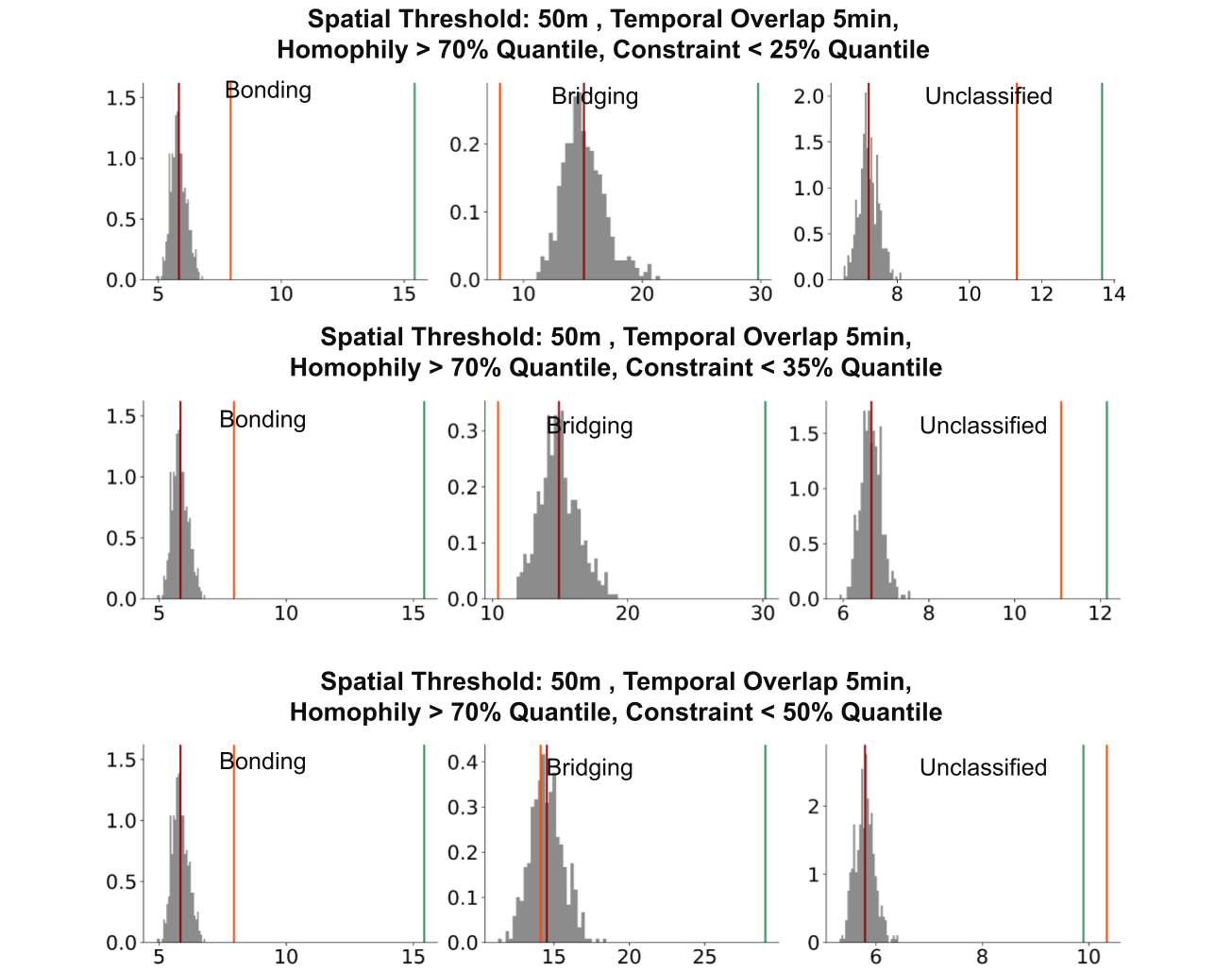}
    \captionsetup{singlelinecheck=off}
    \caption[Robustness of mean weighted degree under a 50\,m spatial threshold, 5-minute temporal overlap, and 70th-percentile homophily cutoff.]{
    \textbf{Robustness of mean weighted degree under a 50\,m spatial threshold, 5-minute temporal overlap, and 70th-percentile homophily cutoff.}
    The figure shows mean weighted degree distributions across 500 behavior-based counterfactual runs for the bonding, refined bridging, and unclassified subnetworks under the specified co-location and homophily thresholds. Histograms represent the counterfactual distributions, while vertical reference lines mark the observed pre-disaster and post-disaster values. The main qualitative conclusion remains unchanged under the broadest spatial threshold and stricter homophily cutoff: bonding and unclassified subnetworks remain more connected than expected, while refined bridging connectivity is weaker than expected.}
    \label{fig:Figure_S_tie_robustness_50x5x70}
\end{figure}

\begin{figure}[h]
    \centering
    \includegraphics[width=0.95\linewidth]{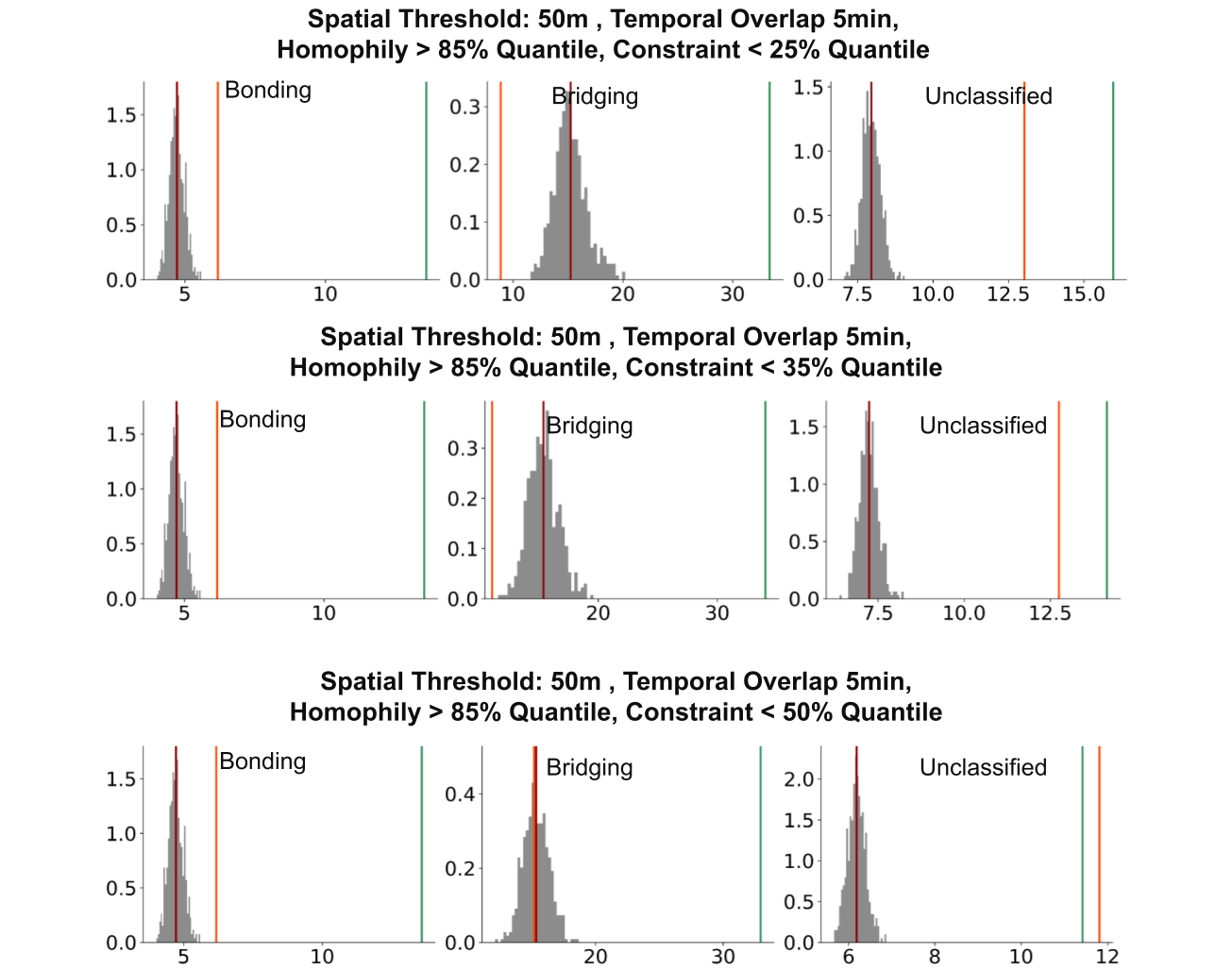}
    \captionsetup{singlelinecheck=off}
    \caption[Robustness of mean weighted degree under a 50\,m spatial threshold, 5-minute temporal overlap, and 85th-percentile homophily cutoff.]{
    \textbf{Robustness of mean weighted degree under a 50\,m spatial threshold, 5-minute temporal overlap, and 85th-percentile homophily cutoff.}
    The figure shows mean weighted degree distributions across 500 behavior-based counterfactual runs for the bonding, refined bridging, and unclassified subnetworks under the specified co-location and homophily thresholds. Histograms represent the counterfactual distributions, while vertical reference lines mark the observed pre-disaster and post-disaster values. Even under this highly selective homophily cutoff, the observed post-disaster pattern remains robust: bonding and unclassified subnetworks are more connected than expected, whereas refined bridging connectivity is weaker than expected.}
    \label{fig:Figure_S_tie_robustness_50x5x85}
\end{figure}

\begin{figure}[h]
    \centering
    \includegraphics[width=0.95\linewidth]{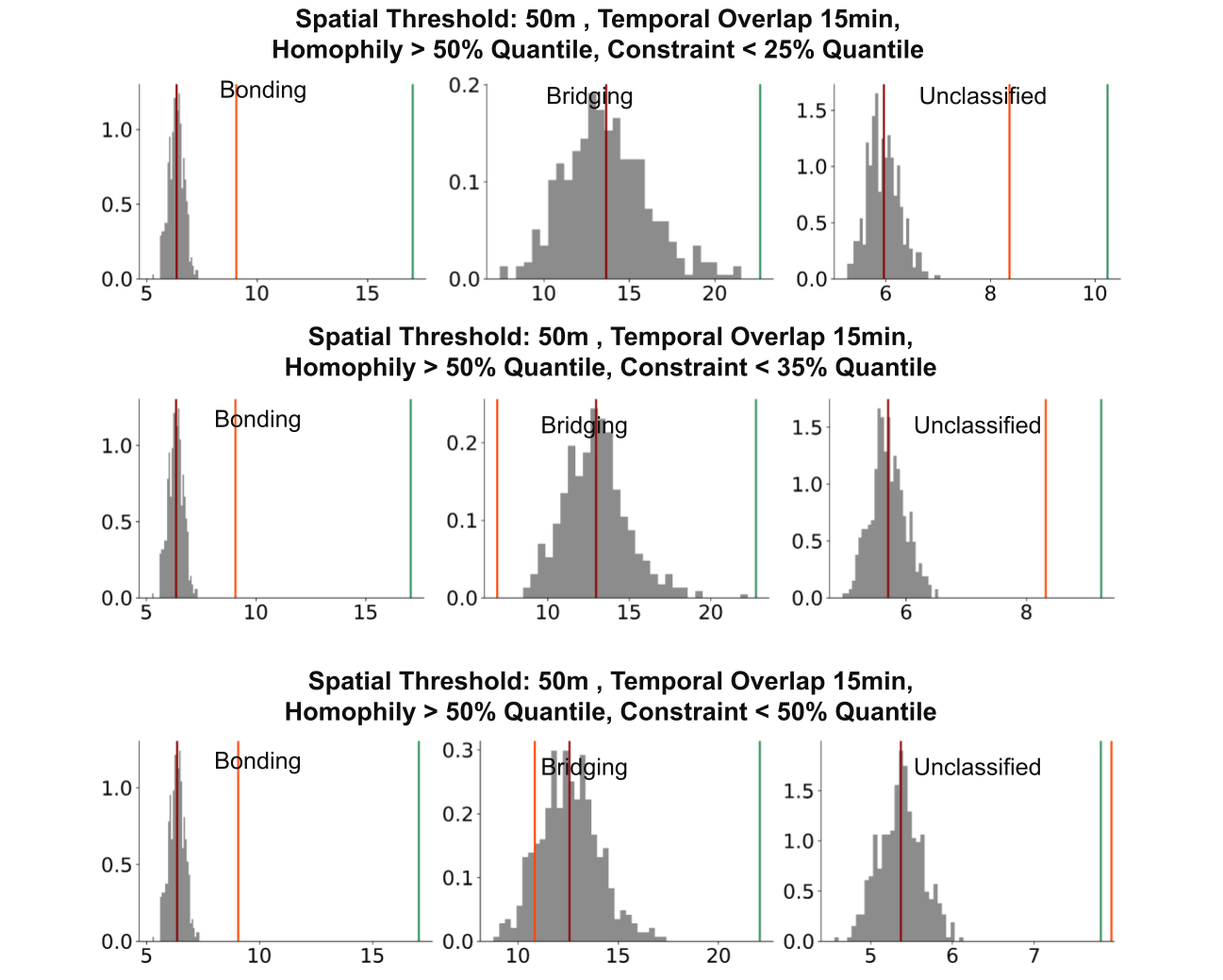}
    \captionsetup{singlelinecheck=off}
    \caption[Robustness of mean weighted degree under a 50\,m spatial threshold, 15-minute temporal overlap, and 50th-percentile homophily cutoff.]{
    \textbf{Robustness of mean weighted degree under a 50\,m spatial threshold, 15-minute temporal overlap, and 50th-percentile homophily cutoff.}
    The figure shows mean weighted degree distributions across 500 behavior-based counterfactual runs for the bonding, refined bridging, and unclassified subnetworks under the specified co-location and homophily thresholds. Histograms represent the counterfactual distributions, while vertical reference lines mark the observed pre-disaster and post-disaster values. The main tie-type structure remains visible under the broadest spatial threshold and stricter temporal overlap rule: bonding and unclassified subnetworks remain more connected than expected, whereas refined bridging connectivity is weaker than expected.}
    \label{fig:Figure_S_tie_robustness_50x15x50}
\end{figure}

\begin{figure}[h]
    \centering
    \includegraphics[width=0.95\linewidth]{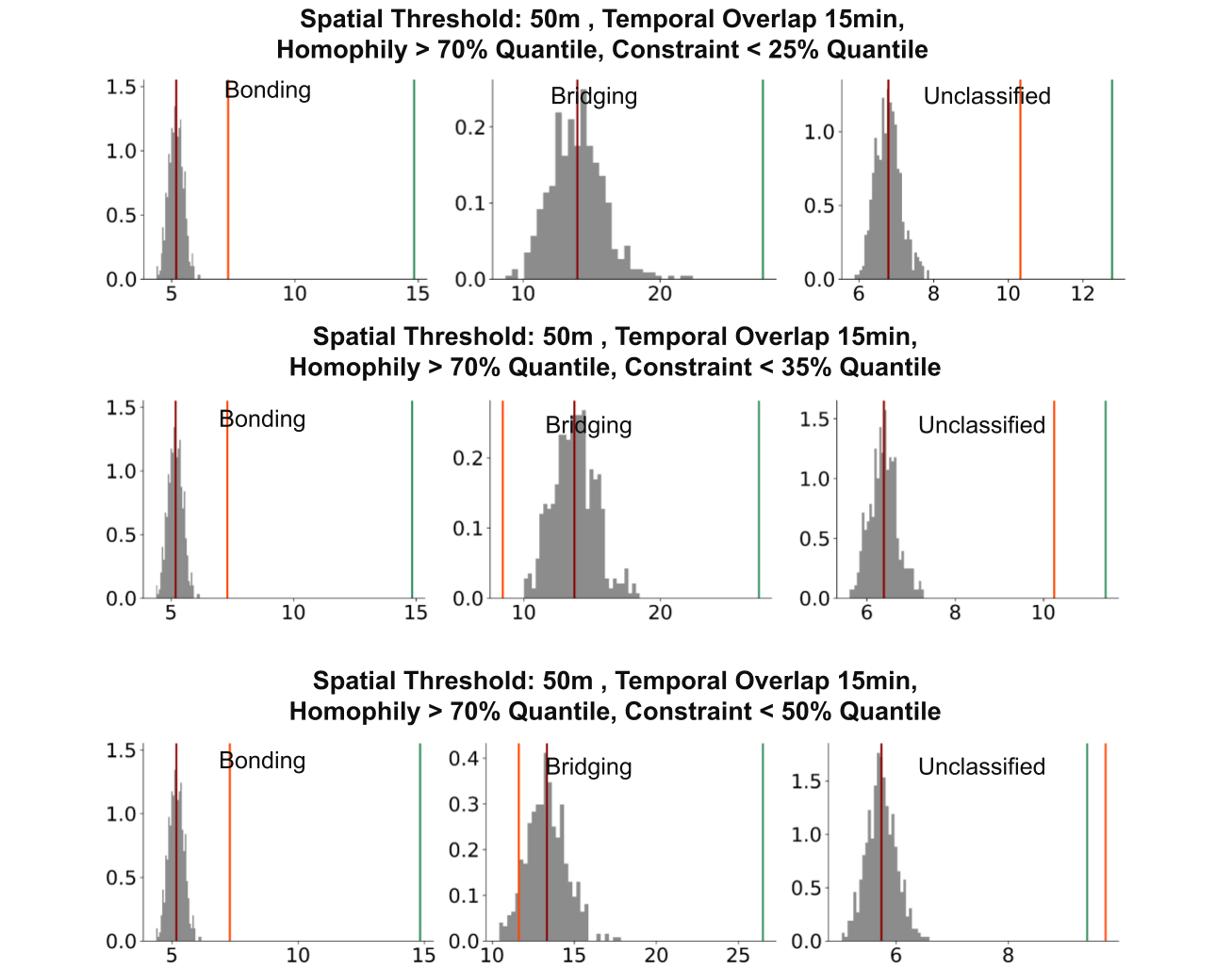}
    \captionsetup{singlelinecheck=off}
    \caption[Robustness of mean weighted degree under a 50\,m spatial threshold, 15-minute temporal overlap, and 70th-percentile homophily cutoff.]{
    \textbf{Robustness of mean weighted degree under a 50\,m spatial threshold, 15-minute temporal overlap, and 70th-percentile homophily cutoff.}
    The figure shows mean weighted degree distributions across 500 behavior-based counterfactual runs for the bonding, refined bridging, and unclassified subnetworks under the specified co-location and homophily thresholds. Histograms represent the counterfactual distributions, while vertical reference lines mark the observed pre-disaster and post-disaster values. The same qualitative pattern persists under the broadest spatial threshold, stricter temporal overlap rule, and stricter homophily cutoff: bonding and unclassified subnetworks remain more connected than expected, while refined bridging connectivity is weaker than expected.}
    \label{fig:Figure_S_tie_robustness_50x15x70}
\end{figure}

\begin{figure}[h]
    \centering
    \includegraphics[width=0.95\linewidth]{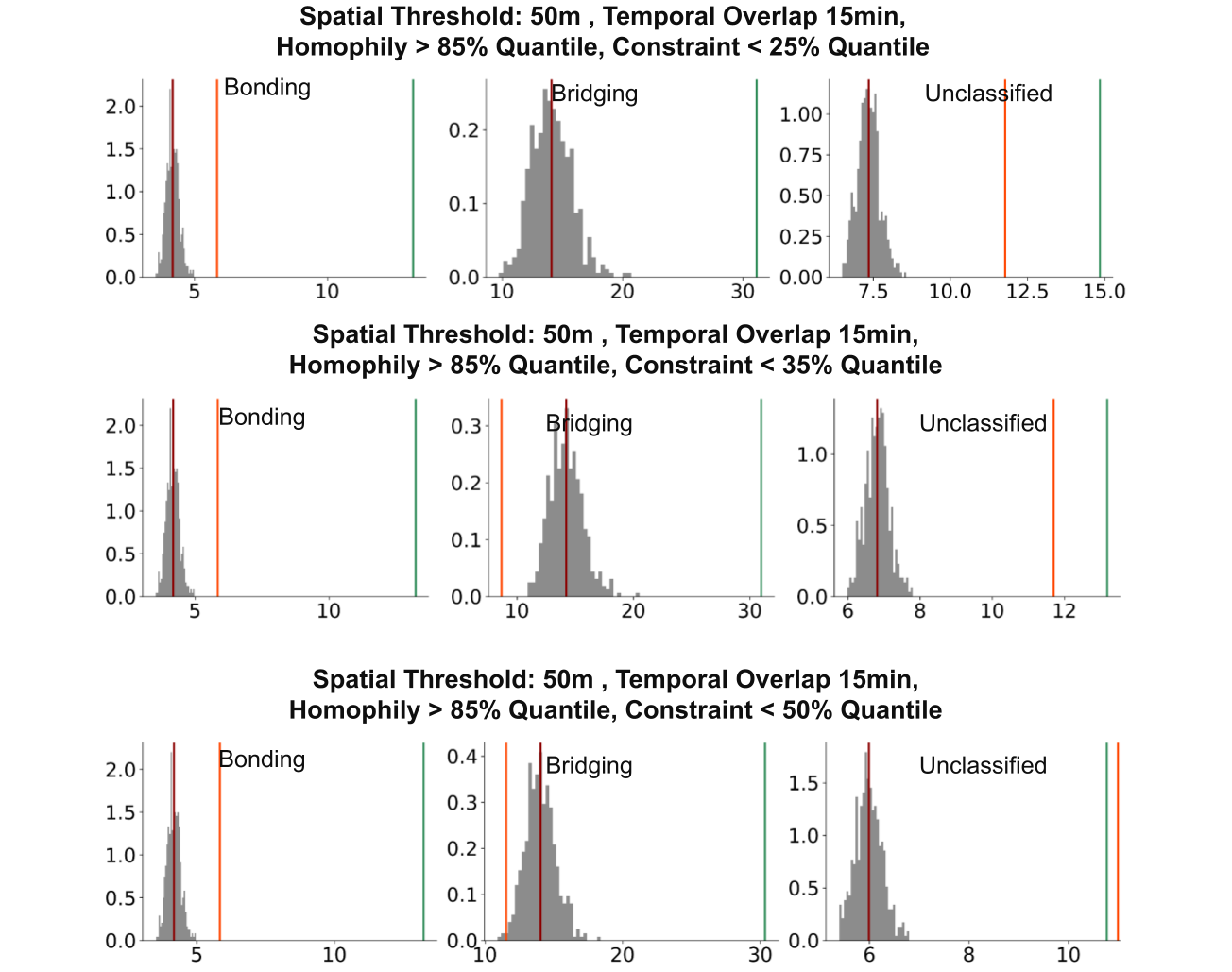}
    \captionsetup{singlelinecheck=off}
    \caption[Robustness of mean weighted degree under a 50\,m spatial threshold, 15-minute temporal overlap, and 85th-percentile homophily cutoff.]{
    \textbf{Robustness of mean weighted degree under a 50\,m spatial threshold, 15-minute temporal overlap, and 85th-percentile homophily cutoff.}
    The figure shows mean weighted degree distributions across 500 behavior-based counterfactual runs for the bonding, refined bridging, and unclassified subnetworks under the specified co-location and homophily thresholds. Histograms represent the counterfactual distributions, while vertical reference lines mark the observed pre-disaster and post-disaster values. Even under this conservative specification, the result is reproduced: bonding and unclassified subnetworks remain more connected after the disaster than expected under the behavior-informed counterfactual, whereas refined bridging connectivity is weaker than expected.}
    \label{fig:Figure_S_tie_robustness_50x15x85}
\end{figure}

\begin{center}
\scriptsize
\setlength{\tabcolsep}{3pt}
\renewcommand{\arraystretch}{0.88}
\begin{adjustbox}{max width=\textwidth,center}
\begin{minipage}{\textwidth}
\begin{longtable}{p{2.8cm}cccc}
\caption{Regression Results Across Tie Types}
\label{tab:zlog_regression_all_panels}\\
\toprule
& \multicolumn{4}{c}{\textit{Dependent variable: Standardized residual change in interaction intensity}} \\
\cmidrule(lr){2-5}
& \textbf{All ties} & \textbf{Bonding} & \textbf{Bridging} & \textbf{Unclassified} \\
\midrule
\endfirsthead
\toprule
& \multicolumn{4}{c}{\textit{Dependent variable: Standardized residual change in interaction intensity (continued)}} \\
\cmidrule(lr){2-5}
& \textbf{All ties} & \textbf{Bonding} & \textbf{Bridging} & \textbf{Unclassified} \\
\midrule
\endhead
\midrule
\multicolumn{5}{r}{\textit{Continued on next page}} \\
\midrule
\endfoot
\bottomrule
\endlastfoot
Pre-disaster interactions & \mbox{0.002*** ($p=0.000$)} & \mbox{0.003 ($p=0.154$)} & \mbox{0.018*** ($p=0.001$)} & \mbox{-0.012*** ($p=0.000$)} \\
& \mbox{[0.002, 0.003]} & \mbox{[-0.001, 0.007]} & \mbox{[0.008, 0.028]} & \mbox{[-0.018, -0.006]} \\
Median rent & \mbox{0.271** ($p=0.025$)} & \mbox{0.508*** ($p=0.000$)} & \mbox{1.02*** ($p=0.000$)} & \mbox{0.901*** ($p=0.000$)} \\
& \mbox{[0.033, 0.508]} & \mbox{[0.26, 0.755]} & \mbox{[0.635, 1.406]} & \mbox{[0.612, 1.189]} \\
Median income (log) & \mbox{-0.047 ($p=0.701$)} & \mbox{-0.183 ($p=0.172$)} & \mbox{-1.019*** ($p=0.000$)} & \mbox{-0.486*** ($p=0.002$)} \\
& \mbox{[-0.288, 0.194]} & \mbox{[-0.445, 0.08]} & \mbox{[-1.489, -0.548]} & \mbox{[-0.792, -0.179]} \\
White population share & \mbox{-0.059 
($p=0.588$)} & \mbox{0.192* ($p=0.093$)} & \mbox{0.181 ($p=0.175$)} & \mbox{0.011 ($p=0.929$)} \\
& \mbox{[-0.275, 0.156]} & \mbox{[-0.032, 0.416]} & \mbox{[-0.082, 0.444]} & \mbox{[-0.229, 0.25]} \\
Median age & \mbox{0.331*** ($p=0.008$)} & \mbox{0.141 ($p=0.265$)} & \mbox{0.134 ($p=0.325$)} & \mbox{0.072 ($p=0.582$)} \\
& \mbox{[0.087, 0.575]} & \mbox{[-0.107, 0.39]} & \mbox{[-0.135, 0.403]} & \mbox{[-0.186, 0.331]} \\
Population & \mbox{0.249** ($p=0.041$)} & \mbox{0.229* ($p=0.090$)} & \mbox{-0.514** ($p=0.018$)} & \mbox{0.011 ($p=0.942$)} \\
& \mbox{[0.01, 0.489]} & \mbox{[-0.036, 0.495]} & \mbox{[-0.937, -0.092]} & \mbox{[-0.274, 0.295]} \\
Mean edge--POI distance & \mbox{0.375** ($p=0.026$)} & \mbox{0.812*** ($p=0.000$)} & \mbox{-0.338 ($p=0.284$)} & \mbox{1.301*** ($p=0.000$)} \\
& \mbox{[0.044, 0.707]} & \mbox{[0.471, 1.154]} & \mbox{[-0.96, 0.284]} & \mbox{[0.979, 1.623]} \\
Distance from fire & \mbox{0.131 ($p=0.653$)} & \mbox{-0.246 ($p=0.433$)} & \mbox{0.458* ($p=0.057$)} & \mbox{-0.708** ($p=0.035$)} \\
& \mbox{[-0.441, 0.703]} & \mbox{[-0.861, 0.37]} & \mbox{[-0.013, 0.928]} & \mbox{[-1.365, -0.05]} \\
Zoos \& botanical gardens & \mbox{3.683 ($p=0.153$)} &  &  & \mbox{3.367 ($p=0.140$)} \\
& \mbox{[-1.377, 8.743]} &  &  & \mbox{[-1.114, 7.848]} \\
Theaters \& dinner theaters & \mbox{3.126*** ($p=0.002$)} & \mbox{1.883* ($p=0.071$)} &  & \mbox{2.948*** ($p=0.007$)} \\
& \mbox{[1.14, 5.111]} & \mbox{[-0.159, 3.926]} &  & \mbox{[0.827, 5.069]} \\
Snack \& beverage bars & \mbox{2.713*** ($p=0.000$)} & \mbox{2.909*** ($p=0.000$)} & \mbox{-0.133 ($p=0.843$)} & \mbox{2.258*** ($p=0.004$)} \\
& \mbox{[1.263, 4.163]} & \mbox{[1.407, 4.411]} & \mbox{[-1.456, 1.191]} & \mbox{[0.735, 3.782]} \\
Personal care services & \mbox{2.699*** ($p=0.000$)} & \mbox{3.241*** ($p=0.000$)} & \mbox{-0.297 ($p=0.709$)} & \mbox{2.579*** ($p=0.001$)} \\
& \mbox{[1.212, 4.186]} & \mbox{[1.689, 4.793]} & \mbox{[-1.865, 1.272]} & \mbox{[1.013, 4.146]} \\
Nature parks & \mbox{2.524*** ($p=0.002$)} & \mbox{2.091** ($p=0.017$)} & \mbox{0.132 ($p=0.868$)} & \mbox{2.28*** ($p=0.009$)} \\
& \mbox{[0.939, 4.109]} & \mbox{[0.378, 3.805]} & \mbox{[-1.44, 1.704]} & \mbox{[0.574, 3.986]} \\
Museums & \mbox{4.207*** ($p=0.002$)} &  &  & \mbox{3.346** ($p=0.024$)} \\
& \mbox{[1.513, 6.902]} &  &  & \mbox{[0.448, 6.245]} \\
Film/video production & \mbox{3.653 ($p=0.155$)} & \mbox{1.811 ($p=0.421$)} &  &  \\
& \mbox{[-1.383, 8.689]} & \mbox{[-2.606, 6.228]} &  &  \\
Movie theaters & \mbox{3.306*** ($p=0.002$)} & \mbox{1.714* ($p=0.082$)} & \mbox{2.162** ($p=0.019$)} & \mbox{2.545** ($p=0.019$)} \\
& \mbox{[1.261, 5.351]} & \mbox{[-0.218, 3.646]} & \mbox{[0.355, 3.968]} & \mbox{[0.423, 4.667]} \\
Limited-service restaurants & \mbox{2.714*** ($p=0.000$)} & \mbox{2.846*** ($p=0.000$)} & \mbox{-0.409 ($p=0.554$)} & \mbox{2.711*** ($p=0.000$)} \\
& \mbox{[1.271, 4.157]} & \mbox{[1.346, 4.345]} & \mbox{[-1.776, 0.957]} & \mbox{[1.193, 4.228]} \\
Hobby/toy/game stores & \mbox{2.897*** ($p=0.007$)} & \mbox{3.328*** ($p=0.003$)} &  & \mbox{2.219* ($p=0.052$)} \\
& \mbox{[0.786, 5.007]} & \mbox{[1.103, 5.553]} &  & \mbox{[-0.016, 4.454]} \\
Historical sites & \mbox{0.851 ($p=0.741$)} & \mbox{2.18 ($p=0.335$)} &  &  \\
& \mbox{[-4.2, 5.902]} & \mbox{[-2.258, 6.618]} &  &  \\
Golf/country clubs & \mbox{5.104*** ($p=0.007$)} & \mbox{-3.495 ($p=0.124$)} &  & \mbox{-3.57 ($p=0.118$)} \\
& \mbox{[1.402, 8.806]} & \mbox{[-7.953, 0.963]} &  & \mbox{[-8.046, 0.906]} \\
Full-service restaurants & \mbox{2.595*** ($p=0.000$)} & \mbox{2.801*** ($p=0.000$)} & \mbox{-0.512 ($p=0.443$)} & \mbox{2.166*** ($p=0.004$)} \\
& \mbox{[1.201, 3.989]} & \mbox{[1.348, 4.255]} & \mbox{[-1.83, 0.806]} & \mbox{[0.695, 3.636]} \\
Fitness \& recreation centers & \mbox{2.452*** ($p=0.001$)} & \mbox{2.185*** ($p=0.005$)} & \mbox{-0.138 ($p=0.840$)} & \mbox{2.397*** ($p=0.003$)} \\
& \mbox{[0.974, 3.93]} & \mbox{[0.648, 3.723]} & \mbox{[-1.495, 1.218]} & \mbox{[0.833, 3.962]} \\
Drive-in theaters & \mbox{3.524** ($p=0.026$)} & \mbox{1.762 ($p=0.215$)} & \mbox{1.502 ($p=0.130$)} & \mbox{1.674 ($p=0.240$)} \\
& \mbox{[0.42, 6.629]} & \mbox{[-1.026, 4.55]} & \mbox{[-0.45, 3.455]} & \mbox{[-1.12, 4.469]} \\
Bars & \mbox{3.713*** ($p=0.000$)} & \mbox{2.485*** ($p=0.007$)} & \mbox{-0.034 ($p=0.973$)} & \mbox{3.122*** ($p=0.001$)} \\
& \mbox{[2.024, 5.401]} & \mbox{[0.69, 4.28]} & \mbox{[-2.028, 1.96]} & \mbox{[1.232, 5.011]} \\
Bowling centers & \mbox{3.415 ($p=0.193$)} & \mbox{1.707 ($p=0.460$)} &  &  \\
& \mbox{[-1.726, 8.557]} & \mbox{[-2.83, 6.244]} &  &  \\
Book stores & \mbox{2.292* ($p=0.050$)} & \mbox{2.579** ($p=0.032$)} &  & \mbox{2.609** ($p=0.045$)} \\
& \mbox{[-0.001, 4.584]} & \mbox{[0.226, 4.932]} &  & \mbox{[0.06, 5.158]} \\
Arcades & \mbox{3.422* ($p=0.069$)} & \mbox{3.314** ($p=0.047$)} & \mbox{-0.15 ($p=0.916$)} & \mbox{3.321** ($p=0.048$)} \\
& \mbox{[-0.273, 7.117]} & \mbox{[0.041, 6.588]} & \mbox{[-2.978, 2.678]} & \mbox{[0.029, 6.613]} \\
Specialty food stores & \mbox{2.735** ($p=0.015$)} & \mbox{2.807** ($p=0.013$)} & \mbox{0.044 ($p=0.964$)} & \mbox{2.678** ($p=0.025$)} \\
& \mbox{[0.536, 4.934]} & \mbox{[0.584, 5.029]} & \mbox{[-1.875, 1.963]} & \mbox{[0.332, 5.025]} \\
Distance To Nearest Evac Center Km Full & \mbox{-0.155 ($p=0.600$)} & \mbox{-0.011 ($p=0.973$)} & \mbox{0.113 ($p=0.692$)} & \mbox{0.416 ($p=0.236$)} \\
& \mbox{[-0.734, 0.425]} & \mbox{[-0.63, 0.609]} & \mbox{[-0.451, 0.677]} & \mbox{[-0.273, 1.104]} \\
\midrule
Observations &  &  &  &  \\
$R^2$ & 0.247 & 0.237 & 0.486 & 0.291 \\
\midrule
\multicolumn{5}{l}{\scriptsize Note: $^{*}p<0.1$; $^{**}p<0.05$; $^{***}p<0.01$.} \\
\end{longtable}
\end{minipage}
\end{adjustbox}
\end{center}
\bibliographystyle{plainnat}
\bibliography{sample}